\documentclass[prd,superscriptaddress,preprintnumbers,nofootinbib,11pt]{revtex4}
\pdfoutput=1
\usepackage{amsmath,amssymb,graphicx}
\graphicspath{{figs/}}
\usepackage{epsf,verbatim}
\usepackage{hyperref}
\usepackage{comment}
\usepackage{color}
\usepackage{slashed}
\usepackage{subfigure}
\usepackage[usenames,dvipsnames]{xcolor}
\usepackage{comment}
\usepackage{mdwlist, paralist}
\usepackage{rotating}
\usepackage{multirow}
\usepackage[utf8]{inputenc}
\usepackage[page,title,titletoc,header]{appendix}
\DeclareMathOperator{\sign}{sign}

\def\to{\rightarrow}

\begin{document}

\title{Two-Higgs-Doublet Model with Soft CP-violation Confronting Electric Dipole Moments and Colliders}

\preprint{NCTS-PH/2004}

\affiliation{Physics Division, National Center for Theoretical Sciences, Hsinchu, Taiwan 300}
\affiliation{Department of Physics, National Tsing Hua University, Hsinchu, Taiwan 300}
\affiliation{Division of Quantum Phases and Devices, School of Physics, Konkuk University, Seoul 143-701, Republic of Korea}
\affiliation{School of Physics and Astronomy, University of Southampton, Southampton, SO17 1BJ, United Kingdom}

\author{Kingman Cheung}
\thanks{cheung@phys.nthu.edu.tw}
\affiliation{Physics Division, National Center for Theoretical Sciences, Hsinchu, Taiwan 300}
\affiliation{Department of Physics, National Tsing Hua University, Hsinchu, Taiwan 300}
\affiliation{Division of Quantum Phases and Devices, School of Physics, Konkuk University, Seoul 143-701, Republic of Korea}

\author{Adil Jueid}
\thanks{adiljueid@konkuk.ac.kr}
\affiliation{Division of Quantum Phases and Devices, School of Physics, Konkuk University, Seoul 143-701, Republic of Korea}

\author{Ying-nan Mao}
\thanks{ynmao@cts.nthu.edu.tw}
\affiliation{Physics Division, National Center for Theoretical Sciences, Hsinchu, Taiwan 300}

\author{Stefano Moretti}
\thanks{s.moretti@soton.ac.uk}
\affiliation{School of Physics and Astronomy, University of Southampton, Southampton, SO17 1BJ, United Kingdom}

\begin{abstract}
We analyze CP-violating effects in both Electric Dipole Moment (EDM) measurements and future analyses at the Large Hadron Collider
(LHC) assuming a Two-Higgs-Doublet Model (2HDM) with ``soft'' CP-violation. Our analysis of EDMs and current LHC constraints shows
that, in the case of a 2HDM Type II and Type III, an $\mathcal{O}(0.1)$ CP-violating phase in the Yukawa interaction between
$H_1$ (the $125~\textrm{GeV}$ Higgs boson) and fermions is still allowed. For these scenarios, we study CP-violating effects in the
neutron EDM and $t\bar{t}H_1$ production at the LHC. Our analysis shows that such an $\mathcal{O}(0.1)$ CP-violating phase can be
easily confirmed or excluded by future neutron EDM tests, with LHC data providing a complementary cross-check.
\end{abstract}

\maketitle

\setcounter{equation}{0} \setcounter{footnote}{0}

\section{Introduction}
\label{sec:intro}
CP-violation was first discovered in 1964 through the
$K_L\rightarrow\pi\pi$ rare decay channel
\cite{Christenson:1964fg}. Later, more CP-violation effects were
discovered in the  K-, B-, and D-meson sectors
\cite{Tanabashi:2018oca,Aaij:2019kcg} and all the discovered effects
are consistent with the explanation given by the Kobayashi-Maskawa (KM)
mechanism \cite{Kobayashi:1973fv}.
However, the KM mechanism itself cannot generate a  large enough
matter-antimatter asymmetry in the Universe. Therefore,
new CP-violation sources beyond the KM mechanism are needed
to explain the latter
\cite{Cohen:1991iu,Cohen:1993nk,Morrissey:2012db}.

Experimentally, all the discovered effects of CP-violation till now have appeared in flavor physics measurements, yet they can also be tested through other methods. These can generally be divided into two different categories: (a) indirect tests, which can merely probe the existence of  CP-violation but cannot confirm the source(s) behind it; (b) direct tests, which can directly lead us to the actual CP-violation interaction(s).

For indirect tests, there is a typical example that one most often
uses, the Electric Dipole Moment (EDM) measurements
\cite{Khriplovich:1997ga,Pospelov:2005pr,Engel:2013lsa,Yamanaka:2017mef,Safronova:2017xyt,Chupp:2017rkp}. The
reason is that the EDM effective interaction of a fermion is
\begin{equation}
\mathcal{L}_{\textrm{EDM}}=-\frac{\textrm{i}}{2}d_f\bar{f}\sigma^{\mu\nu}\gamma^5fF_{\mu\nu},
\end{equation}
wherein $d_f$ is the EDM of such a fermion $f$, which
leads to P- and CP-violation simultaneously \cite{Pospelov:2005pr}. It is a pure quantum effect, i.e., emerging at
loop level and, in the Standard Model (SM), the electron and neutron
EDMs are predicted to be extremely small \cite{Pospelov:2005pr},
\begin{equation}
|d_e^{\textrm{SM}}|\sim10^{-38}~e\cdot\textrm{cm},\quad|d_n^{\textrm{SM}}|\sim10^{-32}~e\cdot\textrm{cm},
\end{equation}
because they are generated at four- or three-loop level, respectively. Thus, since the SM predictions for these are still far below the recent experimental limits \cite{Andreev:2018ayy,Baker:2006ts,Afach:2015sja,Abel:2020gbr}
\begin{equation}
|d_e|<1.1\times10^{-29}~e\cdot\textrm{cm},\quad|d_n|<1.8\times10^{-26}~e\cdot\textrm{cm},
\end{equation}
both given at $90\%$ Confidence Level (C.L.)\footnote{An earlier
result \cite{Baker:2006ts,Afach:2015sja} is
$|d_n|<3.0\times10^{-26}~e\cdot\textrm{cm}$ while a most recent
measurement by the nEDM group \cite{Abel:2020gbr} set a stricter
constraint $|d_n|<1.8\times10^{-26}~e\cdot\textrm{cm}$,  both at
$90\%$ C.L. At $95\%$ C.L., the latest constraint is then
$|d_n|<2.2\times10^{-26}~e\cdot\textrm{cm}$.}, these EDMs provide
a fertile ground to test the possibility of CP-violation due to  new
physics. In fact, in some Beyond the SM (BSM) scenarios, the EDMs of
the electron and neutron can be generated already at one- or two-loop
level, thus these constructs may be already strictly constrained or
excluded. In measurements of $d_e$ and $d_n$, though, even if we
discover that either or both EDMs are far above the SM predictions, we
cannot determine the exact interaction which constitutes such a
CP-violation.

For direct tests, there are several typical channels to test
CP-violation at colliders. For instance, measuring the final state
distributions from top pair
\cite{Schmidt:1992et,Mahlon:1995zn,BhupalDev:2007ftb,He:2014xla,Boudjema:2015nda,Buckley:2015vsa,AmorDosSantos:2017ayi,Azevedo:2017qiz,Bernreuther:2017cyi,Hagiwara:2017ban,Ma:2018ott,Cepeda:2019klc,Faroughy:2019ird,Cao:2020hhb}
or $\tau$ pair
\cite{Desch:2003rw,Berge:2008wi,Harnik:2013aja,Berge:2013jra,Berge:2014sra,Berge:2015nua,Askew:2015mda,Hagiwara:2016zqz,Jozefowicz:2016kvz}
production enables one to test CP-violating effects entering the
interactions of the fermions with one or more Higgs bosons. The
discovery of the $125~\textrm{GeV}$ Higgs boson
\cite{Aad:2012tfa,Chatrchyan:2012xdj,Aad:2015zhl} makes such
experiments feasible. Indeed, if more (pseudo)scalar or else new
vector states are discovered, one can also try to measure the
couplings amongst (old and new) scalars and vectors themselves to
probe CP-violation entirely from the bosonic sector
\cite{Li:2016zzh,Mao:2017hpp,Mao:2018kfn}. At high energy colliders,
the discovery of some CP-violation effects can lead us directly to the
CP-violating interaction(s), essentially because herein one can
produce final states that can be studied at the differential level,
thanks to the ability of the detectors to reconstruct their (at times,
full) kinematics, which can then be mapped to both cross section and
charge/spin asymmetry observables.

Theoretically, new CP-violation can appear in many new physics models,
for example, those with an extended Higgs sector
\cite{Bento:1991ez,Lee:1973iz,Lee:1974jb,Georgi:1978xz,Branco:2011iw,Weinberg:1976hu}. Among
these, we choose to deal here with the well known Two-Higgs-Doublet
Model (2HDM) \cite{Branco:2011iw}, which we use as a prototypical
source of CP-violation entertaining both direct and indirect tests of it.
In 2HDM, another Higgs doublet brings four additional
scalar degrees of freedom, and two of which are neutral. Thus there are
totally three neutral (pseudo)scalars. In the CP-conserving case,
two of them are scalars and one is a pseudoscalar. While for some parameter choices,
the pseudoscalar can mix with the scalar(s), and then CP-violation happens.
Specifically, the 2HDM with a $Z_2$ symmetry is used here, in
order to avoid large Flavor Changing Neutral Currents (FCNCs),
yet such a symmetry must be softly broken if one wants CP-violation
to arise in this scenario \cite{Branco:2011iw}.

The CP-violation effects in 2HDM were widely studied in recent years.
People carefully calculated the EDMs in 2HDM and discussed their further phenomenology \cite{Barr:1990vd,Chang:1990sf,Leigh:1990kf,Jung:2013hka,Abe:2013qla,Brod:2013cka,Cheung:2014oaa,Chen:2015gaa,Keus:2015hva,Fontes:2015mea,Altmannshofer:2015qra,Chen:2017com,Fontes:2017zfn,Chun:2019oix}. Usually,
the domain contributions come from the two-loop Barr-Zee type diagrams \cite{Barr:1990vd},
and complex Yukawa interactions provide the CP-violation sources. Especially, for the
electron EDM, a cancellation between different contributions may appear in some region \cite{Inoue:2014nva,Mao:2014oya,Bian:2014zka,Mao:2016jor,Bian:2016awe,Bian:2016zba,Egana-Ugrinovic:2018fpy,Fuyuto:2019svr}, thus a relative large CP-phase $\sim\mathcal{O}(0.1)$ in the Yukawa interactions will
still be allowed. Such a CP-phase is helpful to explain the matter-antimatter asymmetry in the Universe
\cite{Shu:2013uua,Bian:2014zka,Fuyuto:2019svr,Hou:2017kmo,Fuyuto:2017ewj}.
However, such cancellation usually does not appear in the same region for
neutron EDM, thus the future measurements on neutron EDM will be helpful
for testing the CP-phases, which will be discussed in details below.
The collider studies for CP-violation were usually performed model-independently,
but the results can be simply applied for 2HDM. We will therefore study
the effects of such a CP-violating 2HDM onto electron and neutron EDMs
as well as processes entering the Large Hadron Collider (LHC),
specifically, those involving the production of a top-antitop pair in
association with the $125$ GeV Higgs boson.

This paper is organized as follows. In \autoref{sec:model}, we review
the construction of the 2HDM with so-called ``soft'' CP-violation with
the four standard types of Yukawa interactions. Then, in
\autoref{sec:EDM},  we discuss the current constraints from electron and
neutron EDMs, show the reason why we eventually choose to pursue
phenomenologically only the 2HDM Type II and Type III for our collider
analysis and discuss the importance of future neutron EDM tests. In
\autoref{sec:const}, we discuss the current constraints from collider
experiments on these two realizations of a 2HDM. In \autoref{sec:phe},
we discuss LHC phenomenology studies on CP-violation effects in
the $t\bar{t}H_1$ associated production process. Finally, we summarize and
conclude in \autoref{sec:dnc}. There are also several appendices which
we use to collect technical details.

\section{Model Set-up}
\label{sec:model}
In this section, we briefly review the 2HDM with a softly broken $Z_2$
symmetry and how CP-violation arises in such a model. We mainly follow
the conventions in \cite{ElKaffas:2006gdt,Osland:2008aw,Arhrib:2010ju}.
The Lagrangian of the scalar sector can be written as
\begin{equation}
\mathcal{L}=\mathop{\sum}_{i=1,2}(D_{\mu}\phi_i)^{\dag}(D^{\mu}\phi_i)-V(\phi_1,\phi_2).
\end{equation}
Under a $Z_2$ transformation, we can have $\phi_1\rightarrow\phi_1$,
$\phi_2\rightarrow-\phi_2$, thus, in the scalar potential, all terms
must contain even numbers of $\phi_i$. However, if the $Z_2$ symmetry is softly
broken, a term $\propto\phi_1^{\dag}\phi_2$ is allowed, thus the
scalar potential becomes
\begin{eqnarray}
V(\phi_1,\phi_2)&=&-\frac{1}{2}\left[m^2_1\phi_1^{\dag}\phi_1+m^2_2\phi_2^{\dag}\phi_2+\left(m^2_{12}\phi_1^{\dag}\phi_2+\textrm{H.c.}\right)\right]
+\frac{1}{2}\left[\lambda_1\left(\phi_1^{\dag}\phi_1\right)^2+\lambda_2\left(\phi_2^{\dag}\phi_2\right)^2\right]\nonumber\\
&&+\lambda_3\left(\phi_1^{\dag}\phi_1\right)\left(\phi_2^{\dag}\phi_2\right)
+\lambda_4\left(\phi_1^{\dag}\phi_2\right)\left(\phi_2^{\dag}\phi_1\right)+\left[\frac{\lambda_5}{2}\left(\phi_1^{\dag}\phi_2\right)^2+\textrm{H.c.}\right].
\end{eqnarray}
Here $\phi_{1,2}$ are $\textrm{SU}(2)$ scalar doublets, which are defined as
\begin{equation}
\phi_1\equiv\left(\begin{array}{c}\varphi_1^+\\ \frac{v_1+\eta_1+\textrm{i}\chi_1}{\sqrt{2}}\end{array}\right),\quad
\phi_2\equiv\left(\begin{array}{c}\varphi_2^+\\ \frac{v_2+\eta_2+\textrm{i}\chi_2}{\sqrt{2}}\end{array}\right).
\end{equation}
The parameters $m^2_{1,2}$ and $\lambda_{1,2,3,4}$ must be real, while
$m^2_{12}$ and $\lambda_5$ can be complex. Further, $v_{1,2}$ are the Vacuum
Expectation Values (VEVs) of the scalar doublets with the relation
$\sqrt{|v_1|^2+|v_2|^2}=246~\textrm{GeV}$. The ratio $v_2/v_1$ may
also be complex\footnote{We can always fix $v_1$ real through gauge
  transformation and $v_2$ may be complex at the same time.}, and we
define $t_{\beta}\equiv|v_2/v_1|$ as usual\footnote{In this paper, we
  denote $s_{\alpha}\equiv\sin\alpha$, $c_{\alpha}\equiv\cos\alpha$,
  and $t_{\alpha}\equiv\tan\alpha$.}.

As was shown in \cite{Branco:2011iw}, CP-violation in the scalar
sector requires a nonzero $m^2_{12}$. For the three possible complex
parameters $m^2_{12}$, $\lambda_5$, and $v_2/v_1$, we can always
perform a field rotation to keep at least one of them real. In this
paper, we choose $v_2/v_1$ to be real (thus both $v_{1,2}$ are real) like in \cite{ElKaffas:2006gdt,Osland:2008aw,Arhrib:2010ju}, and
have the relation
\begin{equation}
\textrm{Im}(m^2_{12})=v_1v_2\textrm{Im}(\lambda_5)
\end{equation}
following the minimization conditions for the scalar potential. If $\textrm{Im}(m^2_{12})$ and $\textrm{Im}(\lambda_5)$ are non-zero, CP-violation occurs in the scalar sector.

We diagonalize the charged components as
\begin{equation}
\left(\begin{array}{c}G^+\\H^+\end{array}\right)=\left(\begin{array}{cc}c_{\beta}&s_{\beta}\\-s_{\beta}&c_{\beta}\end{array}\right)
\left(\begin{array}{c}\varphi_1^+\\ \varphi_2^+\end{array}\right),
\end{equation}
where $H^+$ is the charged Higgs boson and $G^+$ is the charged Goldstone. Similarly, for the CP-odd neutral components,
\begin{equation}
\left(\begin{array}{c}G^0\\A\end{array}\right)=\left(\begin{array}{cc}c_{\beta}&s_{\beta}\\-s_{\beta}&c_{\beta}\end{array}\right)
\left(\begin{array}{c}\chi_1\\ \chi_2\end{array}\right),
\end{equation}
where  $A$ is the physical CP-odd degree of freedom and $G^0$ is the neutral Goldstone. In the CP-conserved case, $A$ is a pseudoscalar boson while, in the CP-violating case, $A$ has further mixing with the CP-even degrees of freedom as
\begin{equation}
\left(\begin{array}{c}H_1\\H_2\\H_3\end{array}\right)=R\left(\begin{array}{c}\eta_1\\ \eta_2\\A\end{array}\right).
\end{equation}
Here $H_{1,2,3}$ are mass eigenstates and we choose $H_1$ as the lightest one with  mass $m_1=125~\textrm{GeV}$, so that it is the discovered SM-like Higgs boson. The rotation matrix $R$ can be parameterized as
\begin{equation}
\label{eq:R}
R=\left(\begin{array}{ccc}1&&\\&c_{\alpha_3}&s_{\alpha_3}\\&-s_{\alpha_3}&c_{\alpha_3}\end{array}\right)\left(\begin{array}{ccc}c_{\alpha_2}&&s_{\alpha_2}\\
&1&\\-s_{\alpha_2}&&c_{\alpha_2}\end{array}\right)
\left(\begin{array}{ccc}c_{\beta+\alpha_1}&s_{\beta+\alpha_1}&\\-s_{\beta+\alpha_1}&c_{\beta+\alpha_1}&\\&&1\end{array}\right).
\end{equation}
When $\alpha_{1,2}\rightarrow0$, $H_1$ becomes the SM Higgs boson. If $m_{1,2}$, $\alpha_{1,2,3}$ and $\beta$ are known, $m_3$ can be expressed as \cite{Arhrib:2010ju,Fontes:2015mea,Fontes:2017zfn}
\begin{equation}
\label{eq:m3}
m^2_3=\frac{(m^2_1-m^2_2s^2_{\alpha_3})c_{2\beta+\alpha_1}/c^2_{\alpha_3}-m^2_2s_{2\beta+\alpha_1}t_{\alpha_3}}
{c_{2\beta+\alpha_1}s_{\alpha_2}-s_{2\beta+\alpha_1}t_{\alpha_3}}.
\end{equation}

In the mass eigenstates, the couplings between neutral scalars and gauge bosons can be parameterized via
\begin{equation}
\mathcal{L}\supset\mathop{\sum}_{1\leq i\leq3}c_{V,i}H_i\left(\frac{2m^2_W}{v}W^{+,\mu}W^-_{\mu}+\frac{m^2_Z}{v}Z^{\mu}Z_{\mu}\right)
+\mathop{\sum}_{i=1}^3\frac{c_{ij}g}{2c_{\theta_W}}Z_{\mu}(H_i\partial^{\mu}H_j-H_j\partial^{\mu}H_i).
\end{equation}
The coefficients are then
\begin{eqnarray}
c_{V,1}&=&c_{23}=c_{\alpha_1}c_{\alpha_2},\\
c_{V,2}&=&-c_{13}=-c_{\alpha_3}s_{\alpha_1}-c_{\alpha_1}s_{\alpha_2}s_{\alpha_3},\\
c_{V,3}&=&c_{12}=s_{\alpha_1}s_{\alpha_3}-c_{\alpha_1}c_{\alpha_3}s_{\alpha_2}.
\end{eqnarray}

Next we turn to the Yukawa sector. Due to the $Z_2$ symmetry, a fermion bilinear can couple to only one scalar doublet, with the form $\bar{Q}_L\phi_iD_R$, $\bar{Q}_L\tilde{\phi}_iU_R$, or $\bar{L}_L\phi_i\ell_R$, thus it is helpful to avoid the FCNC problem \cite{Branco:2011iw}. Here $\tilde{\phi}_i\equiv\textrm{i}\sigma_2\phi^*_i$ and left-handed fermion doublets are defined as $Q_{i,L}\equiv(U_i,D_i)_L^T$ and $L_L\equiv(\nu_i,\ell_i)_L^T$, for the $i$-th generation. Since the scalar potential contains a $\phi_1\leftrightarrow\phi_2$ exchange symmetry, we can set the convention in which $\bar{Q}_LU_R$ always couple to $\phi_2$ so that  there are four standard types of Yukawa couplings \cite{Arhrib:2010ju,Branco:2011iw}:
\begin{equation}
\mathcal{L}\supset\left\{\begin{array}{cl}
-Y_U\bar{Q}_L\tilde{\phi}_2U_R-Y_D\bar{Q}_L\phi_2D_R-Y_{\ell}\bar{L}_L\phi_2\ell_R+\textrm{H.c.},&~(\textrm{Type I}),\\
-Y_U\bar{Q}_L\tilde{\phi}_2U_R-Y_D\bar{Q}_L\phi_1D_R-Y_{\ell}\bar{L}_L\phi_1\ell_R+\textrm{H.c.},&~(\textrm{Type II}),\\
-Y_U\bar{Q}_L\tilde{\phi}_2U_R-Y_D\bar{Q}_L\phi_2D_R-Y_{\ell}\bar{L}_L\phi_1\ell_R+\textrm{H.c.},&~(\textrm{Type III}),\\
-Y_U\bar{Q}_L\tilde{\phi}_2U_R-Y_D\bar{Q}_L\phi_1D_R-Y_{\ell}\bar{L}_L\phi_2\ell_R+\textrm{H.c.},&~(\textrm{Type IV}).
\end{array}\right.
\end{equation}
The fermion mass matrix is $M_f=Y_fvc_{\beta}/\sqrt{2}$ if the fermion couples to $\phi_1$ and $M_f=Y_fvs_{\beta}/\sqrt{2}$ if it couples to $\phi_2$.
We parameterize the Yukawa couplings of  mass eigenstates as
\begin{equation}
\label{eq:para}
\mathcal{L}\supset-\mathop{\sum}_{i,f}\frac{m_f}{v}\left(c_{f,i}H_i\bar{f}_Lf_R+\textrm{H.c.}\right).
\end{equation}
For CP-violating models, $c_{f,i}$ are complex numbers and we list them in \autoref{app:Yuk} for all  four types of Yukawa interactions.
In all these models, $\textrm{Im}(c_{f,1})\propto s_{\alpha_2}$, thus $\alpha_2$ is an important mixing angle which measures the
CP-violating phase in the Yukawa couplings of $H_1$.

\section{Current EDM Constraints and Future Tests}
\label{sec:EDM}
In this section, we analyze the EDM constraints of the electron and
neutron for the four types of 2HDM in some detail. The $b\rightarrow
s\gamma$ decay requires the charged Higgs mass to be
$m_{H^\pm}\gtrsim600~\textrm{GeV}$ for all the four types of Yukawa
couplings when $t_{\beta}\sim1$ \cite{Belle:2016ufb,Hermann:2012fc,Misiak:2015xwa,Misiak:2017bgg}.
If $t_{\beta}$ gets larger, the
constraints will become weaker for Type I and III Yukawa
couplings. The oblique parameters \cite{Peskin:1990zt,Peskin:1991sw}
will then favor the case $m_{H_{2,3}}\gtrsim500~\textrm{GeV}$
\cite{deBlas:2016ojx,Grimus:2007if,Grimus:2008nb,Haber:2010bw}\footnote{When
  $H_1$ is SM-like, the oblique parameter constraints are sensitive
  mainly to the mass splitting between the charged and neutral
  scalars. They are not sensitive to the mixing parameters in
  \autoref{eq:R}.}. With such choices for the scalar masses, the vacuum
stability condition favors
$\mu^2\equiv\textrm{Re}(m^2_{12})/s_{2\beta}\lesssim(450~\textrm{GeV})^2$
\cite{Osland:2008aw}. Notice that $\mu^2$ will modify the charged Higgs couplings
a little, but it is not numerically important to the EDM calculation, so
we fix it at $\mu^2=(450~\textrm{GeV})^2$ in the rest of
this work. More discussions  about the scalar couplings will appear in
\autoref{app:sca}.

An electron EDM measurement places a very strict constraint on the
complex Yukawa couplings in most models. As a rough estimation, if we
consider CP-violation only in the $125~\textrm{GeV}$ Higgs interaction
with the  top quark, the typical constraint is
$\textrm{arg}(c_{t,1})\lesssim10^{-3}$
\cite{Egana-Ugrinovic:2018fpy}. However, some models (including the 2HDM)
allow for  the accidental cancellation among various contributions,
so that  larger $\arg(c_{t,1})$ may still be allowed
\cite{Inoue:2014nva,Mao:2014oya,Bian:2014zka,Mao:2016jor,Bian:2016awe,Bian:2016zba,Egana-Ugrinovic:2018fpy,Fuyuto:2019svr}. In
such cases, neutron EDM constraints will also become important, as shown in the
analysis later in this section.

\subsection{Electron EDM}
A recent electron EDM measurement was performed using the
ThO molecule \cite{Andreev:2018ayy}. The exact constrained quantity is
\begin{equation}
|d_e^{\textrm{eff}}|\equiv|d_e+kC|<1.1\times10^{-29}~e\cdot\textrm{cm}.
\end{equation}
The second term measures the contribution from CP-violating electron-nucleon interactions via
\begin{equation}
\mathcal{L}\supset C\left(\bar{N}N\right)\left(\bar{e}\textrm{i}\gamma^5e\right),
\end{equation}
where the coefficient $C$ is almost the same for proton and
neutron. Here, $k\approx1.6\times10^{-15}~\textrm{GeV}^2
e\cdot\textrm{cm}$, which was obtained for ThO
\cite{Chupp:2014gka,Cesarotti:2018huy}, however, for most other materials
with heavy atoms, this quantity appears to be of the same order
\cite{Pospelov:2005pr,Yamanaka:2014mda}. The contribution from
electron-nucleon interactions is usually sub-leading, though it
can also become important.

\begin{figure}[h]
\caption{Typical Feynman diagrams contributing to the electron EDM in the 2HDM. The blue lines can be $\gamma$ or $Z$ while red lines are neutral Higgses $H_{1,2,3}$. Diagrams (a)-(g) will contribute to $d_e$ directly while diagrams (h)-(i) will contribute to the electron-nucleon interaction term.}\label{fig:eEDM}
\centering
\includegraphics[scale=0.7]{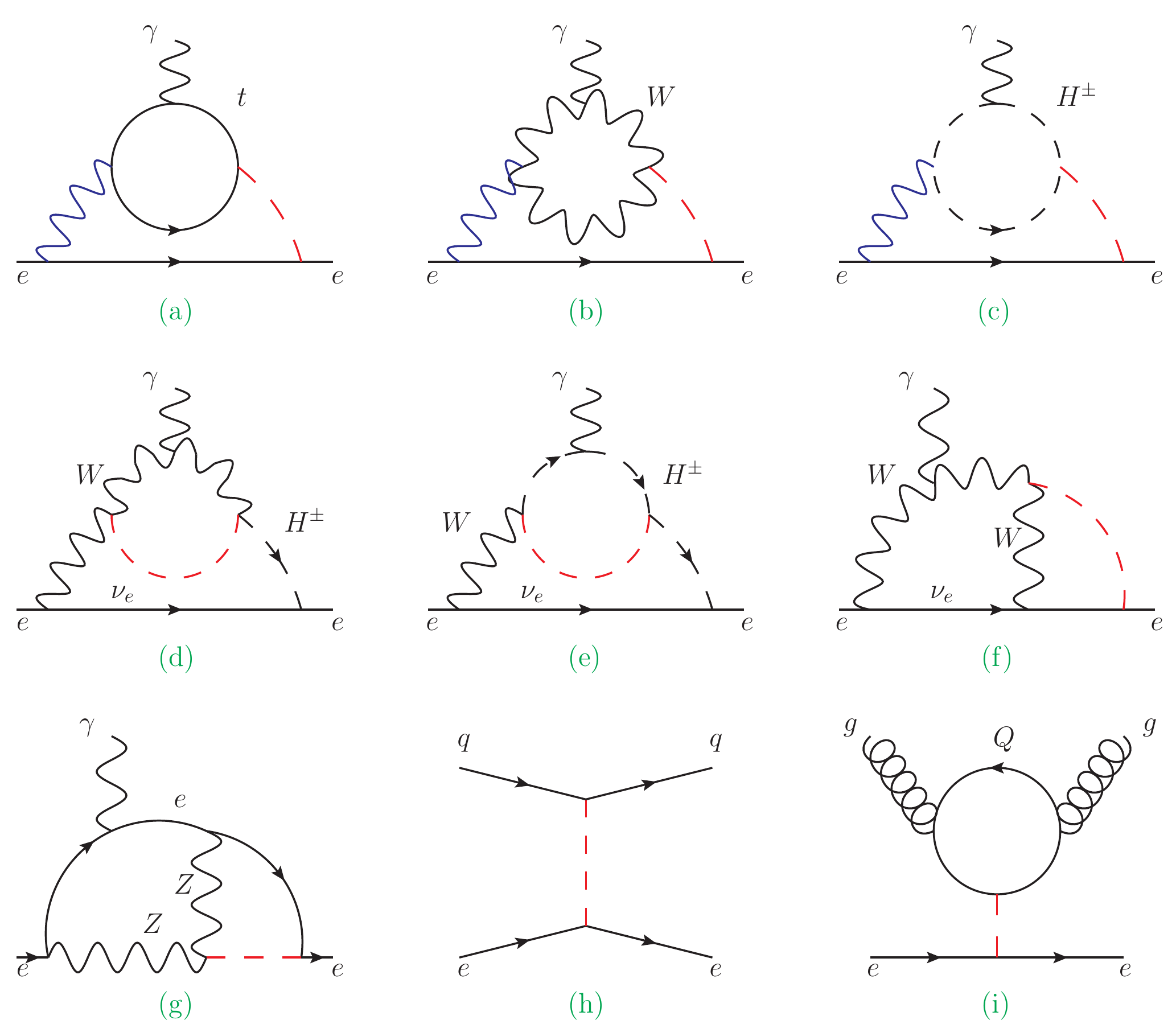}
\end{figure}

The typical Feynman diagrams contributing to the electron EDM in the 2HDM are
listed in \autoref{fig:eEDM}. Diagrams (a)-(e) are Barr-Zee type
diagrams \cite{Barr:1990vd} with the top quark $t$, $W^{\pm}$-boson, or
charged Higgs $H^{\pm}$ in the upper loop, while diagrams (f) and (g) are
non-Barr-Zee type. Such seven diagrams contribute directly to
$d_e$. Diagram (h) shows the contribution through the electron-quark
interaction, while diagram (i) shows the contribution through
the electron-gluon interaction. The contributions can be divided into
eight parts as summarized in \autoref{table:edm}.
\begin{table}[h]
\caption{Different contributions to the electron EDM and the corresponding Feynman diagrams.}\label{table:edm}\vspace*{0.25truecm}
\centering
\begin{tabular}{|c|c|c|c|}
\hline
&Diagram&Contribution&CP-violation vertex\\
\hline
$d_e^{t,\gamma/Z,H_i}$&(a)&Fermion (top) loop&$H_i\bar{e}e$, $H_i\bar{t}t$\\
\hline
$d_e^{W,\gamma/Z,H_i}$&(b)&$W$-loop&$H_i\bar{e}e$\\
\hline
$d_e^{H^{\pm},\gamma/Z,H_i}$&(c)&Charged Higgs $H^{\pm}$ loop&$H_i\bar{e}e$\\
\hline
$d_e^{W,H^{\pm},H_i}$&(d) and (e)&$W^{\pm}$-$H^{\pm}$-loop&$H^{\pm}W^{\mp}H_i$\\
\hline
$\delta d_e^W$&(f)&non-Barr-Zee $W$-loop&$H_i\bar{e}e$\\
\hline
$\delta d_e^Z$&(g)&non-Barr-Zee $Z$-loop&$H_i\bar{e}e$\\
\hline
$d_{e,q,i}^{\textrm{int}}$&(h)&Electron-quark interaction&$H_i\bar{e}e$\\
\hline
$d_{e,g,i}^{\textrm{int}}$&(i)&Electron-gluon interaction&$H_i\bar{e}e$\\
\hline
\end{tabular}
\end{table}

The analytical expressions in the Feynman-'t Hooft gauge are listed below.
For simplicity we denote
\begin{equation}
\delta_0\equiv\frac{\sqrt{2}m_eG_F\alpha_{\textrm{em}}}{(4\pi)^3}=3.1\times10^{-14}~\textrm{GeV}=6.1\times10^{-28}~e\cdot\textrm{cm}
\end{equation}
from now on. For the fermion-loop contribution in which the top quark is
dominant, we have \cite{Barr:1990vd,Chang:1990sf,Leigh:1990kf,Jung:2013hka,Abe:2013qla,Brod:2013cka,Inoue:2014nva,Cheung:2014oaa,Altmannshofer:2015qra,Chun:2019oix}
\begin{eqnarray}
\frac{d_e^{t,\gamma,H_i}}{e}&=&\frac{32}{3}\delta_0\left[f(z_{tH_i})\textrm{Re}\left(c_{t,i}\right)\textrm{Im}\left(c_{e,i}\right)
+g(z_{tH_i})\textrm{Re}\left(c_{e,i}\right)\textrm{Im}\left(c_{t,i}\right)\right],\\
\frac{d_e^{t,Z,H_i}}{e}&=&-\frac{\left(1-\frac{8s^2_{\theta_W}}{3}\right)(-1+4s^2_{\theta_W})}{s^2_{\theta_W}c^2_{\theta_W}}\delta_0\times\nonumber\\
&&\left[F(z_{tH_i},z_{tZ})\textrm{Re}\left(c_{t,i}\right)\textrm{Im}\left(c_{e,i}\right)
+G(z_{tH_i},z_{tZ})\textrm{Re}\left(c_{e,i}\right)\textrm{Im}\left(c_{t,i}\right)\right].
\end{eqnarray}
Here $z_{ij}\equiv m^2_i/m^2_j$ and $\theta_W$ is the weak
mixing angle with $s^2_{\theta_W}=0.23$. The loop integration
functions here and below are all listed in \autoref{app:loop}. For the
electron EDM calculation, the $Z$-mediated contribution is accidentally
suppressed by $-1/2+2s^2_{\theta_W}\sim-0.04$.  For the $W$-loop
contribution, we have
\cite{Barr:1990vd,Chang:1990sf,Leigh:1990kf,Abe:2013qla,Brod:2013cka,Cheung:2014oaa,Altmannshofer:2015qra,Chun:2019oix}
\begin{eqnarray}
\frac{d_e^{W,\gamma,H_i}}{e}&=&-\delta_0\Bigg[12f(z_{WH_i})+23g(z_{WH_i})+3h(z_{WH_i})\nonumber\\
&&+\frac{2}{z_{WH_i}}(f(z_{WH_i})-g(z_{WH_i}))\Bigg]c_{V,i}\textrm{Im}\left(c_{e,i}\right),\\
\frac{d_e^{W,Z,H_i}}{e}&=&\frac{-1+4s^2_{\theta_W}}{s^2_{\theta_W}}\delta_0\left[\frac{5-t^2_{\theta_W}}{2}F(z_{WH_i},c^2_{\theta_W})
+\frac{7-3t^2_{\theta_W}}{2}G(z_{WH_i},c^2_{\theta_W})+\frac{3}{4}h(z_{WH_i})\right.\nonumber\\
&&\left.+\frac{3}{4}g(z_{WH_i})+\frac{1-t^2_{\theta_W}}{4z_{WH_i}}\left(F(z_{WH_i},c^2_{\theta_W})-G(z_{WH_i},c^2_{\theta_W})\right)\right]c_{V,i}\textrm{Im}\left(c_{e,i}\right).
\end{eqnarray}
This contribution will cross zero around $m_i\sim500~\textrm{GeV}$
because of the cancellation between $W$ and Goldstone contributions
and, in the heavy $m_i$ limit, the pure Goldstone diagram has the
behavior $\sim\ln(m^2_i/m^2_W)$. The charged Higgs loop contributions
are \cite{Abe:2013qla}
\begin{eqnarray}
\frac{d_e^{H^{\pm},\gamma,H_i}}{e}&=&-\left(\frac{2\delta_0v^2}{m^2_{\pm}}\right)\left[f(z_{\pm,i})-g(z_{\pm,i})\right]c_{\pm,i}\textrm{Im}\left(c_{e,i}\right),\\
\frac{d_e^{H^{\pm},Z,H_i}}{e}&=&\frac{-1+4s^2_{\theta_W}}{s_{2\theta_W}t_{2\theta_W}}\left(\frac{2\delta_0v^2}{m^2_{\pm}}\right)
\left[F(z_{\pm,i},z_{\pm,Z})-G(z_{\pm,i},z_{\pm,Z})\right]c_{\pm,i}\textrm{Im}\left(c_{e,i}\right).
\end{eqnarray}
Hereafter,
``$\pm$'' is used to denote the charged Higgs boson while  $c_{\pm,i}$ is the coupling constant between the charged and neutral scalars entering via  $\mathcal{L}\supset-c_{\pm,i}vH_iH^+H^-$. The $W^{\pm}$-$H^{\pm}$ associated loop yields \cite{Abe:2013qla}
\begin{equation}
\label{eq:WHG}
\frac{d_e^{W,H^{\pm},H_i}}{e}=-\frac{\delta_0}{2s^2_{\theta_W}}\left[\frac{H^a_i(z_{WH_i})-H^a_i(z_{\pm,i})}{z_{\pm,W}-1}c_{V,i}
-\frac{H^b_i(z_{WH_i})-H^b_i(z_{\pm,i})}{z_{\pm,W}-1}c_{\pm,i}\right]\textrm{Im}\left(c_{e,i}\right).
\end{equation}
The first term corresponds to diagram (d) while the second term corresponds to diagram (e). The non-Barr-Zee type diagrams give  \cite{Leigh:1990kf,Altmannshofer:2015qra}\footnote{We have checked the results in \cite{Leigh:1990kf} and
  \cite{Altmannshofer:2015qra}. In the heavy $m_i$ limit, the loop
  functions should be logarithm enhanced as in \cite{Leigh:1990kf} (just like
  the pure Goldstone contribution in \cite{Chang:1990sf}). However,
  the results in \cite{Altmannshofer:2015qra} have improper power
  enhancement thus this behavior cannot be physical. So we used for validation  the
  result in \cite{Leigh:1990kf}.}
\begin{eqnarray}
\frac{d_e^{W,H_i}}{e}&=&-\frac{\delta_0}{s^2_{\theta_W}}\left(D_{W,i}^a+D_{W,i}^b+D_{W,i}^c+D_{W,i}^d+D_{W,i}^e\right)c_{V,i}\textrm{Im}\left(c_{e,i}\right),\\
\frac{d_e^{Z,H_i}}{e}&=&-4\delta_0t^2_{\theta_W}\left(D_{Z,i}^a+D_{Z,i}^b+D_{Z,i}^c\right)c_{V,i}\textrm{Im}\left(c_{e,i}\right).
\end{eqnarray}
The analytical expressions are too lengthy to present them here so that we list all of them in \autoref{app:loop}.
One-loop contributions to $d_e$ are highly suppressed by $m^3_e$ and thus we ignore them \cite{Mao:2016jor,Crivellin:2013wna}.
The interaction induced effective EDM terms  are \cite{Cesarotti:2018huy,Barr:1991yx,Dekens:2018bci,Cheung:2019bkw}
\begin{eqnarray}
d_{e,q,i}^{\textrm{int}}&=&\frac{\sqrt{2}m_eG_Fk}{m^2_i}\textrm{Im}(c_{e,i})\left[\textrm{Re}(c_{u,i})\left\langle m_u\bar{u}u\right\rangle+\textrm{Re}(c_{d,i})\left(\left\langle m_d\bar{d}d\right\rangle+\left\langle m_s\bar{s}s\right\rangle\right)\right],\\
d_{e,g,i}^{\textrm{int}}&=&-\frac{\sqrt{2}m_eG_Fk}{3m^2_i}\textrm{Im}(c_{e,i})\left[2\textrm{Re}(c_{u,i})+\textrm{Re}(c_{d,i})\right]
\left\langle\frac{\alpha_s}{4\pi}G_{\mu\nu}G^{\mu\nu}\right\rangle.
\end{eqnarray}
The nucleon matrix elements $\left\langle\mathcal{O}\right\rangle\equiv\left\langle N|\mathcal{O}|N\right\rangle$ and their values are similar for proton and neutron. Thus we choose the average values of proton and neutron considering three active quarks $(u,d,s)$ at the hadron scale $\sim1~\textrm{GeV}$ \cite{Ji:1994av,Cheng:2012qr,Hill:2014yxa,Yang:2015uis,Yanase:2018qqq,Yamanaka:2018uud,Cheung:2019bkw}, as listed in \autoref{table:ma}.
\begin{table}[h]
\caption{Nucleon matrix elements in the 3-flavor scheme at the hadron scale $\sim1~\textrm{GeV}$. The lattice calculation of quark matrix elements are a bit different from different groups as summarized in \cite{Yamanaka:2018uud}, and the results in this table were quoted from \cite{Yang:2015uis} which are closed to the averaged values. The gluon matrix element was derived based on \cite{Ji:1994av}.}\label{table:ma}\hspace*{0.25cm}
\centering
\begin{tabular}{|c|c|c|c|}
\hline
$\left\langle m_u\bar{u}u\right\rangle$&$\left\langle m_d\bar{d}d\right\rangle$&$\left\langle m_s\bar{s}s\right\rangle$&$\left\langle\frac{\alpha_s}{4\pi}G_{\mu\nu}G^{\mu\nu}\right\rangle$\\
\hline
$14.5~\textrm{MeV}$&$31.4~\textrm{MeV}$&$40.2~\textrm{MeV}$&$-183~\textrm{MeV}$\\
\hline
\end{tabular}
\end{table}
Summing all parts together, the effective electron EDM is
\begin{eqnarray}
d_e^{\textrm{eff}}&=&d_e+d_e^{\textrm{int}}\nonumber\\
&=&d_e^{t,\gamma,H_i}+d_e^{t,Z,H_i}+d_e^{W,\gamma,H_i}+d_e^{W,Z,H_i}+d_e^{H^{\pm},\gamma,H_i}+d_e^{H^{\pm},Z,H_i}+d_e^{W,H^{\pm},H_i}\nonumber\\
\label{eq:EDM}
&&+\delta d_e^W+\delta d_e^Z+d_{e,q,i}^{\textrm{int}}+d_{e,g,i}^{\textrm{int}}.
\end{eqnarray}
For each part above, $d_e^j\propto m_e$ thus it is suppressed by the  small
electron mass. We can extract $C_e^j\equiv d_e^j/(-m_e)$, which is
independent of the fermion mass. This coefficient is not useful in the
electron EDM calculation, but it will be helpful in order to map the
corresponding part into the quark EDM, which is important in the neutron EDM
calculation below.

\subsection{Neutron EDM}
The neutron EDM calculation is more complex as it
involves more contributions and QCD effects.
\begin{figure}[ht]
  \caption{Various contributions to the neutron EDM: quark EDM,
    quark Color EDM (CEDM) and Weinberg operator.}\label{fig:nEDM}
\centering
\includegraphics[scale=0.7]{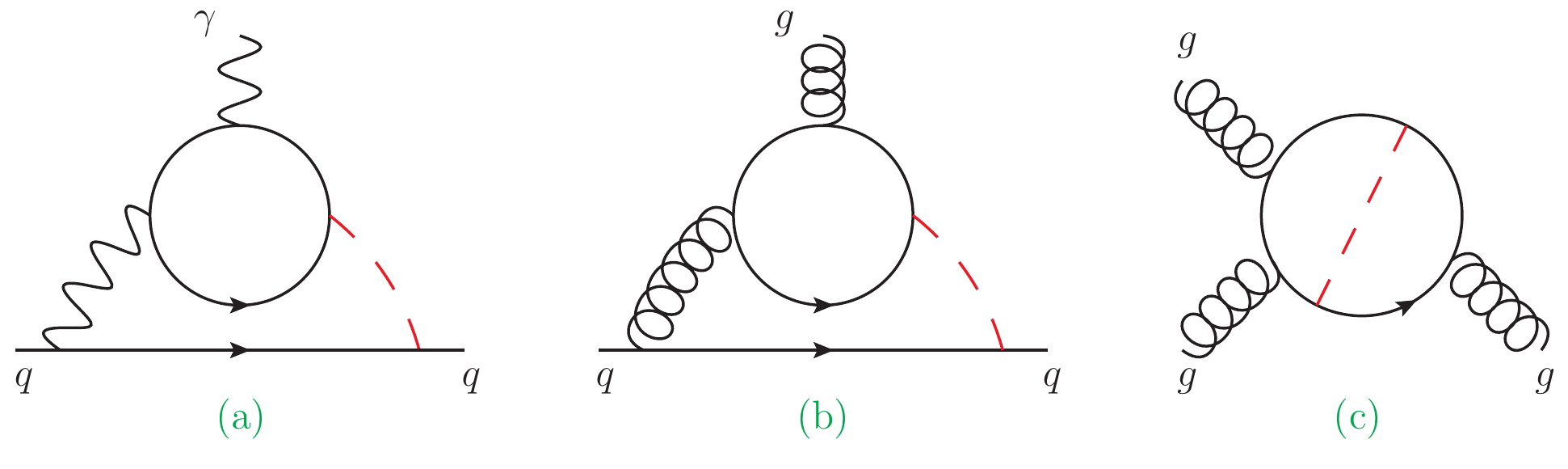}
\end{figure}
As shown in \autoref{fig:nEDM}, there are three types of operators
contributing to the neutron EDM, including the quark EDM operator
$\mathcal{O}_q$, quark CEDM  operator
$\mathcal{\tilde{O}}_q$ and Weinberg operator $\mathcal{O}_g$
\footnote{In 2HDM with $Z_2$ symmetry, there is no CP-violation entering
the $tbH^{\pm}$-vertex, thus we do not need to consider the diagram with
a charged Higgs boson inside the loop for Weinberg operator \cite{Jung:2013hka}.}. They
are chosen as follows \cite{Pospelov:2005pr,Brod:2013cka}:
\begin{eqnarray}
\mathcal{O}_q&=&-\frac{\textrm{i}}{2}eQ_qm_q\bar{q}\sigma^{\mu\nu}\gamma_5qF_{\mu\nu},\\
\mathcal{\tilde{O}}_q&=&-\frac{\textrm{i}}{2}g_sm_q\bar{q}\sigma^{\mu\nu}t^a\gamma_5qG^a_{\mu\nu},\\
\mathcal{O}_g&=&-\frac{1}{3}g_sf^{abc}G^{a}_{\mu\rho}G^{b,\rho}_{\nu}\tilde{G}^{c,\mu\nu},
\end{eqnarray}
where $g_s$ is the QCD coupling constant, $t^a$ is a generator of the QCD group and $f^{abc}$ denotes a QCD structure constant. At a scale $\mu$,
\begin{equation}
\mathcal{L}\supset \mathop{\sum}_{q=u,d}\left(C_q(\mu)\mathcal{O}_q(\mu)+\tilde{C}_q(\mu)\mathcal{\tilde{O}}_q(\mu)\right)+C_g(\mu)\mathcal{O}_g(\mu)
\end{equation}
and
\begin{equation}
\label{eq:dq}
d_q(\mu)/e\equiv Q_qm_q(\mu)C_q(\mu),\quad\tilde{d}_q(\mu)\equiv m_q(\mu)\tilde{C}_q(\mu).
\end{equation}
For convenience we also redefine $w(\mu)\equiv g_s(\mu)C_g(\mu)$. Notice that these EDMs   should be first calculated at the weak scale $\mu_W\sim m_t$.

The calculation methods of $C_u$ and $C_d$ are the same as those for $d_e$ through  diagrams (a)-(g) in \autoref{fig:eEDM}. For the quark EDM, we perform the calculation at the weak scale $\mu_W\approx m_t$ and list the results of the $C_q^j$ evaluation \cite{Abe:2013qla} as follows:
\begin{eqnarray}
\label{eq:repb}
\left(C_q^{t/W/H^{\pm},\gamma,H_i},~\delta C_q^Z\right)&=&\left(\bar{C}_e^{t/W/H^{\pm},\gamma,H_i},~\delta\bar{C}_e^Z\right),\\
C_d^{t/W/H^{\pm},Z,H_i}&=&\frac{-\frac{1}{2}+\frac{2s^2_{\theta_W}}{3}}{-\frac{1}{2}+2s^2_{\theta_W}}\cdot\frac{-1}{Q_d}\bar{C}_e^{t/W/H^{\pm},Z,H_i},\\
C_u^{t/W/H^{\pm},Z,H_i}&=&\frac{\frac{1}{2}-\frac{4s^2_{\theta_W}}{3}}{-\frac{1}{2}+2s^2_{\theta_W}}\cdot\frac{-1}{Q_u}\bar{C}_e^{t/W/H^{\pm},Z,H_i},\\
\left(C_u^{W,H^{\pm},H_i},~\delta C_u^W\right)&=&\left(\frac{1}{Q_u}\bar{C}_e^{W,H^{\pm},H_i},~\frac{1}{Q_u}\delta\bar{C}_e^W\right),\\
\left(C_d^{W,H^{\pm},H_i},~\delta C_d^W\right)&=&\left(\frac{-1}{Q_d}\bar{C}_e^{W,H^{\pm},H_i},~\frac{-1}{Q_d}\delta\bar{C}_e^W\right).
\label{eq:repe}
\end{eqnarray}
Here, each $\bar{C}_e^{j}$ means $C_e^j$ with a replacement $c_{e,i}\rightarrow c_{q,i}$ in the Yukawa couplings. The contributions including the $Z$
boson in the Bar-Zee diagram become important in the quark EDM calculation, because there is no accidental suppression like that in the electron EDM calculation.
For the CEDM terms, only Barr-Zee diagrams with a top loop contribute. The result at the weak scale $\mu_W\sim m_t$ is then \cite{Abe:2013qla,Brod:2013cka}
\begin{equation}
\label{eq:cedm}
\tilde{C}_q(\mu_W)=-\frac{2\sqrt{2}\alpha_s(\mu_W)G_F}{(4\pi)^3}
\mathop{\sum}_{i=1}^3\left[f(z_{tH_i})\textrm{Re}(c_{U,i})\textrm{Im}(c_{q,i})
+g(z_{tH_i})\textrm{Re}(c_{q,i})\textrm{Im}(c_{U,i})\right].
\end{equation}
The coefficient of the Weinberg operator at weak scale is \cite{Pospelov:2005pr,Brod:2013cka}
\begin{equation}
C_g(\mu_W)=\frac{\sqrt{2}\alpha_s(\mu_W)G_F}{4(4\pi)^3}\mathop{\sum}_{i=1}^3W(z_{tH_i})\textrm{Re}(c_{U,i})\textrm{Im}(c_{U,i}),
\end{equation}
and the loop integration $W(z)$ is listed in \autoref{app:loop}.

To calculate the EDM of the neutron, we must consider the Renormalization Group Equation (RGE) running effects to evolve these to the hadron scale $\mu_H\sim1~\textrm{GeV}$. The one-loop running gives \cite{Brod:2013cka,Weinberg:1989dx,Dicus:1989va,Braaten:1990gq,Degrassi:2005zd}
\begin{equation}
\left(\begin{array}{c}C_q(\mu_H)\\ \tilde{C}_q(\mu_H)\\C_g(\mu_H)\end{array}\right)=\left(\begin{array}{ccc}0.42&-0.38&-0.07\\&0.47&0.15\\&&0.20\end{array}\right)\left(\begin{array}{c}C_q(\mu_W)\\ \tilde{C}_q(\mu_W)\\C_g(\mu_W)\end{array}\right).
\end{equation}
There is no quark mass dependence in $C_q$ or $\tilde{C}_q$ and the evolution of $C_g$ is equivalent to $w(\mu_H)=0.41w(\mu_W)$. According to \autoref{eq:dq}, we only need the quark mass parameters at $\mu_H\sim1~\textrm{GeV}$ in the final calculation. The one-loop running mass effect is \cite{Tanabashi:2018oca}
\begin{equation}
m_q(1~\textrm{GeV})/m_q(2~\textrm{GeV})=1.38
\end{equation}
and, with the lattice results at $2~\textrm{GeV}$ \cite{Tanabashi:2018oca,Aoki:2016frl,Aoki:2019cca}, we have
\begin{equation}
m_u(1~\textrm{GeV})\simeq3.0~\textrm{MeV},\quad m_d(1~\textrm{GeV})\simeq6.5~\textrm{MeV}.
\end{equation}
The hadron scale estimation was performed based on QCD sum rules \cite{Pospelov:2005pr,Brod:2013cka,Hisano:2012sc,Demir:2002gg,Haisch:2019bml}\footnote{The contributions from quark EDM are consistent with a recent lattice calculation with better uncertainty \cite{Yamanaka:2018uud}, while the lattice calculation on the contributions from quark CEDM and Weinberg operator are still ongoing \cite{Yoon:2020soi}.}
\begin{equation}
\frac{d_n}{e}\simeq(22~\textrm{MeV})w(\mu_H)+0.65\frac{d_d(\mu_H)}{e}-0.16\frac{d_u(\mu_H)}{e}+0.48\tilde{d}_d(\mu_H)+0.24\tilde{d}_u(\mu_H),
\end{equation}
with an uncertainty of about $50\%$. The light quark condensation is chosen as $\langle\bar{q}q\rangle(1~\textrm{GeV})=-(254~\textrm{MeV})^3$ \cite{McNeile:2012xh}, which is a bit larger than that from \cite{Pospelov:2005pr,Hisano:2012sc}\footnote{Ref.~\cite{McNeile:2012xh} presents the lattice result $\langle\bar{q}q\rangle(2~\textrm{GeV})=-(283~\textrm{MeV})^3$ and also shows the RGE running effect as $d\langle m_q\bar{q}q\rangle(\mu)/d\ln\mu\propto m_q^4$, which is negligible for $u$ and $d$ quarks. Thus we have $\langle\bar{q}q\rangle(1~\textrm{GeV})/\langle\bar{q}q\rangle(2~\textrm{GeV})=m_q(2~\textrm{GeV})/m_q(1~\textrm{GeV})=0.73$.}. Combining all these results above, we have
\begin{eqnarray}
\frac{d_n}{e}&=&m_d(\mu_H)\left(0.27Q_dC_d(\mu_W)+0.31\tilde{C}_d(\mu_W)\right)\nonumber\\
\label{eq:dn}
&&+m_u(\mu_H)\left(-0.07Q_uC_u(\mu_W)+0.16\tilde{C}_u(\mu_W)\right)+\left(9.6~\textrm{MeV}\right)w(\mu_W).
\end{eqnarray}

\subsection{Numerical Analysis for the 2HDMs}
In this section we analyze the 2HDM with soft CP-violation, including
all the four types of Yukawa interactions. For the electron EDM, the Type I
and IV models give the same results, while the Type II and III models
give the same results\footnote{During the calculation of diagram (a) in \autoref{fig:eEDM},
we consider only top quark in the upper loop and ignore the small contributions from other
fermions. Such approximation is good enough when $t_{\beta}$ is not too large, for example,
$\lesssim10$. In cases with larger $t_{\beta}$, contributions from bottom quark or $\tau$
in the loop will become important.}. In the calculation of electron EDM, Diagrams (a) and (b) in
\autoref{fig:eEDM} usually contribute dominantly.

For Type I and IV models, numerical results show that there is no
cancellation among various contributions to the electron EDM, thus the
CP-violating phase is strictly constrained. The reason is that
in these two models, both $\sum_i\left(d_e^{t,\gamma/Z,H_i}\right)$
and $\sum_i\left(d_e^{W,\gamma/Z,H_i}\right)$
have the behavior $\propto-s_{2\alpha_2}/t_{\beta}$, and thus
$d_e$ cannot go close to zero when keeping the CP-violation phases.
This behavior is consistent with the results in which only the contribution
from $H_1$ is considered \cite{Egana-Ugrinovic:2018fpy},
because in most cases, the $H_1$ contribution is dominant comparing with the heavy scalars
if $t_{\beta}$ is not too large, such as $\lesssim10$.

We take $m_{2,3}\sim500~\textrm{GeV}$ and $m_{\pm}\sim600~\textrm{GeV}$ as a
benchmark point and find
\begin{equation}
d_e^{\textrm{I},\textrm{IV}}\simeq-6.7\times10^{-27}\left(\frac{s_{2\alpha_2}}{t_{\beta}}\right)~e\cdot\textrm{cm}
\end{equation}
in the region $t_{\beta}\lesssim10$ and $s_{2\alpha_2}\ll1$. This result is not sensitive to
$\alpha_{1,3}$ and gives\footnote{$\alpha_2\simeq\pi/2$ is not allowed by other experiments,
thus we consider only the case $\alpha_2\ll1$.}
$|s_{\alpha_2}/t_{\beta}|\lesssim8.2\times10^{-4}$, which means the
CP-phase $|\arg(c_{f,1})|\lesssim8.2\times10^{-4}$ for
$f=\ell_i,U_i$. This is extremely small and would not be able
to produce interesting CP-violating effects, so in the rest of this work, we do
not discuss further on these two 2HDM realizations.

For Type II and III models, in contrast, numerical results
show significant
cancellation behavior for some parameter regions in the electron EDM
calculation and thus $\alpha_2$ is allowed to reach
$\mathcal{O}(0.1)$. The reason is that different terms
depend differently on $t_{\beta}$. As shown above, we can divide $d_e^{t,\gamma/Z,H_i}$ into two
parts as $d_{e,(a)}^{t,\gamma/Z,H_i}\propto\textrm{Re}(c_{t,i})\textrm{Im}(c_{e,i})$ and $d_{e,(b)}^{t,\gamma/Z,H_i}\propto\textrm{Re}(c_{e,i})\textrm{Im}(c_{t,i})$.
Then based on the behavior $\sum_i\left(d_e^{W,\gamma/Z,H_i}+d_{e,(a)}^{t,\gamma/Z,H_i}\right)
\propto s_{2\alpha_2}t_{\beta}$ and $\sum_i\left(d_{e,(b)}^{t,\gamma/Z,H_i}\right)\propto-s_{2\alpha_2}/t_{\beta}$,
we confirm that there is always some region in which different contributions to electron EDM almost cancel with
each other, and thus a large $|\alpha_2|\sim\mathcal{O}(0.1)$ can be allowed. Other contributions may shift mildly
the exact location where cancellation happens, but do not modify the cancellation behavior.
For these two models, we can discuss two different
scenarios: (a) the heavy neutral scalars $H_{2,3}$ are
close in mass and $\alpha_3$ can be changed in a wide range; (b) $H_2$
and $H_3$ have large mass splitting, and thus $\alpha_3$ must be close
to $0$ or $\pi/2$.

\begin{figure}[h]
\caption{Cancellation behavior between $\beta$ and $\alpha_1$ in
  scenario (a) of a Type II and III 2HDM. As an example, the fixed parameters
  are listed in \autoref{table:bmpa}. The solid lines are the
  boundaries with $|d_e|=1.1\times10^{-29}~e\cdot\textrm{cm}$ and the
  regions between solid lines are allowed by the ACME experiment while the
  dashed lines mean $d_e=0$. In the left plot, we choose a Type II
  model. The blue, orange, and red lines are shown for
  $\alpha_2=0.05,0.1,0.15$, respectively. In the right plot, we fix
  $\alpha_2=0.1$ and show the comparison between the Type II and Type III
  models. The orange lines are for the Type II model while the cyan lines
  are for the Type III model.}\label{fig:cancel} \centering
\includegraphics[scale=0.55]{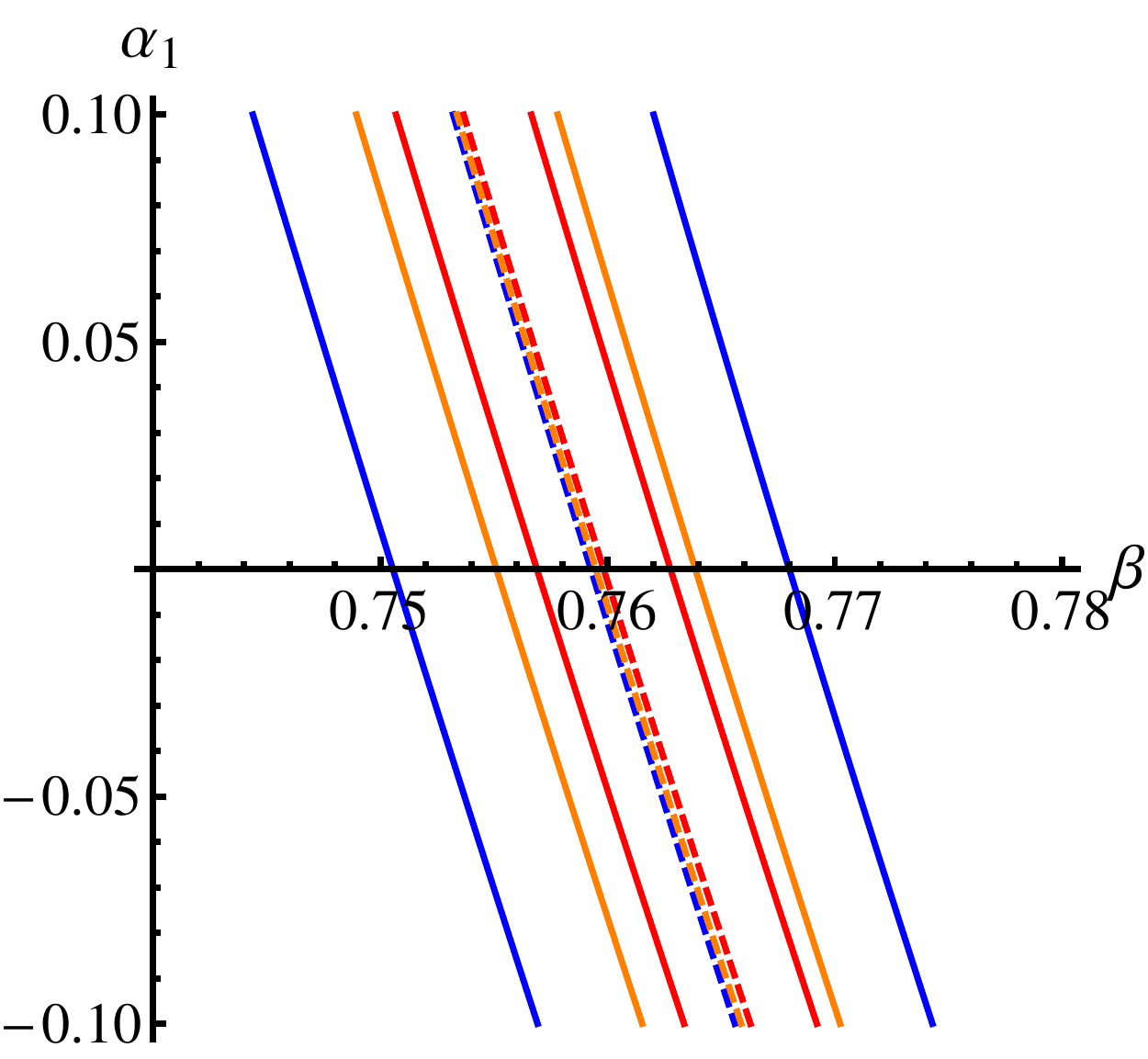}
\includegraphics[scale=0.55]{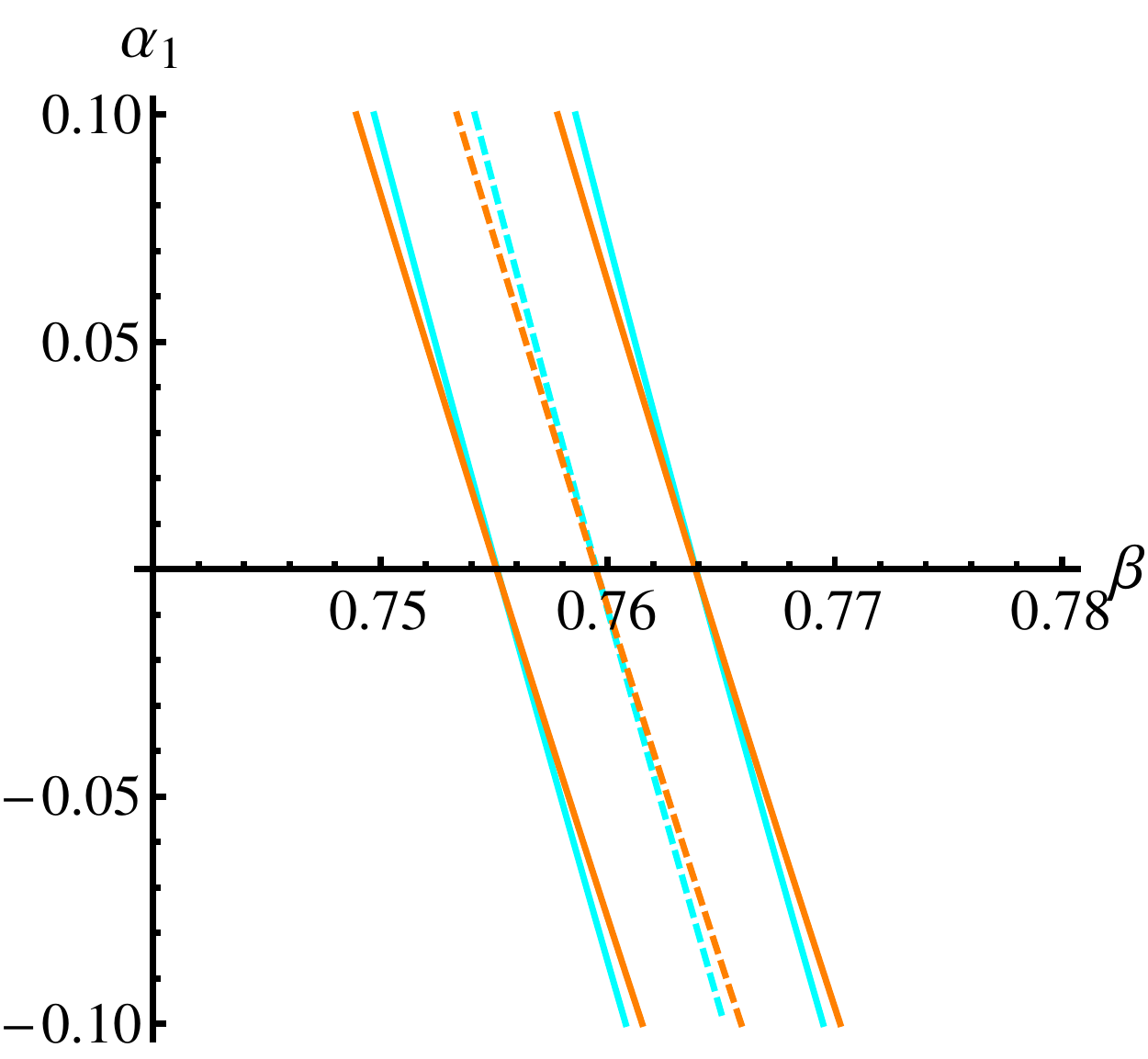}
\end{figure}
\begin{table}[h]
\caption{Fixed parameters of Scenario (a) to discuss the cancellation behavior. With these parameters and $\beta,\alpha_{1,2}$,
we can calculate $m_3$ through \autoref{eq:m3}, and calculate the couplings through the equations
in \autoref{app:Yuk} and \autoref{app:sca}.}\label{table:bmpa}\hspace*{0.25cm}
\centering
\begin{tabular}{|c|c|c|c|c|}
\hline
$m_1$&$m_2$&$m_{\pm}$&$\mu^2$&$\alpha_3$\\
\hline
$125~\textrm{GeV}$&$500~\textrm{GeV}$&$600~\textrm{GeV}$&$(450~\textrm{GeV})^2$&$0.8$\\
\hline
\end{tabular}
\end{table}

\begin{figure}[ht]
\caption{Mass dependence in the cancellation region in the Type II model. Choosing $m_{\pm}-m_2=100~\textrm{GeV}$, $\alpha_3=0.8$, $\alpha_2=0.1$,
 $\alpha_1=0$, and $\mu^2=(450~\textrm{GeV})^2$ as an example, the black line shows the value of $\beta$ satisfying $d_e=0$ while the dark blue region satisfies $|d_e|<1.1\times10^{-29}~e\cdot\textrm{cm}$, which is allowed by the ACME experiment at $90\%$ C.L. If we set $|\alpha_1|<0.1$, the light blue region is allowed. Results in the Type III model are almost the same and thus we do not show these.}\label{fig:massdp}
\centering
\includegraphics[scale=0.7]{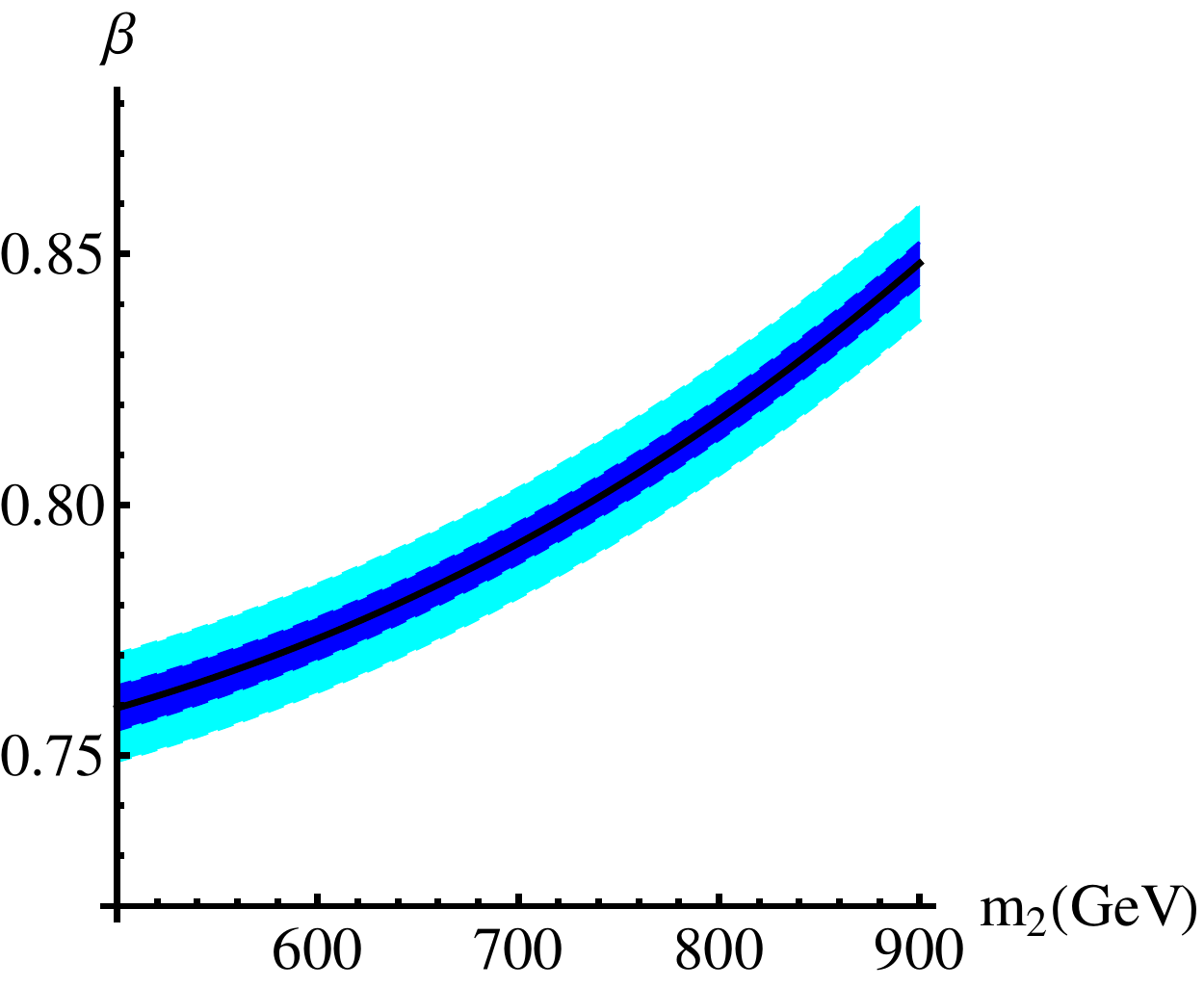}
\end{figure}

We first consider Scenario (a). In this scenario, the cancellation behavior is
not sensitive to $\alpha_3$ in a wide region (for example, $0.2\lesssim\alpha_3\lesssim1.4$),
because the $H_{2,3}$ are close in mass and thus the dependence on $\alpha_3$
from $H_2$ and $H_3$ contributions almost cancel with each other.
Thus, we choose $\alpha_3=0.8$, $m_2=500~\textrm{GeV}$ and $m_{\pm}=600~\textrm{GeV}$
as a benchmark point. We focus on the lower $t_{\beta}$ region, which can generate relative large
CP-violation phase in $ht\bar{t}$ vertex\footnote{As pointed in \cite{Inoue:2014nva},
another cancellation region is around $t_{\beta}\simeq(10-20)$. However, $\arg(c_{t,1})\propto t^{-1}_{\beta}$
thus it is suppressed and difficult to be tested at colliders in this scenario. So we will not discuss the
large $t_{\beta}$ scenario in this paper.}.
In the region with $\alpha_1\sim0$ and $t_{\beta}\sim1$, we have
\begin{equation}
\label{eq:de2}
d_e^{\textrm{II},\textrm{III}}\simeq3.4\times10^{-27}s_{2\alpha2}
\left(t_{\beta}-\frac{0.904}{t_{\beta}}\right)~e\cdot\textrm{cm},
\end{equation}
which means the cancellation appears around $t_{\beta}\simeq0.95$, or equivalently, $\beta\simeq0.76$.
Different from the Type I and IV models, a large mixing angle $|\alpha_2|\sim\mathcal{O}(0.1)$
(and hence a CP-phase $|\arg(c_{f,1})|\sim\mathcal{O}(0.1)$ for
$f=\ell_i,U_i$) can be allowed due to the cancellation.
We show the cancellation behavior of the electron EDM in the $\beta$-$\alpha_1$
plane in \autoref{fig:cancel} for Type II and III models.
The electron EDM sets a strict constraint which behaves as a strong
correlation between $\beta$ and $\alpha_1$. Numerical analysis shows
that, with fixed heavy scalar masses, the location where the cancellation
happens is not sensitive to $\alpha_2$, which is consistent with the
result in \autoref{eq:de2}, but the width of the allowed
region is almost proportional to $1/s_{2\alpha_2}$. We show this
behavior for the Type II model in the left plot of \autoref{fig:cancel},
for $\alpha_2=0.05,0.1,0.15$, using blue, orange and red lines,
respectively. The cancellation behavior in the Type III model is similar
to that in the Type II model, because the Barr-Zee diagram with a bottom
quark loop is negligible and thus the only difference comes from the
electron-nucleon interaction part. In the right plot of
\autoref{fig:cancel}, with fixed $\alpha_2=0.1$, we show the
comparison results between the Type II model (orange lines) and Type III
model (cyan lines), finding that they are almost the same.
When $m_{2,3}$ increases, the location where the cancellation happens will
also change slowly and we show the corresponding results in
\autoref{fig:massdp}.
When $m_2$ increases from $500~\textrm{GeV}$ to
$900~\textrm{GeV}$, the cancellation location also moves slowly
from about $\beta\simeq0.76$ to $\beta\simeq0.84$. The width of the allowed region is almost
independent of the heavy scalar masses, as it is sensitive only to
$\alpha_2$. The cancellation behavior leads to the conclusion that there is
always a narrow region which is allowed by the electron EDM measurement,
thus we cannot set a definite constraint on the CP-violation mixing angle
$\alpha_2$ only through the electron EDM, such as in the ACME experiment.

In contrast, the neutron EDM calculation does not
involve such a cancellation behavior in the same region
as the electron one, thus it can be used to set
direct constraints on the CP-violating mixing angle $\alpha_2$. In the
parameter region allowed by the electron EDM constraints, the CEDM of the $d$
quark contributes dominantly to the neutron EDM. Numerical analysis shows
that the neutron EDM $d_n\propto s_{2\alpha_2}$ and it is not
sensitive to $\alpha_{1,3}$.
\begin{figure}[h]
\caption{In the left plot, we show the $d_n/s_{2\alpha_2}$ dependence on
  $m_2$ in the Type II (blue) and Type III (orange) models using the central
  value estimation of \autoref{eq:dn} in the parameter region allowed by
  ACME experiment. We choose $\alpha_1=0$ and $\alpha_3=0.8$ as an
  example, but the modification due to these two angles is less than
  percent level, which is far smaller than the uncertainty in the
  theoretical estimation (about $50\%$ level). In the right plot, we
  show the limit on $\alpha_2$ in the Type II (blue) and III (orange)
  models. The solid lines are obtained through the estimation of
  central value and the dashed lines are the boundaries considering the
  theoretical uncertainty. If theoretical uncertainties are taken into
  account, we cannot set any limit on $\alpha_2$ in the Type III model
  through neutron EDM measurements.}\label{fig:nedmc} \centering
\includegraphics[scale=0.6]{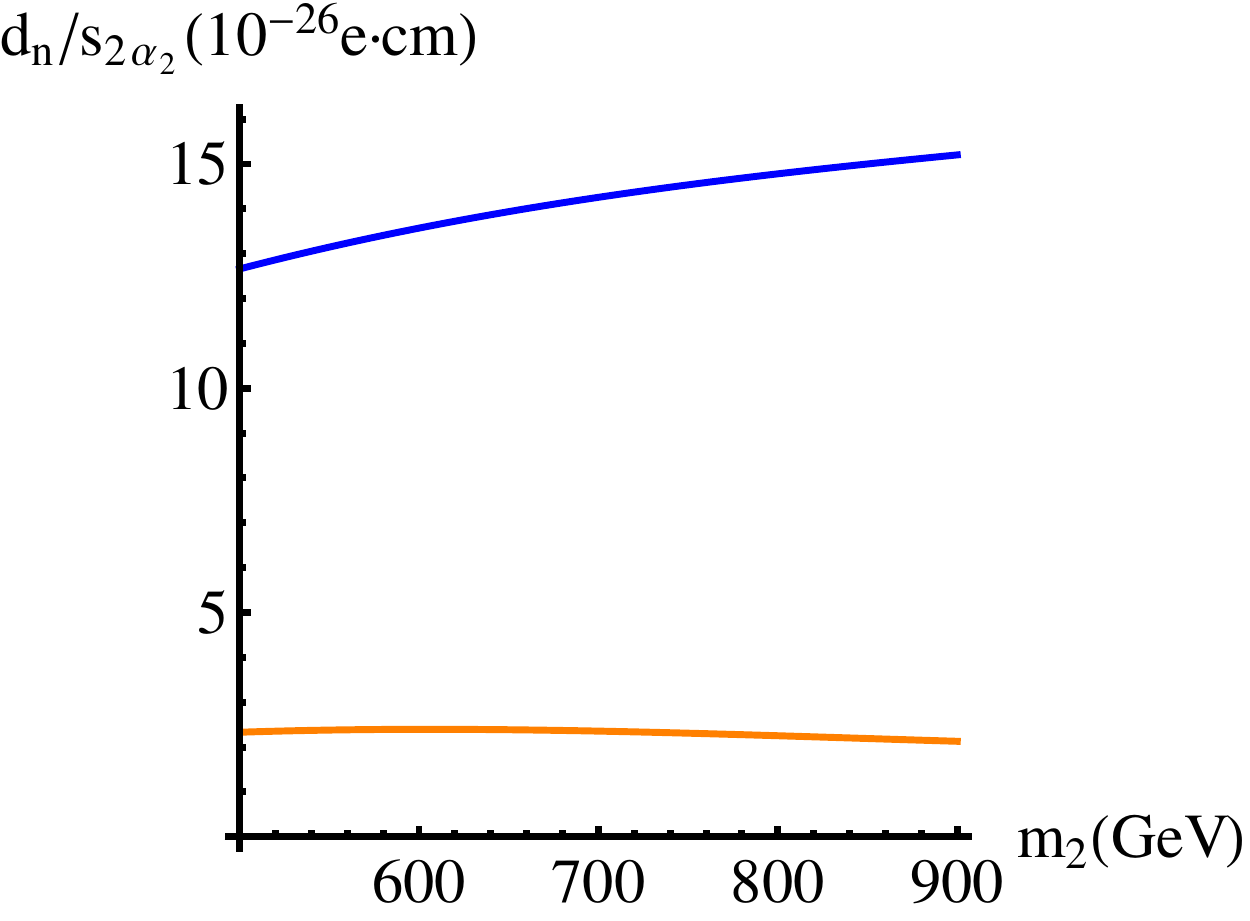}\includegraphics[scale=0.54]{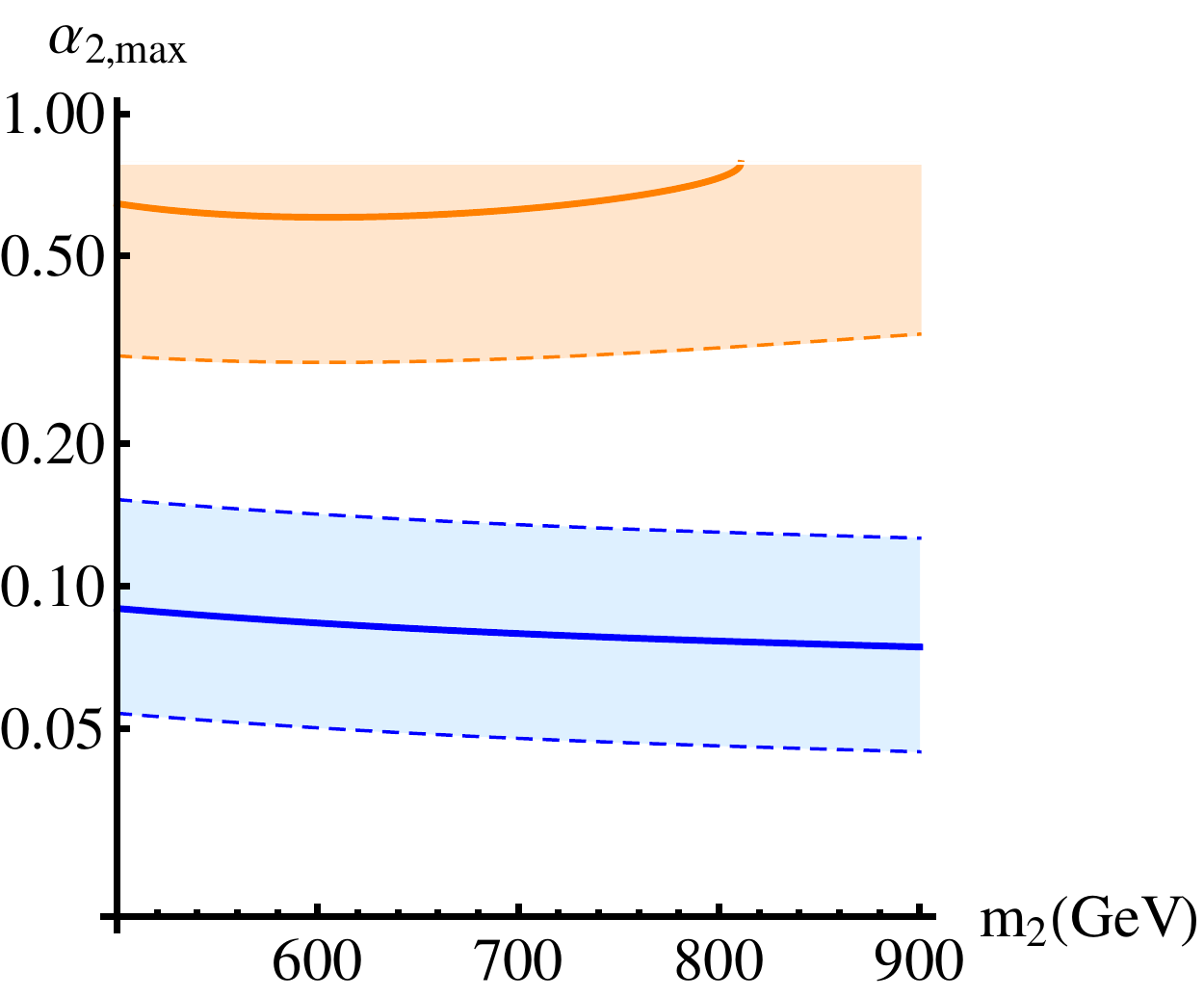}
\end{figure}
We calculate its dependence on $m_2$ in the Type II and III models using
the central value estimated in \autoref{eq:dn} and show the results
in the left plot of \autoref{fig:nedmc}. In the Type II model, $\alpha_2$
is constrained by the neutron EDM
(the latest result is $|d_n|<2.2\times10^{-26}~e\cdot\textrm{cm}$ at $95\%$
C.L. \cite{Abel:2020gbr}). Using the central value
estimation in \autoref{eq:dn}, $|\alpha_2|\lesssim(0.073-0.088)$ if $m_2$ changes in the
range $(500-900)~\textrm{GeV}$, as shown in the right plot of
\autoref{fig:nedmc}. Considering the uncertainty in the neutron EDM
estimation \cite{Hisano:2012sc}, a larger $|\alpha_2|\sim0.15$ can also be allowed\footnote{As
discussed above, here we do not consider the region $\alpha_2$ close to $\pi/2$ since it
corresponds to the case in which $H_1$ is pseudoscalar component dominated,
which can be excluded by other experiments. See more details in the next section.}. In the Type
III model, there is almost no constraint on $\alpha_2$ from the neutron
EDM\footnote{If we consider only the central value of the neutron EDM
  estimation \autoref{eq:dn}, the constraint is about
  $|s_{2\alpha_2}|\lesssim0.9$, meaning that $\alpha_{2,\textrm{max}}$ is
  already close to $\pi/4$. However, if the large theoretical
  uncertainty in the neutron EDM estimation is also taken into account, we
  cannot exclude any value for $|s_{2\alpha_2}|\leq1$, which means no
  constraint on $|\alpha_2|$ can be set in the Type III model.}. That's because
in Type III model, $\textrm{Re}(c_{u,i})\textrm{Im}(c_{d,i})=-\textrm{Re}(c_{d,i})\textrm{Im}(c_{u,i})$,
which is different from the relation in Type II model.
It leads to an accidental partial cancellation between the two terms (see
\autoref{eq:cedm}) in the $d$ quark CEDM contribution, which dominates the
neutron EDM calculation.

\begin{figure}[h]
\caption{Similar to Scenario (a), the electron EDM sets a strict constraint
  which behaves as a strong correlation between $\beta$ and
  $\alpha_1$. As an example, the fixed parameters are listed in \autoref{table:bmpb}.
  We show the cancellation behavior of the Type II model in
  the upper two plots and present the comparison between the Type II and III
  models in the lower two plots. The color notation is the same as that in
  \autoref{fig:cancel}. The left two plots correspond to the
  case $\alpha_3^+$, while the right two plots correspond to the case $\alpha_3^-$.
  We approximately have $\alpha_3^+\simeq\pi/2-1.5\times10^{-2}\alpha_2$ and
  $\alpha_3^-\simeq-0.52\alpha_2$.}\label{fig:cancelsplit}
\centering
\includegraphics[scale=0.55]{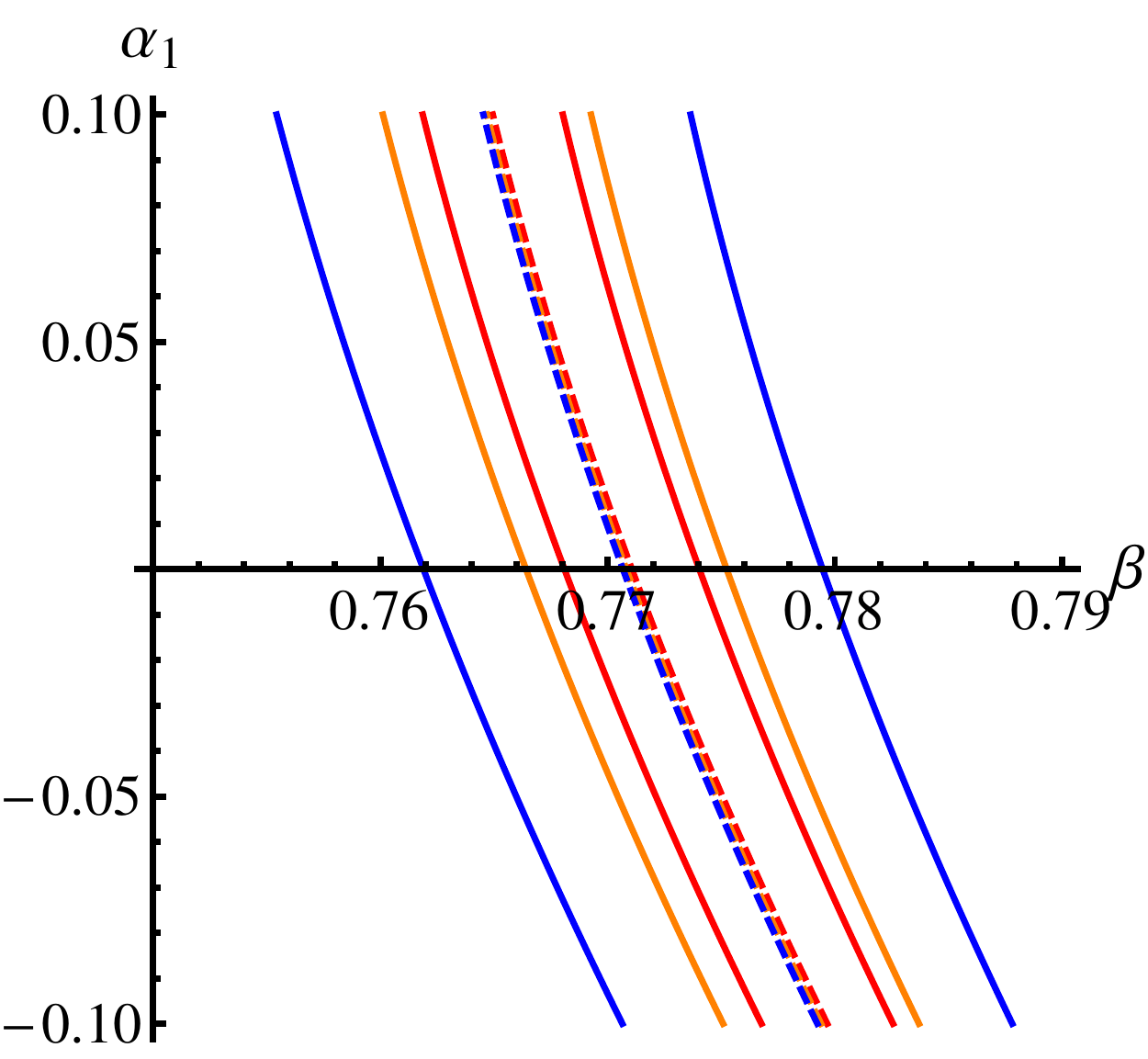}\includegraphics[scale=0.55]{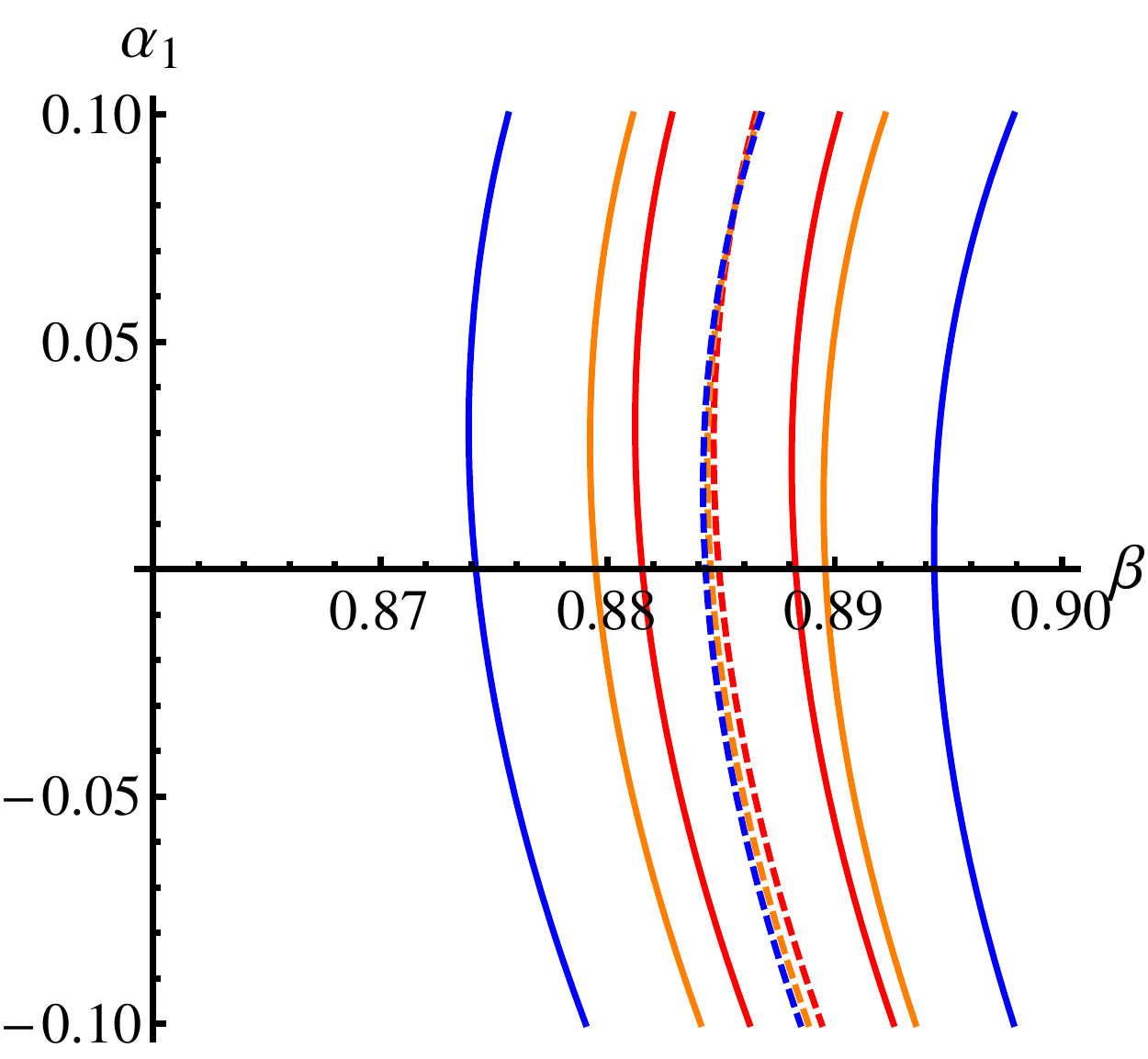}
\includegraphics[scale=0.55]{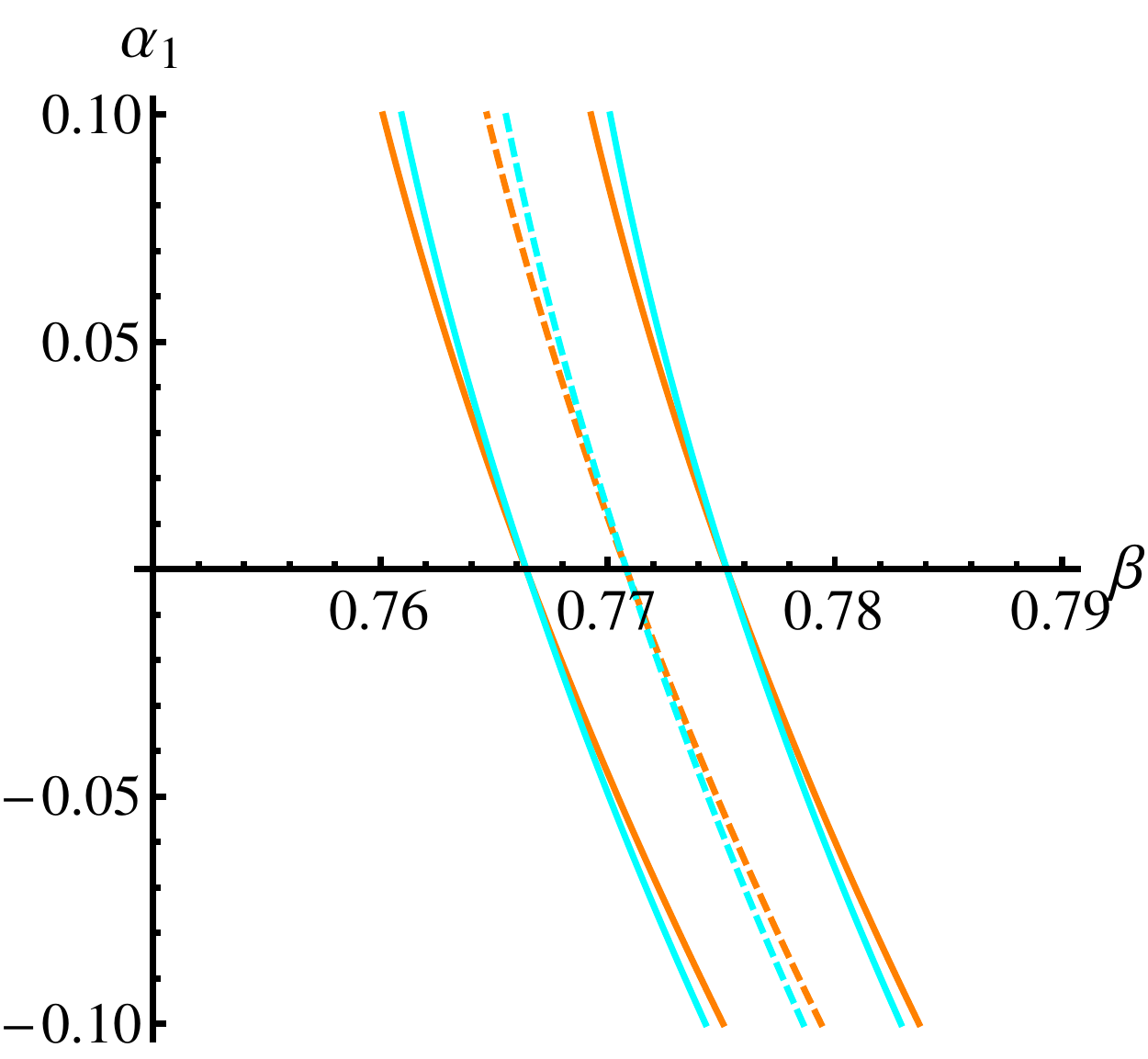}\includegraphics[scale=0.55]{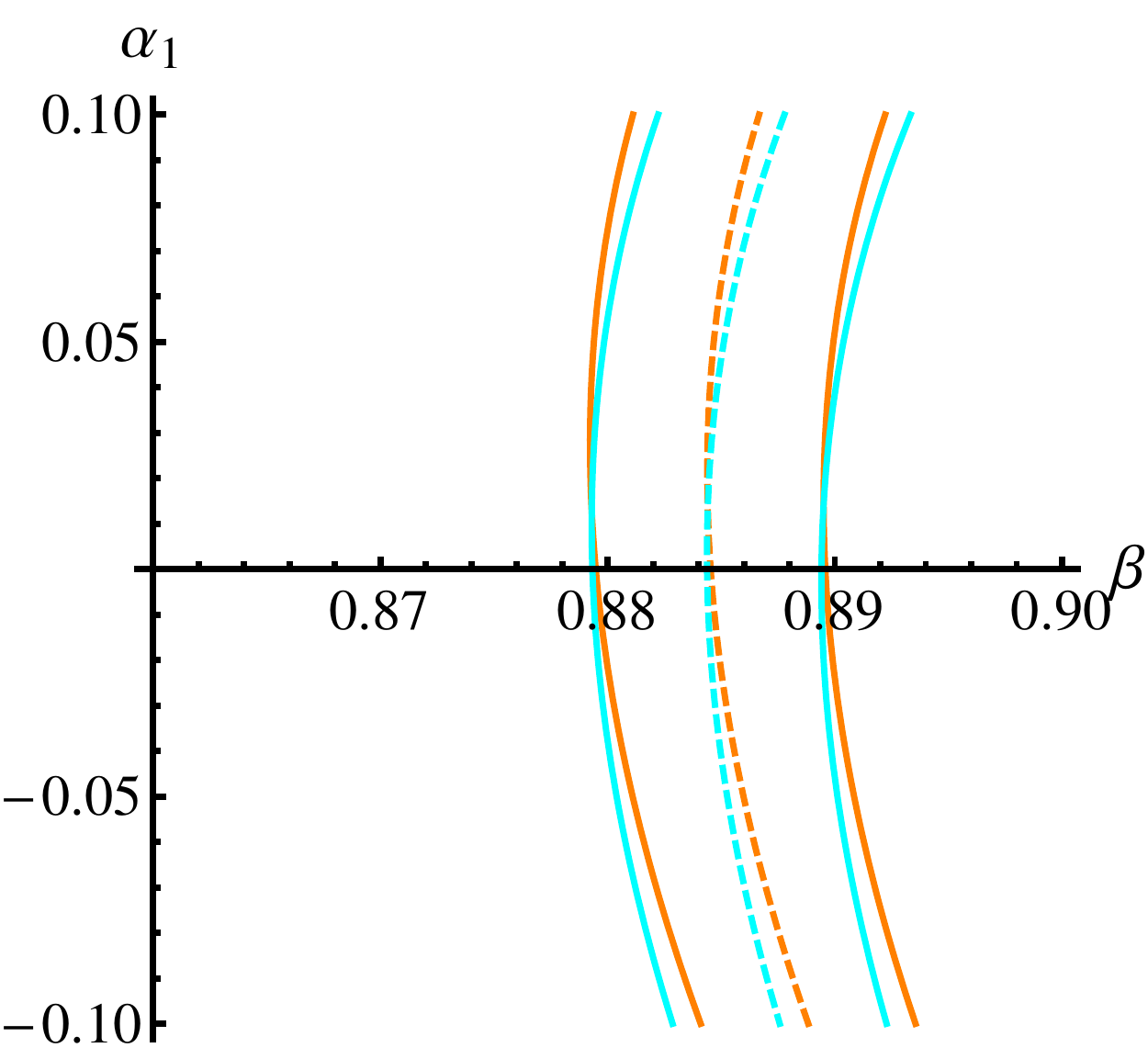}
\end{figure}

Next, we discuss Scenario (b), in which a large mass splitting
exists in $m_{2,3}$, corresponding to the cases in which $\alpha_3$ is close to either
$\pi/2$ or $0$. From \autoref{eq:m3}, we can find two solutions for
$t_{\alpha_3}$ as
\begin{equation}
\label{eq:a3}
t_{\alpha_3^{\pm}}=\frac{\left(m_3^2-m^2_2\right)\pm\sqrt{\left(m^2_3-m^2_2\right)^2s^2_{2\beta+\alpha_1}-4\left(m^2_3-m_1^2\right)\left(m^2_2-m^2_1\right)s^2_{\alpha_2}
c^2_{2\beta+\alpha_1}}}{2\left(m_2^2-m^2_1\right)s_{\alpha_2}c_{2\beta+\alpha_1}}.
\end{equation}
In the large mass splitting scenario, $\alpha_3^+$ is close to
$\pi/2$, and $\alpha_3^-$ is close to $0$. In the $\alpha_3^+$ case, $H_2$ is a
CP-mixed state  in which the pseudoscalar component is dominant, while
$H_3$ is almost a pure scalar. Conversely, in the $\alpha_3^-$ case,
$H_3$ is a CP-mixed state while $H_2$ is almost a pure scalar. In
this scenario, the large mass splitting between $H_{2,3}$ leads to a
significant $H_3\rightarrow H_2Z$ decay, because the coupling is just
$c_{V,1}$, which is not suppressed by mixing angles. Numerical
analysis shows a similar cancellation behavior as Scenario (a) in both
$\alpha_3^{\pm}$ cases.
We show the results of the Type II model in the upper two plots in
  \autoref{fig:cancelsplit}.
Similar to Scenario (a),
the cancellation behavior in the Type III model is almost the same as that
in the Type II model and we show the comparison in the lower two plots in \autoref{fig:cancelsplit}.
\begin{table}[h]
\caption{Fixed parameters of Scenario (b) to discuss the cancellation behavior. With these parameters and $\beta,\alpha_{1,2}$,
we can calculate $\alpha_3^{\pm}$ through \autoref{eq:a3}, and calculate the couplings through the equations
in \autoref{app:Yuk} and \autoref{app:sca}.}\label{table:bmpb}\hspace*{0.25cm}
\centering
\begin{tabular}{|c|c|c|c|c|}
\hline
$m_1$&$m_2$&$m_3$&$m_{\pm}$&$\mu^2$\\
\hline
$125~\textrm{GeV}$&$500~\textrm{GeV}$&$650~\textrm{GeV}$&$700~\textrm{GeV}$&$(450~\textrm{GeV})^2$\\
\hline
\end{tabular}
\end{table}

The behavior of the neutron EDM is also similar to that of Scenario
(a). In the regions allowed  by electron EDM constraint, $d_n$ is only
sensitive to $\alpha_2$ and is almost independent of $\alpha_1$. With the
benchmark points in \autoref{table:bmpb}, and using the indices II/III and $+/-$ to
denote Type II/III models and $\alpha^{+/-}$ cases, we have
\begin{eqnarray}
d_n^{\textrm{II},+}/s_{2\alpha_2}\simeq1.4\times10^{-25}~e\cdot\textrm{cm},\\
d_n^{\textrm{II},-}/s_{2\alpha_2}\simeq1.3\times10^{-25}~e\cdot\textrm{cm},\\
d_n^{\textrm{III},+}/s_{2\alpha_2}\simeq2.4\times10^{-26}~e\cdot\textrm{cm},\\
d_n^{\textrm{III},-}/s_{2\alpha_2}\simeq1.9\times10^{-26}~e\cdot\textrm{cm},
\end{eqnarray}
based on the central value estimation in \autoref{eq:dn}. Thus,
we can obtain the upper limit on $\alpha_2$ in the Type II model as
\begin{equation}
\alpha_2\lesssim\left\{\begin{array}{cc}0.079,&(\alpha_3^+~\textrm{case}),\\0.085,&(\alpha_3^-~\textrm{case}).\end{array}\right.
\end{equation}
There is no constraint on $\alpha_2$ from the  neutron EDM in the Type III model,
due to the same reason as discussed above for Scenario (a).

In both Scenarios (a) and (b), the cancellation can appear around the region $t_{\beta}\simeq1$, thus both $\alpha_2$
and $\arg({c_{t,1}})$ can reach $\mathcal{O}(0.1)$, which lead us to the phenomenological studies of CP-violation in
$t\bar{t}H_1$ production in \autoref{sec:phe}. For this process, both scenarios have similar behaviors. In the future,
if we go deeper into the phenomenology of heavy scalars, differences between these two scenarios will arise.
For example, there will be $H_3\rightarrow ZH_2$ decay in Scenario (b), but such process cannot appear in Scenario (a).

\subsection{Future Neutron EDM Tests}
Several groups are currently planning new measurements on neutron EDM,
to the accuracy of $\mathcal{O}(10^{-27}~e\cdot\textrm{cm})$ or even
better
\cite{Engel:2013lsa,Chupp:2017rkp,Baker:2010zza,Picker:2016ygp,Ayres:2018dbg,Abel:2018yeo,Ahmed:2019dhe}.
Such an order of magnitude
improvement in accuracy would be very helpful to perform
further tests on the 2HDM Type II and III scenarios considered here.

\begin{figure}[h]
\caption{Upper limit on $\alpha_2$ in the Type II and III models when the future limit decreases to $|d_n|<10^{-27}~e\cdot\textrm{cm}$. The color scheme is the same as above: blue for the Type II model and orange for the Type III model. The solid lines are obtained using the central value estimation and, if we consider the current theoretical uncertainty estimation of \cite{Hisano:2012sc}, the boundaries of the limits on $\alpha_2$ are the dashed lines.}\label{fig:futnedm}
\centering
\includegraphics[scale=0.7]{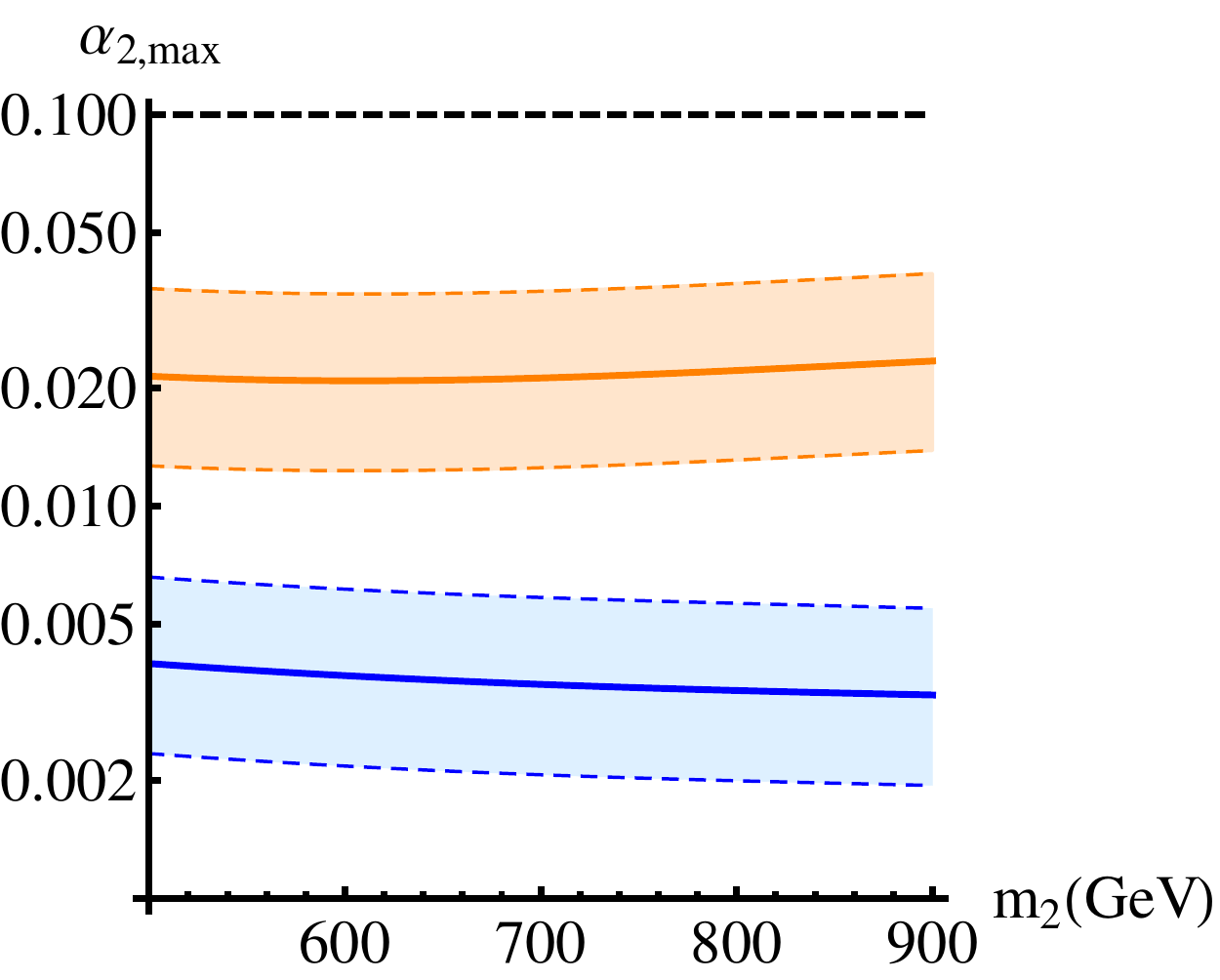}
\end{figure}
If no anomaly is discovered in future neutron EDM measurements, the
upper limit on $d_n$ would improve to about
$10^{-27}~e\cdot\textrm{cm}$, and there would be more stringent limits on
$\alpha_2$ in both Type II and III models, as shown in
\autoref{fig:futnedm} for Scenario (a). With future neutron EDM measurements,
$\alpha_2$ can be constrained to $\mathcal{O}(10^{-2})$ in the Type III model and to
$\mathcal{O}(10^{-3})$ in Type II model. Similar constraints can be
placed in Scenario (b). In current analysis, the expected limit on
$\alpha_2$ still contain large uncertainties (see the colored bands in \autoref{fig:futnedm}),
due to the theoretical uncertainties in the estimation of neutron EDM from sum rules \cite{Hisano:2012sc}.
Future theoretical estimation on neutron EDM from lattice is expected to have better
accuracy, for example, $\sim\mathcal{O}(10\%)$ at lattice \cite{Yamanaka:2018uud,Yoon:2020soi},
thus it will be more effective to obtain the future limit on $\alpha_2$ with smaller
uncertainties\footnote{To obtain an effective limit, the uncertainty
must be not too large, for example, similar to or even larger than the central value.
As a comparison, the EDMs for diamagnetic atoms are also good candidates to probe new CP-violation
\cite{Graner:2016ses,Bishof:2016uqx,Sachdeva:2019rkt}. However, based on the results in
\cite{Jung:2013hka,Inoue:2014nva,Yanase:2018qqq,Pospelov:2001ys}, we have the EDM for $^{199}$Hg atom as
$d_{\textrm{Hg}}^{\textrm{II}}/s_{2\alpha_2}\simeq(-1.7^{+2.3}_{-2.5})\times10^{-27}~e\cdot{\textrm{cm}}$ and
$d_{\textrm{Hg}}^{\textrm{III}}/s_{2\alpha_2}\simeq(-1.3^{+1.9}_{-0.8})\times10^{-27}~e\cdot{\textrm{cm}}$, for
Scenario (a) with $m_2=500~\textrm{GeV}$ in Type II and III models respectively. The results cross zero within
$1\sigma$ level due to large theoretical uncertainties, meaning that it is impossible to set constraints
directly on $\alpha_2$ through the EDM of $^{199}$Hg (or similar atoms) in the low $t_{\beta}$ region.}.

In contrast, if $\alpha_2\sim\mathcal{O}(0.1)$, there will be significant BSM evidence in
future neutron EDM measurements. In the models which contain a similar
cancellation mechanism in electron EDM, the neutron EDM experiments
may be used to find the first evidence of new CP-violation source or set the
strictest limit directly on the CP-violating phase $\alpha_2$.

\subsection{Summary on EDM Tests}
In the previous subsections, we have discussed the electron and neutron EDM tests in the
2HDM with soft CP-violation. There is no cancellation mechanism
in the Type I and IV models and thus the electron EDM can set strict constraints
on the CP-violation angle as $|\arg(c_{f,1})|\simeq
|s_{\alpha_2}/t_{\beta}|\lesssim8.2\times10^{-4}$. However, this value is too small to give any observable CP effects in other
experiments, thus we decided not to have  further discussions on these two 2HDM realizations.
In contrast, cancellations among various contributions to the electron EDM
can occur in the Type II and III models. Here, we
still face stringent constraints but these will
induce a strong correlation between
$\beta$ and $\alpha_1$. We cannot set constraints directly on the
CP-violation mixing angle $\alpha_2$ though. The behavior is the same in the
Type II and III models. In fact, it is also the same in both Scenario (a), in which
$m_{2,3}$ are close to each other, and in Scenario (b), in which $m_{2,3}$
have large splitting. A cancellation generally happens around $t_{\beta}\sim1$ with
the exact location depending weakly on the masses of the heavy (pseudo)scalars.

Current measurements of the neutron EDM can set an upper limit on
$|\alpha_2|\simeq(0.073-0.088)$ in the Type II model, depending on
different scenarios and masses, if we take the central value of the
neutron EDM estimation. Such limits can be weakened to about $0.15$ if we
consider the theoretical uncertainty. But one cannot set limits on
$\alpha_2$ in the Type III model, because the CEDM of the $d$ quark in this
model is suppressed by a partial cancellation. However, $\alpha_2$ in the Type III
model is constrained by collider tests, which will be discussed in the next
section.

Finally, we showed the importance of future neutron EDM measurements in our
models relying on the cancellation mechanism in the electron EDM. For
$\alpha_2\sim\mathcal{O}(0.1)$, there would be significant evidence in
future neutron EDM experiments, which will be more sensitive than any
other experiments. And if there is no evidence of non-zero neutron EDM,
the improved limit on the neutron EDM will set strict constraints on
the CP-violation mixing angle: the upper limit of $|\alpha_2|$ will
reach $\mathcal{O}(10^{-2})$ in the Type III model and $\mathcal{O}(10^{-3})$
in the Type II model. To explain the matter-antimatter asymmetry,
$|\arg(c_{t,1})|\gtrsim10^{-2}$ is required \cite{Fuyuto:2019svr,Shu:2013uua,Hou:2017kmo,Fuyuto:2017ewj}.
Thus if future neutron EDM experiments still show null results to the accuracy
$\sim10^{-27}~e\cdot\textrm{cm}$, the Type II model will not be able to explain the
matter-antimatter asymmetry, due to the very strict constraint on $\alpha_2$.

\section{Current Collider Constraints}
\label{sec:const}
Any BSM model must face LHC tests. In our 2HDM with soft CP-violation, as mentioned, we
treat $H_1$ as the $125~\textrm{GeV}$ Higgs boson. In this scenario then, the latter  mixes with the other
(pseudo)scalar states and its couplings will be modified from the corresponding
SM values. However, these modified  couplings are constrained by global fits on the so-called Higgs
signal-strength measurements. In addition, the scalar sector is extended
in a 2HDM, so that direct searches for these new particles at the LHC will also
set further constraints on this BSM scenario. In this respect, we discuss only
the 2HDM Type II and III, in which the cancellation behavior in the electron EDM
still allow a large CP-phase in Yukawa interactions.

\subsection{Global Fit on Higgs Signal Strengths}
The Higgs boson $H_1$ can be mainly produced at the LHC through four
channels: gluon fusion ($gg$F), vector boson fusion (VBF), associated
production with vector boson ($V+H_1$, here $V=W,Z$) or a top quark pair
($t\bar{t}+H_1$)
\cite{Dittmaier:2011ti,Dittmaier:2012vm,Heinemeyer:2013tqa,deFlorian:2016spz}.
The decay channels $H\rightarrow
b\bar{b},\tau^+\tau^-,\gamma\gamma,WW^*$ and $ZZ^*$  have already been
discovered. Define the signal strength $\mu_{i,f}$ corresponding to
production channel $i$ and decay channel $f$ as follows:
\begin{equation}
\mu_{i,f}\equiv\frac{\sigma_i}{\sigma_{i,\textrm{SM}}}\cdot\frac{\Gamma_f}{\Gamma_{f,\textrm{SM}}}\cdot\frac{\Gamma_{\textrm{tot,SM}}}{\Gamma_{\textrm{tot}}},
\end{equation}
where $\sigma_i$ denotes the production cross section of the production channel $i$ amongst those listed above,
$\Gamma_f$ denotes the decay width of channel $f$ and
$\Gamma_{\textrm{tot}}$ denotes the total decay width of $H_1$. A
quantity with index ``SM'' denotes the value predicted by the
SM. Such signal streengths for different channels have been measured by the ATLAS
\cite{Aad:2019mbh,Cadamuro:2019tcf,ATLAS-CONF-2019-004,Aaboud:2018urx}
and CMS \cite{CMS-PAS-HIG-19-005,Sirunyan:2018koj,Sirunyan:2018kst}
collaborations: we list them in \autoref{tab:stre}.
\begin{table}[h]
\caption{Signal strengths measurements by the ATLAS (left) and CMS (right) collaborations at $\sqrt{s}=13~\textrm{TeV}$. The luminosity is $\leq139~\textrm{fb}^{-1}$ for the ATLAS measurements and $\leq137~\textrm{fb}^{-1}$ for the CMS measurements.}\label{tab:stre}\vspace*{0.25truecm}
\centering
\begin{tabular}{|c|c|c|c|c|}
\hline
&$gg$F&VBF&$V+H$&$t\bar{t}+H$\\
\hline
$H\rightarrow b\bar{b}$&-&$3.01^{+1.67}_{-1.61}$&$1.19^{+0.27}_{-0.25}$&$0.79^{+0.60}_{-0.59}$\\
\hline
$H\rightarrow\tau^+\tau^-$&$0.96^{+0.59}_{-0.52}$&$1.16^{+0.58}_{-0.53}$&-&$1.38^{+1.13}_{-0.96}$\\
\hline
$H\rightarrow \gamma\gamma$&$0.96^{+0.14}_{-0.14}$&$1.39^{+0.40}_{-0.35}$&$1.09^{+0.58}_{-0.54}$&$1.38^{+0.32}_{-0.30}$\\
\hline
$H\rightarrow WW^*$&$1.08^{+0.19}_{-0.19}$&$0.59^{+0.36}_{-0.35}$&-&$1.56^{+0.42}_{-0.40}$\\
\hline
$H\rightarrow ZZ^*$&$1.04^{+0.16}_{-0.15}$&$2.68^{+0.98}_{-0.83}$&$0.68^{+1.20}_{-0.78}$&-\\
\hline
\end{tabular}
\begin{tabular}{|c|c|c|c|c|c|}
\hline
&$gg$F&VBF&$V+H$&$t\bar{t}+H$\\
\hline
$H\rightarrow b\bar{b}$&$2.45^{+2.53}_{-2.35}$&-&$1.06^{+0.26}_{-0.25}$&$1.13^{+0.33}_{-0.30}$\\
\hline
$H\rightarrow\tau^+\tau^-$&$0.39^{+0.38}_{-0.39}$&$1.05^{+0.30}_{-0.29}$&$2.2^{+1.1}_{-1.0}$&$0.81^{+0.74}_{-0.67}$\\
\hline
$H\rightarrow\gamma\gamma$&$1.09^{+0.15}_{-0.14}$&$0.77^{+0.37}_{-0.29}$&-&$1.62^{+0.52}_{-0.43}$\\
\hline
$H\rightarrow WW^*$&$1.28^{+0.20}_{-0.19}$&$0.63^{+0.65}_{-0.61}$&$1.64^{+1.36}_{-1.14}$&$0.93^{+0.48}_{-0.45}$\\
\hline
$H\rightarrow ZZ^*$&$0.98^{+0.12}_{-0.11}$&$0.57^{+0.46}_{-0.36}$&$1.10^{+0.96}_{-0.74}$&$0.25^{+1.03}_{-0.25}$\\
\hline
\end{tabular}
\end{table}

As intimated, in the 2HDM, $H_1$ couplings to SM particles are modified due to the mixing with other (pseudo)scalars and thus the aforementioned signal strengths are modified.
The production cross sections satisfy \cite{Djouadi:2005gi,Djouadi:2005gj,Li:2015kaa}
\begin{eqnarray}
\frac{\sigma_{\textrm{VBF}}}{\sigma_{\textrm{VBF,SM}}}&=&\frac{\sigma_{V+H}}{\sigma_{V+H,\textrm{SM}}}=c_{V,1}^2,\\
\frac{\sigma_{gg\textrm{F}}}{\sigma_{gg\textrm{F,SM}}}&=&\left|\textrm{Re}(c_{t,1})
+\textrm{i}\frac{\mathcal{B}_1\left(\frac{z_{H_1t}}{4}\right)}{\mathcal{A}_1\left(\frac{z_{H_1t}}{4}\right)}\textrm{Im}(c_{t,1})\right|^2\simeq
\left[\textrm{Re}(c_{t,1})\right]^2+2.3\left[\textrm{Im}(c_{t,1})\right]^2,\\
\label{eq:tth}
\frac{\sigma_{t\bar{t}+H}}{\sigma_{t\bar{t}+H,\textrm{SM}}}&\simeq&\left[\textrm{Re}(c_{t,1})\right]^2+0.37\left[\textrm{Im}(c_{t,1})\right]^2,
\end{eqnarray}
while the decay widths satisfy \cite{Djouadi:2005gi,Djouadi:2005gj}
\begin{eqnarray}
\frac{\Gamma_{ZZ^*}}{\Gamma_{ZZ^*,\textrm{SM}}}&=&\frac{\Gamma_{WW^*}}{\Gamma_{WW^*,\textrm{SM}}}=c_{V,1}^2,\\
\frac{\Gamma_{f\bar{f}}}{\Gamma_{f\bar{f},\textrm{SM}}}&=&|c_{f,1}|^2,\quad(f=c,b,\tau),\\
\frac{\Gamma_{gg}}{\Gamma_{gg,\textrm{SM}}}&=&\left|\textrm{Re}(c_{t,1})
+\textrm{i}\frac{\mathcal{B}_1\left(\frac{z_{H_1t}}{4}\right)}{\mathcal{A}_1\left(\frac{z_{H_1t}}{4}\right)}\textrm{Im}(c_{t,1})\right|^2
\simeq\left[\textrm{Re}(c_{t,1})\right]^2+2.3\left[\textrm{Im}(c_{t,1})\right]^2,\\
\label{eq:gaga}
\frac{\Gamma_{\gamma\gamma}}{\Gamma_{\gamma\gamma,\textrm{SM}}}&=&
\left|\frac{\frac{c_{\pm,1}v^2}{2m^2_{\pm}}\mathcal{A}_0(\frac{z_{1,\pm}}{4})+c_{V,1}\mathcal{A}_2(\frac{z_{H_1W}}{4})
+\frac{4}{3}\left[\textrm{Re}(c_{t,1})\mathcal{A}_1(\frac{z_{H_1t}}{4})+\textrm{i}\textrm{Im}(c_{t,1})\mathcal{B}_1(\frac{z_{H_1t}}{4})\right]}
{\frac{4}{3}\mathcal{A}_1(\frac{z_{H_1t}}{4})+\mathcal{A}_2(\frac{z_{H_1W}}{4})}\right|^2\nonumber\\
&\simeq&\left[1.28c_{V,1}-0.28\textrm{Re}(c_{t,1})-0.02\right]^2+0.19\left[\textrm{Im}(c_{t,1})\right]^2.
\end{eqnarray}
The loop functions $\mathcal{A}_{0,1,2}$ and $\mathcal{B}_1$ are
listed in \autoref{app:loopb}. Here, $c_{V,1}=c_{\alpha_1}c_{\alpha_2}$
holds for all types of models, while $c_{f,1}$ which depends on the model
type are listed in \autoref{app:Yuk}. The $t\bar{t}+H_1$ cross section
ratio in \autoref{eq:tth} is only valid for the LHC at
$\sqrt{s}=13~\textrm{TeV}$. For the $\gamma\gamma$ decay
\autoref{eq:gaga}, the charged Higgs loop contribution is small
compared with the top quark and $W$ loops, and we choose the case
$m_{\pm}=600~\textrm{GeV}$ for illustration. The total width satisfies
\begin{equation}
\frac{\Gamma_{\textrm{tot}}}{\Gamma_{\textrm{tot,SM}}}=\mathop{\sum}_f\textrm{BR}_f^{\textrm{SM}}\cdot\frac{\Gamma_f}{\Gamma_{f,\textrm{SM}}}.
\end{equation}
$\textrm{BR}_f^{\textrm{SM}}$ is the SM prediction on the Branching
Ratio (BR) of the SM Higgs boson decay to the final state $f$,  thus all the
modifications are normalized to the SM values. For the 125 GeV SM
Higgs boson, we list the theoretical predictions on the BRs of the main decay channels in \autoref{tab:smhig}
\cite{deFlorian:2016spz}.
\begin{table}[h]
\caption{Predictions of the main  BRs of the SM Higgs boson with mass 125 GeV.}\label{tab:smhig}\vspace*{0.25truecm}
\centering
\begin{tabular}{|c|c|c|c|c|c|}
\hline
&&&&&\\[-0.35cm]
$\textrm{BR}_{b\bar{b}}^{\textrm{SM}}$&$\textrm{BR}_{\tau^+\tau^-}^{\textrm{SM}}$&$\textrm{BR}_{c\bar{c}}^{\textrm{SM}}$&
$\textrm{BR}_{WW^*}^{\textrm{SM}}$&$\textrm{BR}_{ZZ^*}^{\textrm{SM}}$&$\textrm{BR}_{gg}^{\textrm{SM}}$\\
\hline
$58.2\%$&$6.3\%$&$2.9\%$&$21.4\%$&$2.6\%$&$8.2\%$\\
\hline
\end{tabular}
\end{table}

\begin{figure}[!t]
\caption{Allowed regions in the $\alpha_2-\alpha_1$ plane obtained by using the combined results from the ATLAS and CMS collaborations, with fixed $\beta=0.76$ for Type II (left) and Type III (right). Green regions are allowed at $68\%$ C.L. ($\delta\chi^2\leq2.3$) and yellow regions are allowed at $95\%$ C.L. ($\delta\chi^2\leq6.0$).}\label{fig:a1a2}
\centering
\includegraphics[scale=0.55]{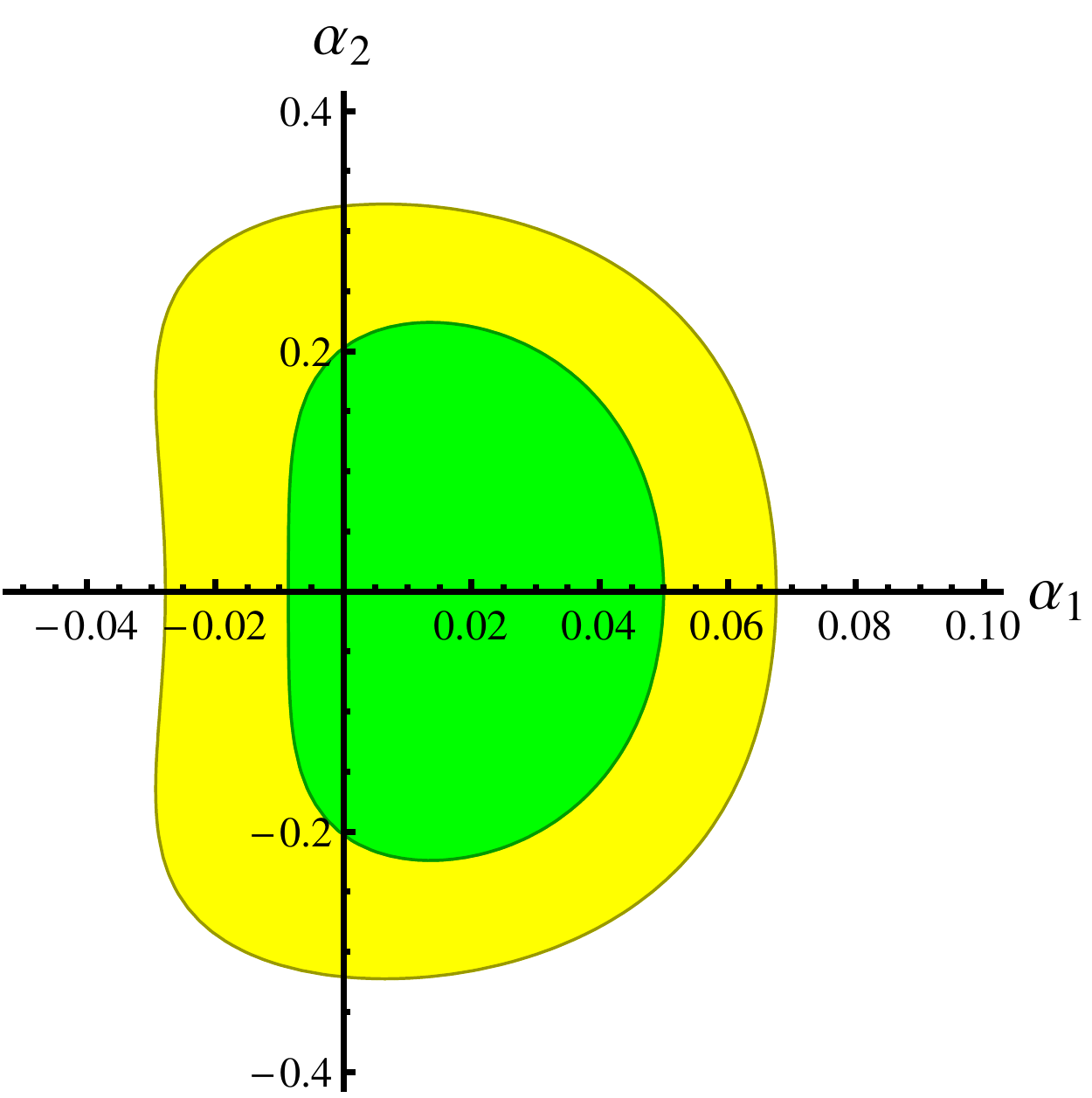}\includegraphics[scale=0.55]{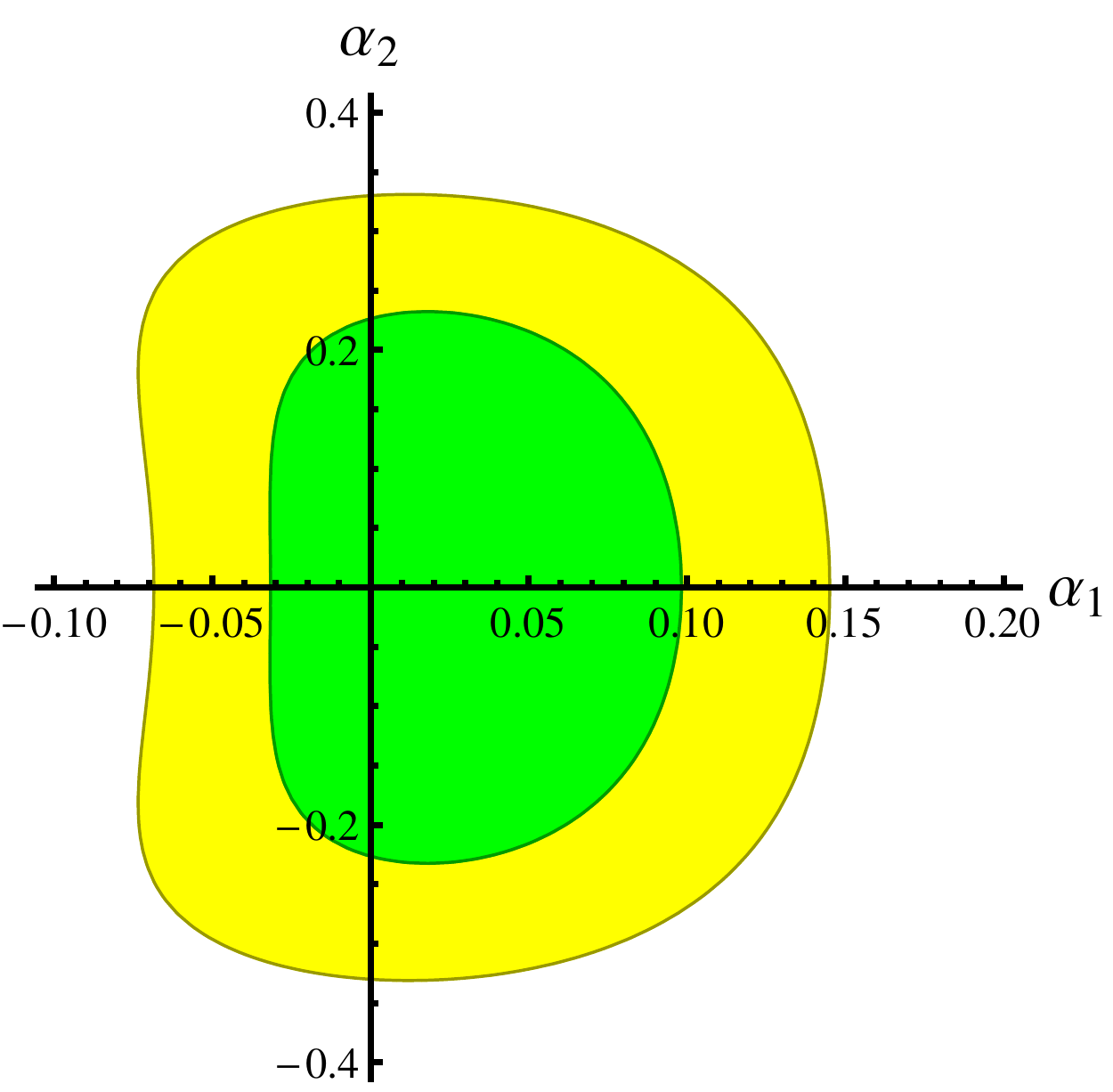}
\end{figure}

We perform $\chi^2$-fits where
\begin{equation}
\chi^2\equiv\mathop{\sum}_{i,f}\left(\frac{\mu_{i,f}^{\textrm{exp}}-\mu_{i,f}^{\textrm{th}}}{\delta\mu^2_{i,f}}\right)^2,
\end{equation}
where $\mu_{i,f}^{\textrm{th}}$ is the theoretically predicted signal
strength, $\mu_{i,f}^{\textrm{exp}}$ is the experimentally measured one and $\delta\mu_{i,f}$ is the associated uncertainty. The
possible small correlations across production and decay channels are ignored. For a 2HDM, $\chi^2$ depends
only on $\beta,\alpha_{1,2}$. We perform global fits for the Type II and
III models, in which $\alpha_2\sim\mathcal{O}(0.1)$ is still
allowed. The minimal $\chi^2$ (denoted by $\chi^2_{\rm min}$) obtained from  ATLAS and CMS data as well as
 the combined one are listed in \autoref{tab:chimin}.
\begin{table}[h]
  \caption{The $\chi^2_{\textrm{min}}/\textrm{d.o.f.}$ for the Type II and Type III
    models using ATLAS data, CMS data and their combination,
    respectively.}\label{tab:chimin}\vspace*{0.25truecm}
\centering
\begin{tabular}{|c|c|c|c|}
\hline
$\chi^2_{\textrm{min}}/\textrm{d.o.f.}$&ATLAS&CMS&ATLAS+CMS\\
\hline
Type II&$11.8/13$&$12.2/15$&$24.2/31$\\
\hline
Type III&$12.7/13$&$11.9/15$&$24.8/31$\\
\hline
\end{tabular}
\end{table}
The fitting, normalized to the degrees of freedom (d.o.f.),  is good enough because the models approach the SM limit when
$\alpha_{1,2}\rightarrow0$. If one then defines
$\delta\chi^2\equiv\chi^2-\chi^2_{\textrm{min}}$, this  is useful to find
the allowed parameter regions of the two 2HDM realizations considered. Our numerical study shows that the results
depend weakly on $\beta$. We choose $\beta=0.76$ (corresponding to
  $m_{2,3}\sim500~\textrm{GeV}$ in Scenario (a)) as an example and
show the allowed region from combined ATLAS and CMS results in the
$\alpha_2-\alpha_1$ plane in \autoref{fig:a1a2}.
For both Type II and III, the global fit requires
$|\alpha_2|\lesssim0.33$ in the region $\beta\sim(0.7-1)$. Forthe  Type II
model, this constraint is weaker when compared with that from the neutron
EDM. Nevertheless, it can set a new constraint on $|\alpha_2|$ for the
Type III model. The allowed range for $|\alpha_1|$ in the latter
is wider than the one in the Type II model, in fact. In both models, $\alpha_1$ is
favored when close to $0$, thus,  in the following discussion, we usually fix
$\alpha_1=0.02$, a value which is not far from the best fit points in most
cases.
\begin{figure}[!t]
\caption{Allowed regions in the $\alpha_1-\beta$ plane obtained by using the combined results from the ATLAS and CMS collaborations, with fixed $\alpha_2=0.1$ (left) and $0.2$ (right), in the Type III model. Green regions are allowed at $68\%$ C.L. ($\delta\chi^2\leq2.3$) and yellow regions are allowed at $95\%$ C.L. ($\delta\chi^2\leq6.0$).}\label{fig:a1b}
\centering
\includegraphics[scale=0.55]{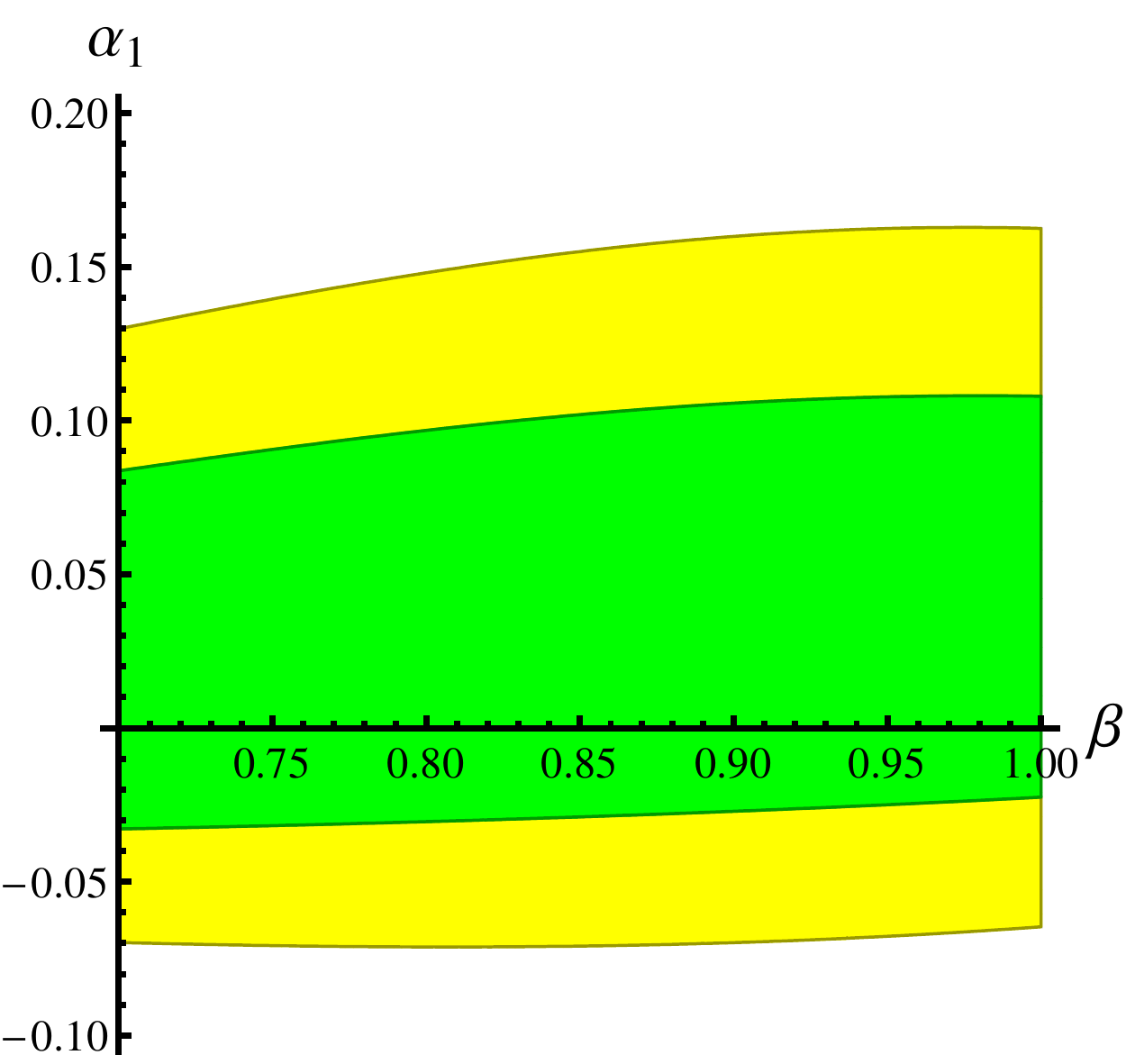}\includegraphics[scale=0.55]{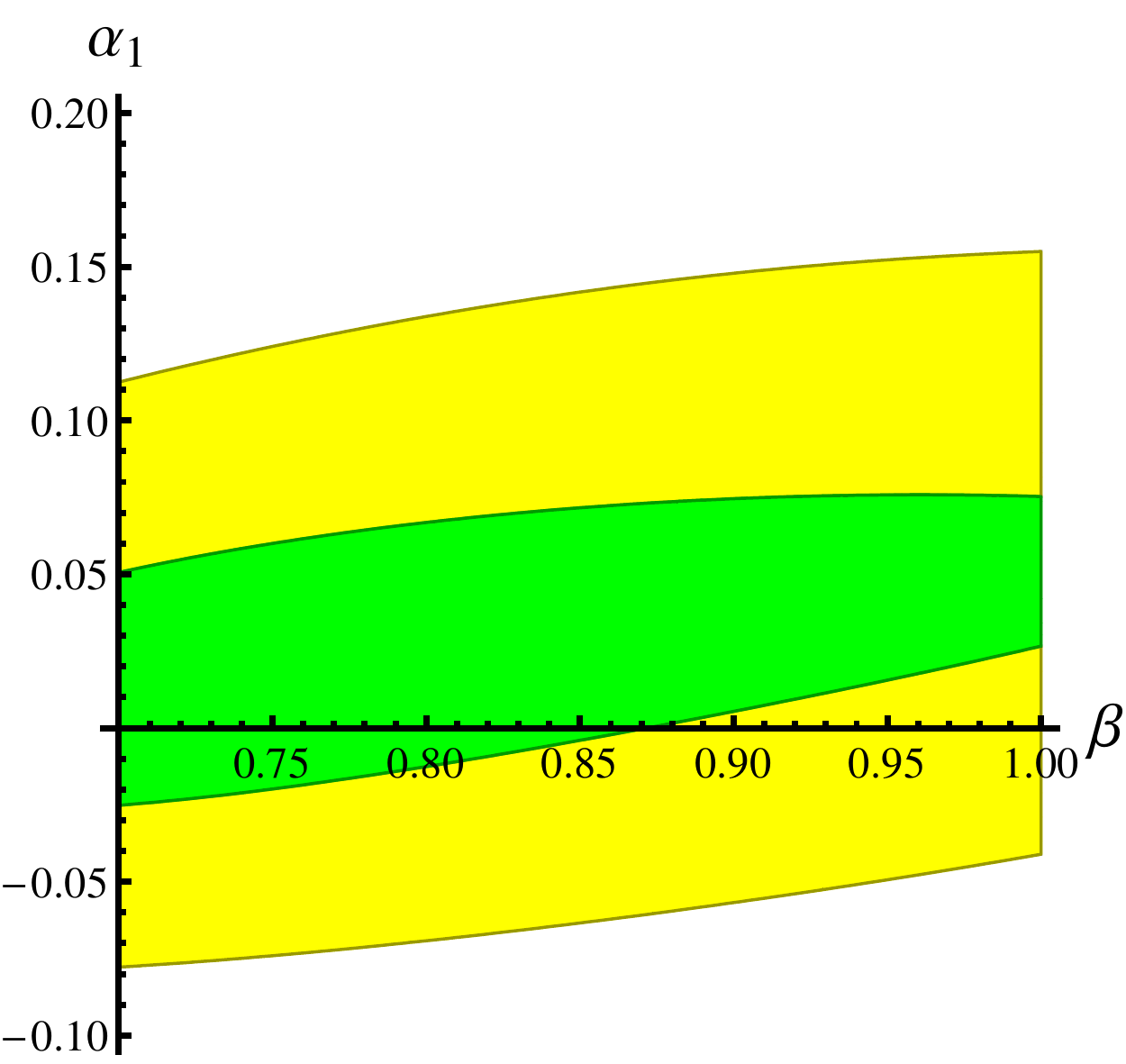}
\end{figure}
In \autoref{fig:a1b}, we show instead the allowed regions in the $\alpha_1-\beta$ plane for fixed $\alpha_2=0.1,0.2$ in the Type III model. The dependence on $\beta$ is indeed weak, but it increases somewhat when $\alpha_2$ gets larger, as shown in the figure.

\subsection{LHC Direct Searches for Heavy Scalars: through $ZZ$ Final State}
In the 2HDM, there are four additional scalars, $H_{2,3}$ and
$H^{\pm}$, beyond the SM-like one $H_1$. Thus, we must also check the direct searches for these
(pseudo)scalars at the LHC. Notice that $H_{2,3}$ decay to $t\bar{t}$ dominantly and we show
their decay widths and BRs in
\autoref{app:heavy}. The $H_{2,3}\rightarrow 2H_1$ decays are
ignored because such channels are suppressed in the allowed parameter region isolated so far.
In Scenario (b), $H_3\rightarrow ZH_2$ decay is also open if
$m_3-m_2>m_Z$. In addition, $H^-$ decays to $\bar{t}b$ dominantly.

In this section we discuss the process $gg\rightarrow H_{2,3}\rightarrow
ZZ$. Theoretically, this process is sensitive to the couplings between
$H_{2,3}$ and the gauge vector bosons, hence sensitive to
$\alpha_2$. Experimentally, this process is the most sensitive channel
in searching for heavy neutral scalars. The current LHC limit for
$m_2=500~\textrm{GeV}$ is $\sigma_{gg\rightarrow H_{2,3}\rightarrow ZZ}\lesssim0.1~\textrm{pb}$ at $95\%$
C.L. \cite{ATLAS-CONF-2017-058} at $\sqrt{s}=13~\textrm{TeV}$ with about 40 fb$^{-1}$ of luminosity.
We first consider the resonance cross section of $gg\rightarrow H_{2,3}\rightarrow ZZ$ process. For Scenario (a),
in which $H_{2,3}$ are close in mass such that $|m_2-m_3|\simeq\mathcal{O}(\textrm{GeV})\ll\Gamma_{2,3}\simeq20~\textrm{GeV}$
for $m_2\simeq500~\textrm{GeV}$ (where we have denoted by $\Gamma_{2,3}$ the widths of the two heavy Higgs states),
we must consider the interference between the $H_2$ and $H_3$ production processes. To the one-loop
order, we have for the resonance cross section
\begin{equation}
\sigma^{\textrm{res}}_{ZZ}=\sigma_S+\sigma_P,
\end{equation}
where $\sigma_S$ is the contribution from $\textrm{Re}(c_{t,2,3})$ corresponding to the CP-conserving part, and $\sigma_P$ is the contribution from $\textrm{Im}(c_{t,2,3})$ corresponding to the CP-violation part. Their $ZZ$ invariant mass distributions are then separately given by
\begin{eqnarray}
\frac{d\sigma_S}{dq}&=&\int dx_1dx_2f_g(x_1)f_g(x_2)\delta\left(x_1x_2-\frac{q^2}{s}\right)\hat{\sigma}_{S}(q)\nonumber\\
&&\times\frac{2q^3m_2\Gamma_0(q)}{\pi s}\left|\mathop{\sum}_{i=2,3}\frac{c_{V,i}\textrm{Re}(c_{t,i})}{q^2-m^2_i-\textrm{i}m_i\Gamma_i}\right|^2,\\
\frac{d\sigma_P}{dq}&=&\int dx_1dx_2f_g(x_1)f_g(x_2)\delta\left(x_1x_2-\frac{q^2}{s}\right)\hat{\sigma}_{P}(q)\nonumber\\
&&\times\frac{2q^3m_2\Gamma_0(q)}{\pi s}\left|\mathop{\sum}_{i=2,3}\frac{c_{V,i}\textrm{Im}(c_{t,i})}{q^2-m^2_i-\textrm{i}m_i\Gamma_i}\right|^2.
\end{eqnarray}
In the equations above, $f_g(x)$ denotes the gluon Parton Distribution Function (PDF), which, in our numerical study, is chosen to be the
MSTW2008 set \cite{Martin:2009iq}. The function \cite{Djouadi:2005gi,Djouadi:2005gj}
\begin{equation}
\Gamma_0(q)=\frac{q^3}{32\pi v^2}\left(1-\frac{4m^2_Z}{q^2}\right)\left(1-\frac{4m^2_Z}{q^2}+\frac{12m^4_Z}{q^4}\right),
\end{equation}
is the decay width to the $ZZ$ final state of a would-be SM Higgs boson
with mass $q$. The functions \cite{Djouadi:2005gi,Djouadi:2005gj}
\begin{eqnarray}
\hat{\sigma}_{S}(q)&=&\frac{G_F\alpha_s^2}{288\sqrt{2}\pi}\left|\frac{3}{4}\mathcal{A}_1\left(\frac{q^2}{4m^2_t}\right)\right|^2,\\
\hat{\sigma}_{P}(q)&=&\frac{G_F\alpha_s^2}{288\sqrt{2}\pi}\left|\frac{3}{4}\mathcal{B}_1\left(\frac{q^2}{4m^2_t}\right)\right|^2,
\end{eqnarray}
are the parton-level cross sections of a pure scalar(pseudoscalar) state with couplings $c_t=1(\textrm{i})$. The loop functions $\mathcal{A}_1$ and $\mathcal{B}_1$ are listed in \autoref{app:loopb}.

The SM $gg\rightarrow ZZ$ production arise through the box diagrams, which leads to the interference effects with the resonance production. We denote
$\sigma^{\textrm{int}}_{ZZ}$ as the cross section induced by interference between resonance and SM background. To one-loop level, our numerical
calculation show that if $|\alpha_2|\sim\mathcal{O}(0.1)$, we have $|\sigma^{\textrm{int}}_{ZZ}/\sigma^{\textrm{res}}_{ZZ}|\sim\mathcal{O}(10^{-3})$
\cite{Glover:1988rg,Pilaftsis:1997dr,Berger:1998vx,Kauer:2012ma}, meaning that we can safely ignore the interference effects and consider only the resonance production\footnote{Different from the $t\bar{t}$ production below, here the interference effects in $ZZ$ production is very small comparing with the resonance production. The reasons are: (i) the SM amplitude generated through box diagrams contains a loop suppression; (ii) the interference only happens between the SM and the CP-conserving part of resonance amplitude $\mathcal{A}_S\propto s_{\alpha_2}^2$ in both scenarios ($\sigma_S\propto|\mathcal{A}_S|^2$). While for the CP-violation part, the amplitude $\mathcal{A}_P\propto s_{\alpha_2}c_{\alpha_2}$, thus $\sigma_P\propto|\mathcal{A}_P|^2$ contributes dominantly to the total
cross section $\sigma_{ZZ}^{\textrm{tot}}\approx\sigma_{ZZ}^{\textrm{res}}$.}. Thus, the total cross section (without SM background) is approximately the resonance cross section
\begin{equation}
\sigma_{ZZ}^{\textrm{tot}}\approx\sigma^{\textrm{res}}_{ZZ}=\mathop{\int}_{m_{2,3}-\Delta q/2}^{m_{2,3}-\Delta q/2}dq\left(\frac{d\sigma_S}{dq}+\frac{d\sigma_P}{dq}\right).
\end{equation}
For $m_2\simeq500~\textrm{GeV}$, we choose $\Delta q=50~\textrm{GeV}$ as the mass window where interference between $H_{2,3}$ is accounted for.

\begin{figure}[h]
\caption{Cross sections $\sigma_{ZZ}^{\textrm{tot}}\approx\sigma_{ZZ}^{\textrm{res}}$
  as a function of the mixing angles $\alpha_{2,3}$ in the Type III model. In the
  left plot, we show the cross section depending on $\alpha_3$ in
  Scenario (a), fixing $\beta=0.76$ and $\alpha_1=0.02$. From top to
  bottom, the four lines show results with
  $\alpha_2=0.33,0.27,0.2,0.14$, respectively. In the right plot, we show
  the cross section depending on $\alpha_2$ in Scenario (b), fixing
  $\beta=0.76$ and $\alpha_1=0.02$, with $\alpha_3$  chosen as
  $\alpha_3^+(\simeq1.5\times10^{-2}\alpha_2)$ which corresponds to
  $m_3\simeq650~\textrm{GeV}$. }\label{fig:csx} \centering
\includegraphics[scale=0.55]{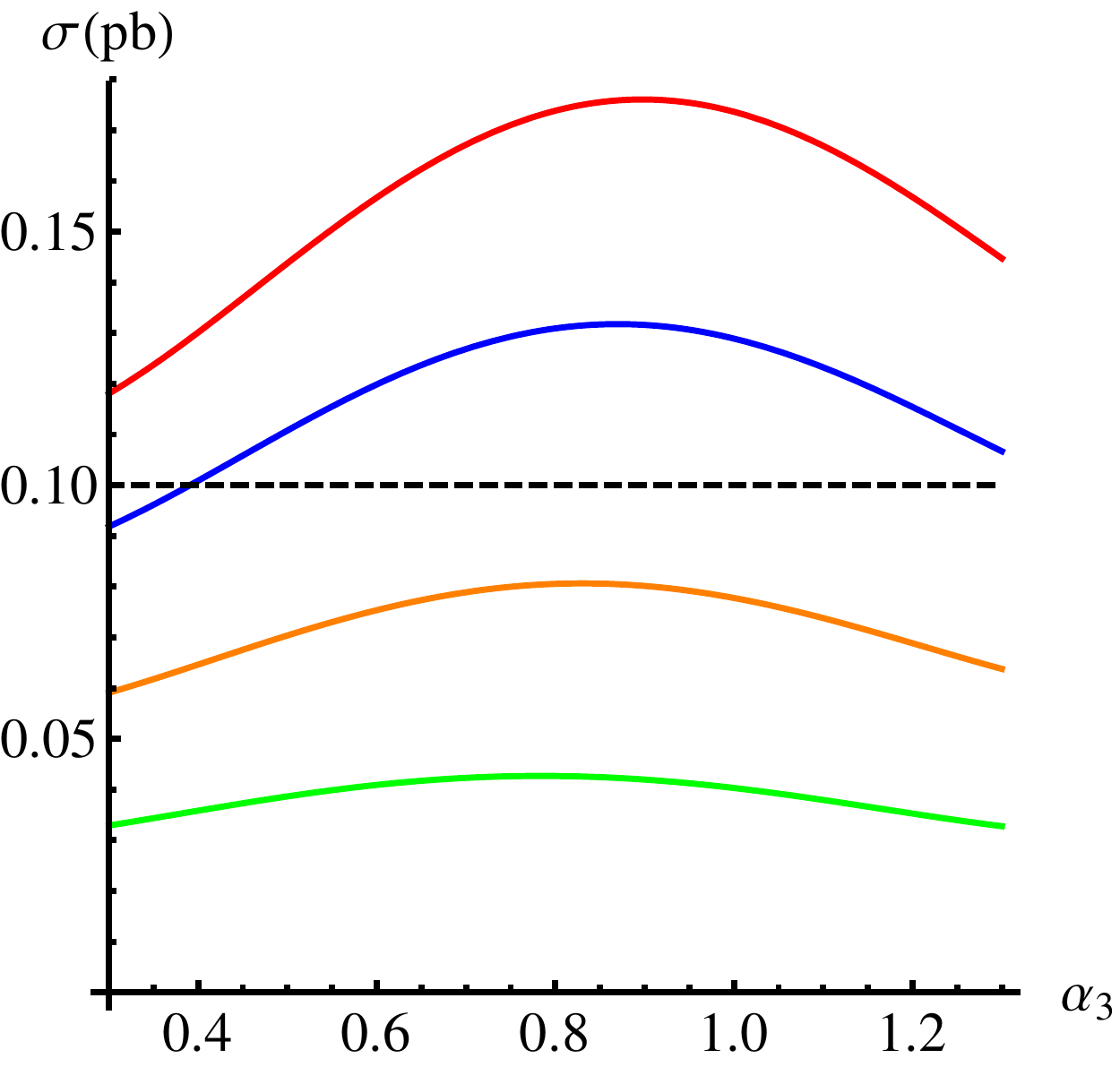}\includegraphics[scale=0.55]{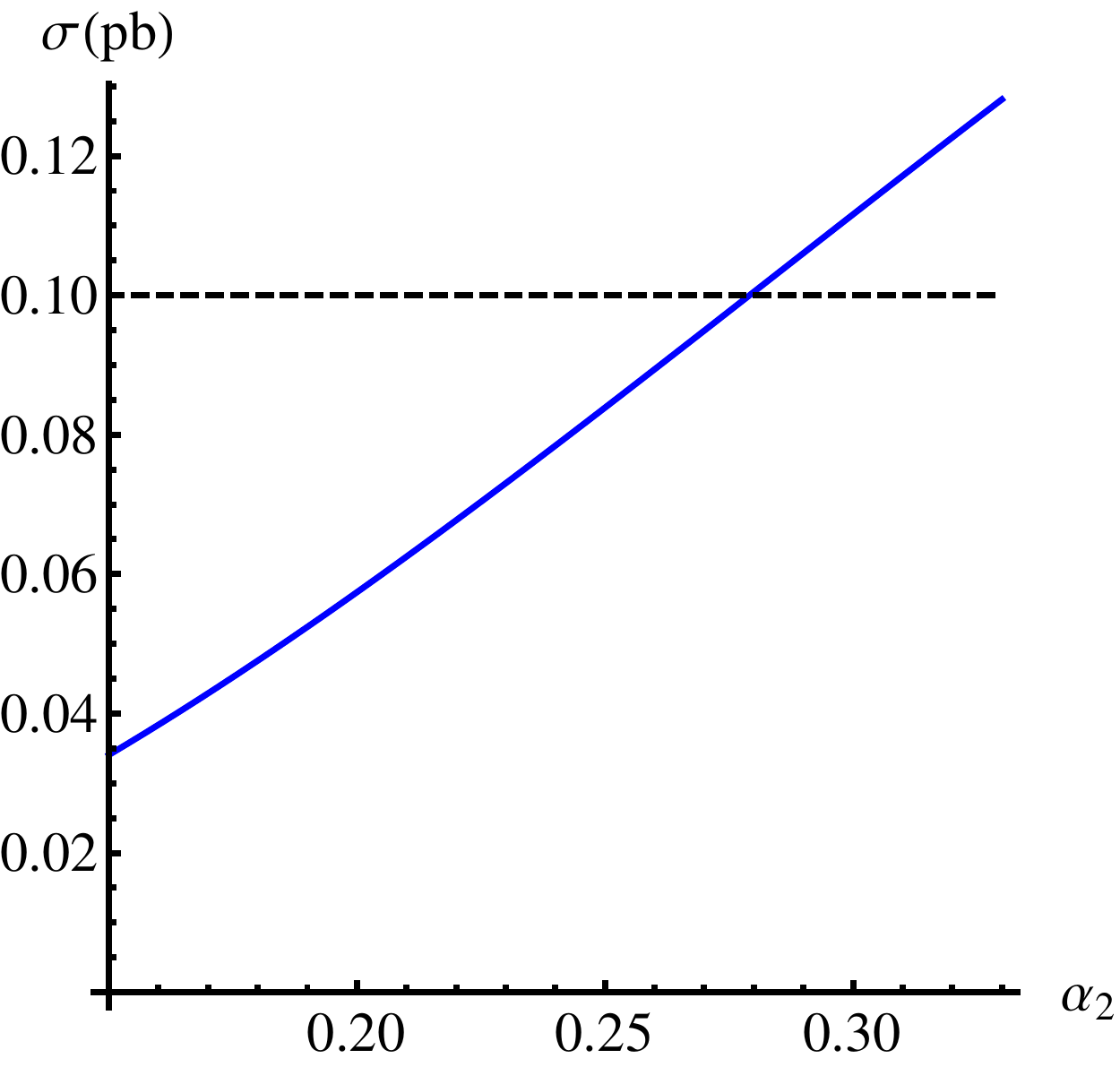}
\end{figure}

Numerically, we show the cross sections depending on the mixing angles
in \autoref{fig:csx} by fixing $m_2=500~\textrm{GeV}$ in the Type III model.
The left plot is for Scenario (a) and the right plot is for Scenario
(b) for the $\alpha_3^+$ case. In both scenarios, we can see that
$\alpha_2\lesssim0.27$ is favored when $m_2=500~\textrm{GeV}$.  For
Scenario (a), when we choose $\alpha_2=0.27$, $\alpha_3\lesssim0.4$ or
$\gtrsim1.2$ is favored, which still keeps $H_{2,3}$ nearly degenerate
in mass. While in the cases $\alpha_2\lesssim0.2$ or $m_2\gtrsim600~\textrm{GeV}$,
there is no further constraints on $\alpha_3$. For Scenario (b), $\alpha_3$
is fixed by other parameters. In the $\alpha_3^-$ case, $c_{2,V}$ is
suppressed (close to $\alpha_1$), thus it faces no further constraints
here. In the Type II model, we can obtain the same cross section as that
in the Type III model with the same parameters. In the Type II model, due to
the stricter neutron EDM constraint on $\alpha_2$, the considered parameter space is
always allowed by the collider data. Thus, in the following
phenomenological analysis, we generally choose $\alpha_2=0.27$
(unless stated otherwise) as a benchmark point, corresponding to the largest allowed
CP-violation effects in Type III model with $m_2=500~\textrm{GeV}$.

\subsection{LHC Direct Searches for Heavy Scalars: through $t\bar{t}$ and other Final States}
As mentioned,  $H_{2,3}$ decay dominantly to a $t\bar{t}$ final state and the current
LHC limit for $m_2=500~\textrm{GeV}$ is about $\sigma_{pp\rightarrow
  H_{2,3}\rightarrow t\bar{t}}\lesssim7~\textrm{pb}$ at $95\%$
C.L. \cite{Aaboud:2018mjh} at $\sqrt{s}=13~\textrm{TeV}$ and 36 fb$^{-1}$ of luminosity.
In contrast to the $ZZ$ channel, the
interference with SM background is very important
in the $t\bar{t}$ channel \cite{Dicus:1994bm,Djouadi:2019cbm},
which strongly decreases the
signal cross section compared with the pure resonance production
cross section, so long that non-resonant Higgs diagrams can be subtracted \cite{Moretti:2012mq}. The total cross section can be divided into
\begin{equation}
\sigma_{gg\rightarrow t\bar{t}}=\sigma_{\textrm{SM}}+\sigma_{\textrm{res}}+\sigma_{\textrm{int}}=\sigma_{\textrm{SM}}+\delta\sigma_{t\bar{t}}.
\end{equation}
Here, $\sigma_{\textrm{SM}}$ denotes the SM background cross section of $gg\rightarrow t\bar{t}$ process; while $\sigma_{\textrm{res}}$ and $\sigma_{\textrm{int}}$ denote the resonant and interference cross section, separately. Furthermore,
$\delta\sigma_{t\bar{t}}$ is the cross section difference between 2HDM and SM, i.e.,
\begin{equation}
\delta\sigma_{t\bar{t}}\equiv\sigma_{\textrm{res}}+\sigma_{\textrm{int}}=\int dx_1dx_2f_g(x_1)f_g(x_2)\left(\hat{\sigma}_{\textrm{res}}+\hat{\sigma}_{\textrm{int}}\right),
\end{equation}
where $\hat{\sigma}$ denotes the parton-level cross section as a
function of the $t\bar{t}$ invariant mass $q$. Following the results in
\cite{Dicus:1994bm,Djouadi:2019cbm}, we have
\begin{eqnarray}
\hat{\sigma}_{\textrm{res}}&=&\hat{\sigma}_{\textrm{res},S}+\hat{\sigma}_{\textrm{res},P}\nonumber\\
&=&\frac{3\alpha_s^2G_F^2m^2_tq^4}{4096\pi^3}
\left[\beta_t^3\left(\left|\mathop{\sum}_{i=2,3}\frac{[\textrm{Re}(c_{t,i})]^2\mathcal{A}_1\left(\frac{q^2}{4m^2_t}\right)}{q^2-m_i^2-\textrm{i}m_i\Gamma_i}\right|^2
+\left|\mathop{\sum}_{i=2,3}\frac{[\textrm{Re}(c_{t,i})\textrm{Im}(c_{t,i})]\mathcal{B}_1\left(\frac{q^2}{4m^2_t}\right)}{q^2-m_i^2-\textrm{i}m_i\Gamma_i}\right|^2\right)\right.\nonumber\\
&&+\left.\beta_t\left(\left|\mathop{\sum}_{i=2,3}\frac{[\textrm{Re}(c_{t,i})\textrm{Im}(c_{t,i})]
\mathcal{A}_1\left(\frac{q^2}{4m^2_t}\right)}{q^2-m_i^2-\textrm{i}m_i\Gamma_i}\right|^2+\left|\mathop{\sum}_{i=2,3}\frac{[\textrm{Im}(c_{t,i})]^2
\mathcal{B}_1\left(\frac{q^2}{4m^2_t}\right)}{q^2-m_i^2-\textrm{i}m_i\Gamma_i}\right|^2\right)\right],\\
\hat{\sigma}_{\textrm{int}}&=&\hat{\sigma}_{\textrm{int},S}+\hat{\sigma}_{\textrm{int},P}\nonumber\\
&=&-\mathop{\int}_{-1}^1dc_{\theta}\frac{\alpha_sG_Fm^2_t}{64\sqrt{2}\pi(1-\beta_t^2c^2_{\theta})}\nonumber\\
&&\times\textrm{Re}\left[\beta_t^3\mathop{\sum}_{i=2,3}\frac{[\textrm{Re}(c_{t,i})]^2\mathcal{A}_1\left(\frac{q^2}{4m^2_t}\right)}{q^2-m_i^2-\textrm{i}m_i\Gamma_i}
+\beta_t\mathop{\sum}_{i=2,3}\frac{[\textrm{Im}(c_{t,i})]^2\mathcal{B}_1\left(\frac{q^2}{4m^2_t}\right)}{q^2-m_i^2-\textrm{i}m_i\Gamma_i}\right].
\end{eqnarray}
Here, $q^2=x_1x_2s$, $\beta_t=\sqrt{1-4m^2_t/q^2}$ is the velocity of
the top quark in the $t\bar{t}$ center-of-mass frame. In our numerical
study, we set $q$ in the range $m_2-\Delta' q/2<q<m_2+\Delta' q/2$,
where we choose the mass window $\Delta'q=100~\textrm{GeV}$ for
$m_2=500~\textrm{GeV}$. We choose the MSTW2008 PDF
\cite{Martin:2009iq} as above. We show the cross sections for some
benchmark points in both Scenario (a) and Scenario (b) in
\autoref{tab:ttbar}.
\begin{table}[h]
\caption{Cross sections $\delta\sigma_{t\bar{t}}$ at the LHC with
  $\sqrt{s}=13~\textrm{TeV}$, fixing $m_2=500~\textrm{GeV}$ and
  $\alpha_1=0.02$. Further, for the Type II model (denoted as
  $\delta\sigma_{t\bar{t}}^{\textrm{II}}$) we fix $\alpha_2=0.14$ while
  for the Type III model (denoted as
  $\delta\sigma_{t\bar{t}}^{\textrm{III}}$) we fix
  $\alpha_2=0.27$. The left table is for Scenario (a), in which we fix
  $\beta=0.76$ and choose $\alpha_3=0.4,0.8,1.2$ from top to
  bottom. The right table is for Scenario (b), in which we fix
  $m_3=650~\textrm{GeV}$, considering two cases: $\beta=0.77$,
  $\alpha_3=\alpha_3^+$ and $\beta=0.885$, $\alpha_3=\alpha_3^-$, again, from top to bottom.}\label{tab:ttbar}\vspace*{0.25truecm} \centering
\begin{tabular}{|c|c|c|}
\hline
$\alpha_3$&$\delta\sigma_{t\bar{t}}^{\textrm{II}}~(\textrm{pb})$&$\delta\sigma_{t\bar{t}}^{\textrm{III}}~(\textrm{pb})$\\
\hline
$0.4$&$0.04$&$-0.40$\\
\hline
$0.8$&$0.39$&$-0.11$\\
\hline
$1.2$&0.25&$-0.07$\\
\hline
\end{tabular}
\begin{tabular}{|c|c|c|}
\hline
$\alpha_3$&$\delta\sigma_{t\bar{t}}^{\textrm{II}}~(\textrm{pb})$&$\delta\sigma_{t\bar{t}}^{\textrm{III}}~(\textrm{pb})$\\
\hline
$\alpha_3^+$&$-0.43$&$-0.67$\\
\hline
$\alpha_3^-$&$0.72$&$0.53$\\
\hline
\end{tabular}
\end{table}
The numerical results show that, for all benchmark points we
consider, the interference with the SM background significantly breaks
the resonance structure of $H_{2,3}$ and decreases the cross sections
to around (even below) $0$, which means the $t\bar{t}$ resonant
search at the  LHC cannot set limits on this model\footnote{In some
  experimental analysis \cite{Aaboud:2017hnm,Sirunyan:2019wph}, the
  interference effects between (pseudo)scalar resonance and the SM background
  were taken into account. Yet, the results cannot be simply rescaled
  to our CP-violating scenario, because the existence of CP-violation
  will modify the shape of the $t\bar{t}$ invariant mass compared with
  the CP-conserving case. We still need further studies on such
  scenarios.}.

The $H_{2,3}$ states can also be produced in association with a $t\bar{t}$ pair at the LHC,
thus we should also check this constraint for our favored benchmark
points. Since $H_{2,3}$ mainly decay into a $t\bar{t}$ pair, the whole production and decay process  will
modify the cross section of the $pp\rightarrow t\bar{t}t\bar{t}$ process (which we denote by $\sigma_{4t}$),
which current LHC limit is about $22.5~\textrm{fb}$ at $95\%$
C.L. \cite{CMS-PAS-TOP-18-003} at $\sqrt{s}=13~\textrm{TeV}$ with 137 fb$^{-1}$ of integrated luminosity by CMS collaboration\footnote{Recently ATLAS collaboration also presented their measurement $\sigma_{4t}=24^{+7}_{-6}~\textrm{fb}$ at $\sqrt{s}=13~\textrm{TeV}$ with 137 fb$^{-1}$ of integrated luminosity \cite{Aad:2020klt}. It is consistent with the CMS result and SM prediction within $2\sigma$ level, but
the constraint is a bit weaker as $\sigma_{4t}\lesssim38~\textrm{fb}$ at $95\%$ C.L.}.
The interference effects between SM and BSM contributions are expected to be significant \cite{Cao:2019ygh}.
We estimate this cross section in the 2HDM considering all interference
effects by  using \textsc{Madgraph5\_aMC@NLO}
\cite{Alwall:2011uj,Alwall:2014hca}. We  then  show the numerical results in
\autoref{tab:4t} for some benchmark points, all allowed by current LHC limits.
\begin{table}[h]
\caption{Cross sections $\sigma_{4t}$ at the LHC with $\sqrt{s}=13~\textrm{TeV}$, fixing $m_2=500~\textrm{GeV}$ and $\alpha_1=0.02$. Further, for the Type II model (denoted as $\sigma_{4t}^{\textrm{II}}$) we fix $\alpha_2=0.14$ while for the Type III model (denoted as $\sigma_{4t}^{\textrm{III}}$) we fix $\alpha_2=0.27$. The left table is for Scenario (a), in which we fix $\beta=0.76$ and choose $\alpha_3=0.4,0.8,1.2$ from top to bottom. The right table is for Scenario (b), in which we fix $m_3=650~\textrm{GeV}$, considering two cases: $\beta=0.77$, $\alpha_3=\alpha_3^+$ and $\beta=0.885$, $\alpha_3=\alpha_3^-$,  again, from top to bottom.}\label{tab:4t}\vspace*{0.25truecm}
\centering
\begin{tabular}{|c|c|c|}
\hline
$\alpha_3$&$\sigma_{4t}^{\textrm{II}}~(\textrm{fb})$&$\sigma_{4t}^{\textrm{III}}~(\textrm{fb})$\\
\hline
$0.4$&$19.9$&$17.9$\\
\hline
$0.8$&$20.8$&$18.7$\\
\hline
$1.2$&20.8&$19.3$\\
\hline
\end{tabular}
\begin{tabular}{|c|c|c|}
\hline
$\alpha_3$&$\sigma_{4t}^{\textrm{II}}~(\textrm{fb})$&$\sigma_{4t}^{\textrm{III}}~(\textrm{fb})$\\
\hline
$\alpha_3^+$&$15.9$&$14.3$\\
\hline
$\alpha_3^-$&$10.4$&$9.4$\\
\hline
\end{tabular}
\end{table}

Finally, we should also check the direct LHC limits on the charged Higgs
boson $H^{\pm}$. As mentioned above, $b\rightarrow s\gamma$ decay
favors a heavy $H^{\pm}$ state with  mass $m_{\pm}\gtrsim600~\textrm{GeV}$
\cite{Belle:2016ufb,Misiak:2017bgg}. For $m_{\pm}=600~\textrm{GeV}$,
the current LHC limit is about $0.1~\textrm{pb}$ at $95\%$
C.L. \cite{Sirunyan:2019arl,CMS-PAS-HIG-18-015} at
$\sqrt{s}=13$ with some 36 fb$^{-1}$ of luminosity~\textrm{TeV}. For large $t_{\beta}$, the interference
effect is negligible \cite{Arhrib:2019ykh}. However, in the Type II and
Type III models with CP-violation as considered above,
$t_{\beta}\sim1$ is favored. For $m_{\pm}\simeq600~\textrm{GeV}$, its
width $\Gamma_{\pm}\gtrsim30~\textrm{GeV}$, which leads to significant
interference effects. Again, we estimate the cross section considering
all interference effects using \textsc{Madgraph5\_aMC@NLO}
\cite{Alwall:2011uj,Alwall:2014hca}. If we denote by $\delta\sigma_{\pm}$ the
cross section modification (including both the resonant and interference
effects) to SM $t\bar{t}b\bar{b}$ process, our numerical estimation
show that
\begin{equation}
\delta\sigma_{\pm}=-0.38~\textrm{pb}<0
\end{equation}
for $m_{\pm}=600~\textrm{GeV}$ and $\beta=0.76$. That means that
the interference effect significantly decreases the $H^{\pm}$ production
cross section in this parameter region, thus the latter is
not constrained by current LHC experiments.

\subsection{Summary on Collider Constraints}
The $125~\textrm{GeV}$ Higgs ($H_1$) signal strength measurements lead
to a constraint $|\alpha_2|\lesssim0.33$, which depends weakly on
$\beta$. The LHC direct searches for heavy neutral scalars decaying to
the $ZZ$ final state set a stricter constraint $|\alpha_2|\lesssim0.27$
for $m_2=500~\textrm{GeV}$ in both Scenario (a) and (b). When
$m_2\gtrsim(550-600)~\textrm{GeV}$, the constraint from direct
searches becomes weaker than that from the global fit ton the $H_1$  signal
strengths. In further analysis, we prefer to choose $\alpha_2=0.27$,
which is the largest allowed value for $m_2=500~\textrm{GeV}$\footnote{Recently
CMS collaboration presented the latest direct constraint on CP-violation in $\tau^+\tau^-H_1$
interaction \cite{CMS-PAS-HIG-20-006}. They obtain the CP-phase $|\arg(c_{\tau,1})|\lesssim0.6$
at $95\%$ C.L., corresponding to $|\alpha_2|\lesssim0.6$ in 2HDM Type II or III with
$t_{\beta}\simeq1$. This is much weaker than our indirect constraint.}. We
have also checked the constraints from $t\bar{t}$, $t\bar tt\bar t$ and charged
Higgs boson searches, in which the interference effects are very important. All
benchmark points that we have considered are allowed by current LHC
measurements. In the remainder of this work, we focus on the phenomenology of
CP-violation in $t\bar{t}H_1$ associate production. We will instead
consider the production and decay  phenomenology of the heavy (pseudo)scalars $H_{2,3}$ in a forthcoming
paper.

%%%%%%%%%%%%%%%%%%%%%%%%%%%%%%%%%%%%%%%%%%%%%%%%
\section{LHC Phenomenology of CP-violation in $t\bar{t}H_1$ Production}
\label{sec:phe}
%%%%%%%%%%%%%%%%%%%%%%%%%%%%%%%%%%%%%%%%%%%%%%%%
In this section, we study the production of the lightest neutral Higgs boson
$H_1$ in association with a $t\bar{t}$ pair at the LHC. We start by discussing the phenomenological set-up used in our analysis, discuss the observables which can be used to probe of the $CP$-nature of the $t\bar{t} H_1$ coupling, both inclusive and differential; and close the section by demonstrating the sensitivity results for a selected benchmark point for the final state consisting of two charged leptons, $n \geq 4$ jets and missing transverse energy $E_{T}^{\mathrm{miss}}$ which is associated with neutrinos from $W$-decays.

%%%%%%%%%%%%%%%%%%%%%%%%%%%%%%%%%%%%%%%%%%%%%%
\subsection{Phenomenological Set-up}
\label{sec:setup}
%%%%%%%%%%%%%%%%%%%%%%%%%%%%%%%%%%%%%%%%%%%%%%%

Events are generated at Leading Order (LO) using the
\textsc{Madgraph5\_aMC@NLO} \cite{Alwall:2011uj,
  Alwall:2014hca}. Cross sections of signal processes are calculated
using a \textsc{UFO} model file \cite{Degrande:2011ua} corresponding
to a Type II 2HDM\footnote{It can also be use for Type III 2HDM.} with flavor-conservation \cite{Degrande:2014vpa} slightly modified to
account for CP-violation effects in vertices involving both the neutral ($H_i$, with $i=1,2,3$) and charged ($H^\pm$) Higgs boson states. Here, we employ the LO version of the \textsc{Mmhtlo68cl} PDF sets \cite{Harland-Lang:2014zoa}. For both the signal and  background processes, we have used the nominal value for the (identical)
renormalization and factorization scales to be equal to half the scalar sum of the transverse mass of all final state particles on an
event-by-event basis, i.e.:
\begin{eqnarray}
 \mu_R=\mu_F = \frac{1}{2} \sum_{i=1}^N \sqrt{m_i^2 + p_{T,i}^2}.
\end{eqnarray}
In the computation of the parton level cross sections, we have employed
the $G_\mu$-scheme, where the input parameters are
$G_F, \alpha_\textrm{em}$ and $m_Z$, the numerical values of which are given by
\begin{equation}
G_F = 1.16639\times10^{-5}~\text{GeV}^{-2},~\alpha_\textrm{em}^{-1}(0)=137,~\mathrm{and}~m_Z=91.188~\text{GeV}.
\end{equation}
The values for $m_W$ and $\sin^2\theta_W$ are computed from the above inputs. For the pole masses of the fermions, we have taken
\begin{equation}
m_t = 172.5~\mathrm{GeV}, \quad m_b = 4.7 ~\mathrm{GeV}.
\end{equation}

Uncertainties due to the scale and PDF variations are computed using \textsc{SysCalc}\cite{Kalogeropoulos:2018cke}. In order to keep full
spin correlations at both the production and decay stages of the top
quarks, we have employed \textsc{MadSpin}
\cite{Artoisenet:2012st}. \textsc{Pythia8} \cite{Sjostrand:2014zea}
is used to perform parton showering and hadronization, albeit without
including Multiple Parton Interactions (MPIs), to the events,
eventually producing a set of event files in \textsc{HepMC} format
\cite{Dobbs:2001ck}. The \textsc{HepMC} files are passed to
\textsc{Rivet} (version 2.7.1) \cite{Buckley:2010ar} for a particle
level analysis. In the latter, jets are clustered using the
anti-$k_T$ algorithm using
\textsc{FastJets}
\cite{Cacciari:2008gp,Cacciari:2011ma}\footnote{Results were found to be stable if replacing \textsc{Pythia8} with \textsc{Herwig6.5}
  \cite{Corcella:2002jc,Corcella:2001wc,Corcella:2001pi,Corcella:2000bw}
  and the anti-$k_T$ algorithm with the Cambridge-Aachen one
  \cite{Wobisch:1998wt,Dokshitzer:1997in}.}.

The particle level events are selected if they contain two charged leptons, high jet multiplicity of $4$-$6$ jets, where at least four of them are $b$-tagged, and missing transverse energy which corresponds to the SM neutrino from the decays of $W$-gauge bosons. Only prompt electrons and muons directly connected to the $W$ boson are accepted,
i.e., we do not select those coming from $\tau$ decays. Electrons are
selected if they pass the basic selection requirement of $p_T^e > 30$ GeV and $|\eta^e| < 2.5$
(excluding the ones that fall in the end-cap or transition regions of the calorimeter, i.e. with
$1.37 < |\eta^e| < 1.52$) while muons are selected if they satisfy the
conditions $p_T^\mu > 27$ GeV and $|\eta^\mu| < 2.4$. Jets are clustered  with
jet radius $D=0.4$ and selected if they satisfy $p_T^j > 30$
GeV and $|\eta^j| < 2.4$. For $b$-tagging, we use the so-called
ghost-association technique \cite{Cacciari:2007fd,Cacciari:2008gn}. In this method, a jet is $b$-tagged if all the jet
particles $i$ within $\Delta R (\mathrm{jet}, i) < 0.3$ of a given anti-$k_T$ jet satisfy $p_T^i > 5$ GeV. We assume a $b$-tagging efficiency of $80\%$ independent of the transverse momentum of the jet. For top quark reconstruction, we use the \textsc{PseudoTop} definition \cite{Collaboration:2267573} (more details along with validation plots can be found in \autoref{app:toprec}). Finally, we require that the invariant mass of $b\bar{b}$ system forming the $H_1$ candidate is around the $H_1$ mass, $|m_{b\bar{b}}-m_1|<15~\textrm{GeV}$, and the transverse energy of the $b\bar{b}$ system forming the $H_1$ candidate is larger than $50~\textrm{GeV}$.

%%%%%%%%%%%%%%%%%%%%%%%%%%%%%%%%%%%%%%%%%%%%%%%%%%%%%%%
\subsection{Inclusive $t\bar{t}H_1$ Cross Section}
\begin{figure}[!t]
\caption{Representative Feynman diagrams corresponding to $t\bar{t}H_1$ production at LO. They consist of production through $q\bar{q}$ annihilation [diagram (a)] and through $gg$F [diagrams (b) and (c)].}
\label{fig:diagrams}
\centering
\includegraphics[width=0.89\linewidth]{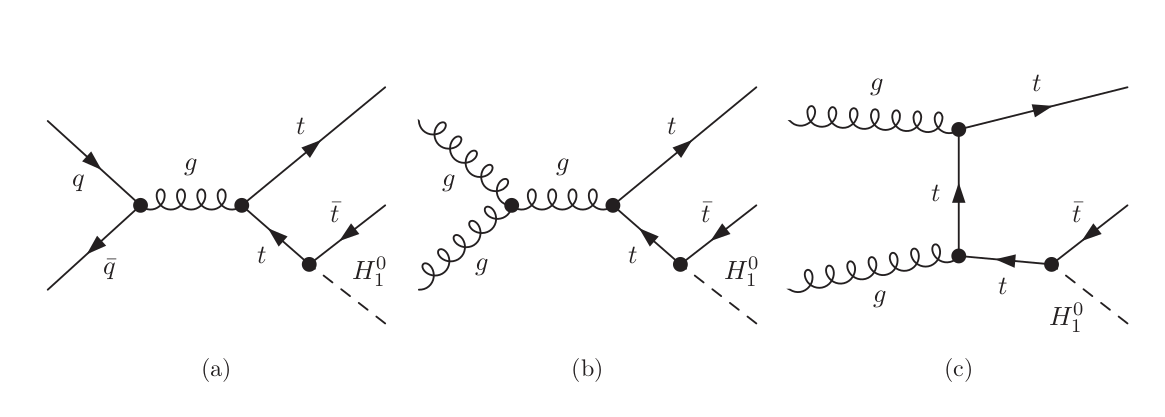}
\end{figure}

%%%%%%%%%%%%%%%%%%%%%%%%%%%%%%%%%%%%%%%%%%%%%%%%%%%%%%%
The parton level Feynman for $t\bar{t} H_1$ production at Leading-Order (LO) are depicted in
\autoref{fig:diagrams}. The cross section has two contributions: (i) from $q\bar{q}$ annihilation,diagram (a) in \autoref{fig:diagrams}, which is expected to dominate in the region of medium and large $x = \tilde{p}_i/P$ with $\tilde{p}_i$ and $P$ are the longitudinal momenta of the parton $i$ and the proton respectively; and (ii) from $gg$ fusion ($gg$F), diagrams (b) and (c) in \autoref{fig:diagrams}, dominating at low $x$. For the calculation of the cross section, we employ \textsc{Madgraph5\_amc@nlo} \cite{Alwall:2011uj,Alwall:2014hca} with
the \textsc{Mmhtlo68cl} and \textsc{Mmhtnlo68cl} PDF sets
\cite{Harland-Lang:2014zoa} in the 4-flavor scheme. Systematic
uncertainties are divided into two categories:  scale and PDF ones. The scale uncertainties are obtained by varying the
renormalization and  factorization scales by a factor of two around
their nominal value, i.e.,
\begin{equation}
(\mu_R, \mu_F) = \{(1,1), (1,0.5), (1,2), (0.5,1), (0.5,0.5), (0.5,2), (2,1), (2,0.5), (2,2)\} (\mu_R^0, \mu_F^0),
\end{equation}
with
\begin{equation}
\mu_F^0 = \mu_R^0 = \frac{1}{2} \sum_i \sqrt{p_{T,i}^2 + m_i^2}.
\end{equation}
Furthermore, PDF uncertainties are estimated using the \textsc{Hessian} method \cite{Pumplin:2001ct}.  \\

\begin{table}[!t]
\caption{Parton level cross sections for the production of $t\bar{t}H_1$ final states at the LHC at LO and Next-to-Leading Order (NLO). The results are shown along with the theoretical uncertainties due to scale variations (first errors) and PDF uncertainties (second errors). The cross sections were computed for the cases of no cuts on the Higgs boson $p_T$ (first row), for $p_T^H > 50$ GeV (second row) and for $p_T^H > 200$ GeV (third row).}
\label{tab:xsecSM}\vspace*{0.5truecm}
\setlength\tabcolsep{18pt}
\begin{center}
\begin{tabular}{ c  c  c  }
%\hline\hline
\cline{1-3}
\cline{1-3}
 & $\sigma_\textrm{LO}~$ [fb] & $\sigma_\textrm{NLO}~$ [fb] \\ \hline
No cuts & $398.9^{+32.7\%}_{-22.9\%}~(\textrm{scale})^{+1.91\%}_{-1.54\%}~(\textrm{PDF}$) & $470.6^{+5.8\%}_{-9.0\%}~(\textrm{scale})^{+2.2\%}_{-2.1\%}~(\textrm{PDF}$) \\ %\hline
$p_T^H > 50$ GeV & $325.2^{+32.8\%}_{-22.9\%}~(\textrm{scale})^{+1.96\%}_{-1.56\%}~(\textrm{PDF}$) & $382.8^{+5.4\%}_{-8.8\%}~(\textrm{scale})^{+2.3\%}_{-2.1\%}~(\textrm{PDF}$) \\ %\hline
$p_T^H > 200$ GeV & $55.6^{+33.9\%}_{-23.5\%}~(\textrm{scale})^{+2.44\%}_{-1.81\%}~(\textrm{PDF}$) & $69.8^{+8.3\%}_{-10.6\%}~(\textrm{scale})^{+2.9\%}_{-2.6\%}~(\textrm{PDF}$) \\
\cline{1-3}
\cline{1-3}
%\hline
%\hline
\end{tabular}
\end{center}
\end{table}
In \autoref{tab:xsecSM}, we show the results of the cross section both at LO and NLO in the SM. We can see that the NLO corrections imply a  $K$-factor of about $1.17$ in the case when  no cuts are applied on the Higgs boson transverse momentum and for the case where $p_T^H > 50$ GeV. The $K$-factor slightly increase to $1.25$ when a more stringent cut ($p_T^H > 200$ GeV) is applied. Furthermore, the theoretical uncertainties are dominated by those associated to scale variations which significantly decrease when we go from LO to NLO. PDF uncertainties are subleading and mildly dependent on the Higgs $p_T$ cut. Finally, we notice that the $gg$F contribution is dominant accounting for  $\simeq 68$~($\simeq 71.5\%$), at LO~(NLO), of the total cross section in the case of $p_T^H > 50$ GeV and slightly decreasing to $\simeq 59\%$~($\simeq 67\%$) for the $p_T^H > 200$ GeV case.

\begin{figure}[!t]
\caption{The Real ({left}) and  imaginary ({right}) parts of the ratio $c_{t,1}^*/c_{t,1}$ projected on the mixing angles $\alpha_1$ and $\alpha_2$ upon fixing $\beta=0.76$. The solid, dashed, dotted and dot-dashed lines show the contours where $\sigma_{\rm 2HDM}(pp\to t\bar{t}H_1)/\sigma_{\textrm{SM}}(pp\to t\bar{t}H_1)$ is $0.01, 0.1, 1~\mathrm{and}~2$, respectively.}
\label{fig:gLgR}
\centering
\includegraphics[width=0.495\linewidth]{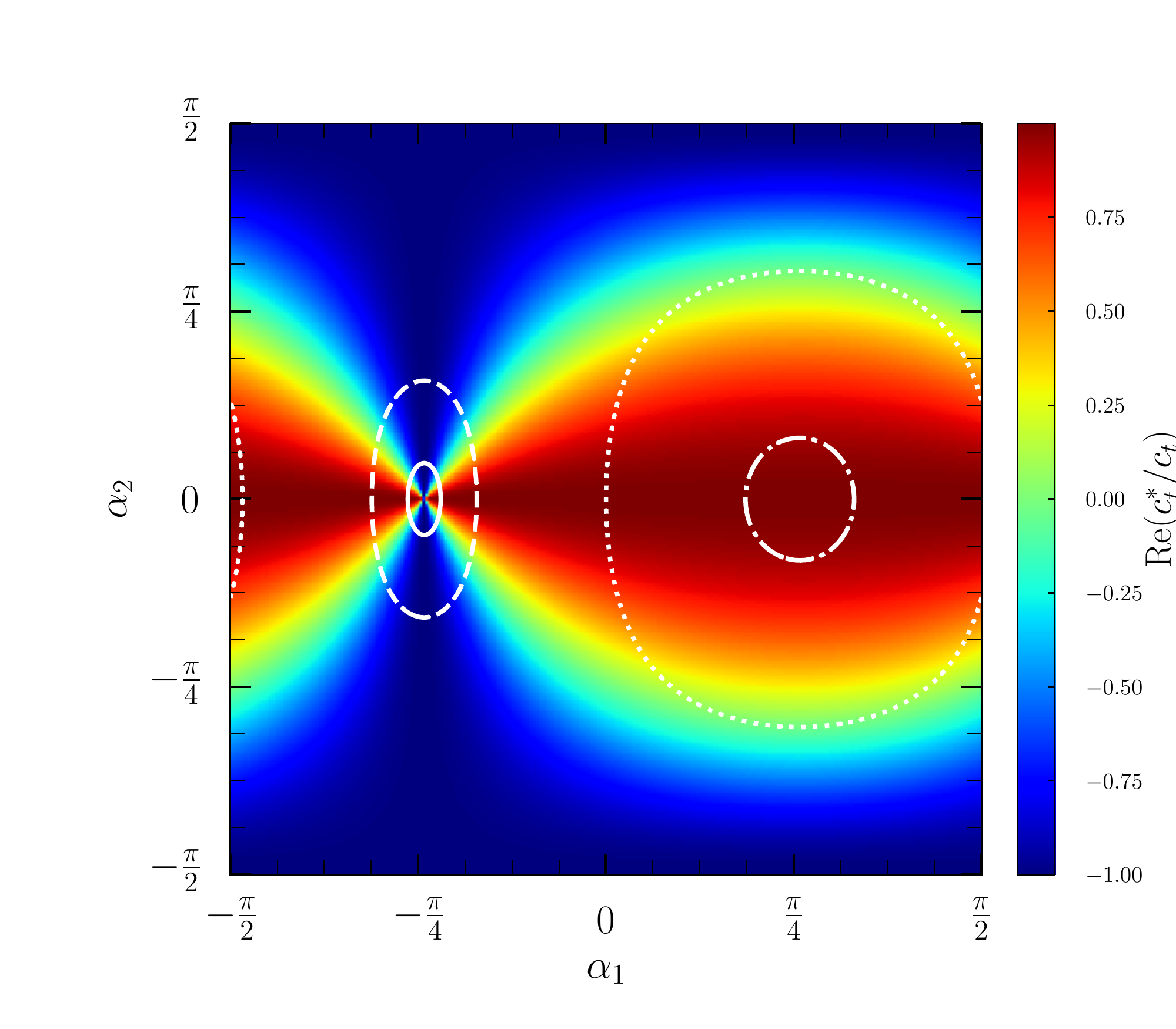}
\hfill
\includegraphics[width=0.495\linewidth]{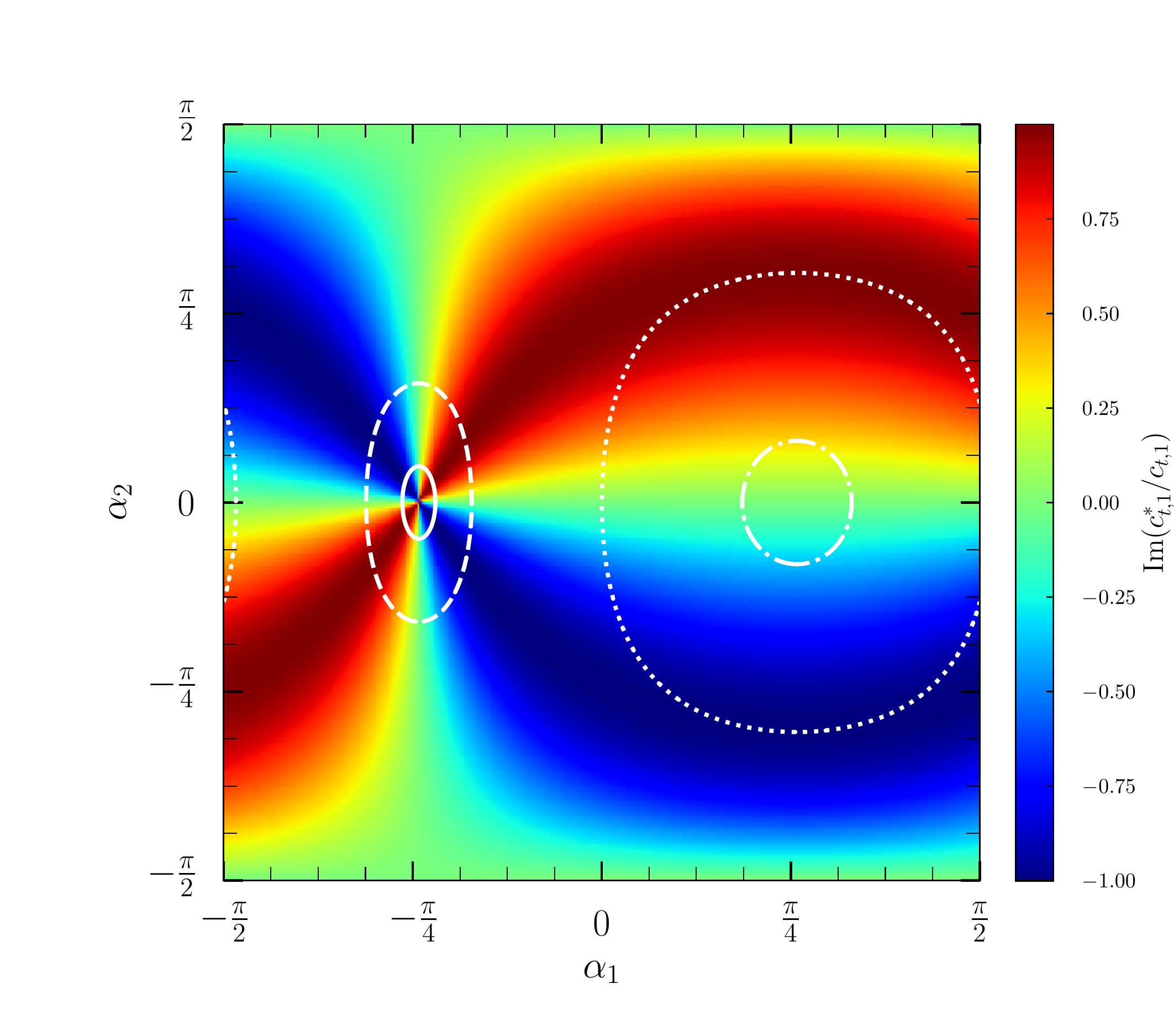}
\end{figure}

In the complex 2HDM, the $t\bar{t}H_1$ coupling is given by
\begin{equation}
\mathcal{L}_{t\bar{t}H_1} = -\frac{m_t}{v}\left(c_{t,1}\bar{t}_Lt_R H_1 +\textrm{H.c.}\right),
\label{eq:ttH}
\end{equation}
with $c_{t,1}=c_{\alpha_2}s_{\beta+\alpha_1}/s_{\beta}-\textrm{i}s_{\alpha_2}/t_{\beta}$ is the $t\bar{t}H_1$ coupling modifier which is independent on the Yukawa-realisation of the 2HDM. % (see \autoref{eq:cu} for all four types of models for the other couplings).
The $t\bar{t}H_1$ production cross section behaves as shown in \autoref{eq:tth}. The presence of the pseudoscalar part in the $t\bar{t}H_1$ coupling can drastically changes the value of the cross section as can be seen in \autoref{fig:gLgR}.

%%%%%%%%%%%%%%%%%%%%%%%%%%%%%%%%%%%%
\subsection{CP-violation Observables in the $t\bar{t}H_1$ Channel}
%%%%%%%%%%%%%%%%%%%%%%%%%%%%%%%%%%%%

In this section, we give an overview of the different observables that we have used in this study to pin-down the spin and CP properties of the SM-like Higgs boson produced in association with a $t\bar{t}$ pair.

First, one can study directly the spin-spin correlations of the $t\bar{t}$ pair by measuring the
differential distribution in $\cos\theta_{\ell^a} \cos\theta_{\ell^b}$ of the emerging leptons,
\begin{equation}
   \frac{1}{\sigma} \frac{\text{d}^2\sigma}{\text{d}\cos\theta_{\ell^a} \text{d}\cos\theta_{\ell^b}} =
   \frac{1}{4}\bigg(1 + \alpha_{\ell^a} P_{a} \cos\theta_{\ell^a} + \alpha_{\ell^b} P_b \cos\theta_{\ell^b} + \alpha_{\ell^a} \alpha_{\ell^b} C_{ab} \cos\theta_{\ell^a} \cos\theta_{\ell^b}\bigg),
 \label{thetadouble}
\end{equation}
where $\alpha_{\ell}$ is the spin analyzing power of the  charged lepton and $\theta_{\ell^{a,b}} = \measuredangle (\hat{\ell}^{a,b}, \hat{S}_{a,b})$, with $\hat{\ell}^{a,b}$ being the direction of flight of the charged lepton in the top quark rest frame and $\hat{S}_{a,b}$ the spin quantization axis in the basis $a$. Furthermore, $C_{ab}$ is the correlation coefficient which is related to the expectation value of $\cos\theta_{\ell^a} \cos\theta_{\ell^b}$ using  \autoref{thetadouble}. In the following, we consider three different bases: the helicity basis ($a=k$), the transverse basis ($a=n$) and the $r$-basis, see, e.g., \cite{Mahlon:1995zn, Bernreuther:2013aga} for more details about the definitions of the spin bases and \cite{Aaboud:2016bit, Sirunyan:2019lnl} for reported measurements of these observables in $t\bar{t}$ production. It was found that the $t\bar{t}$ spin-spin correlations in the transverse and  $r$-bases are good probes of CP-violation, e.g., through the anomalous chromomagnetic and chromoelectric top quark couplings \cite{Bernreuther:2013aga}\footnote{In $t\bar{t}H_1$ production, the contribution of $gg$F is about $70\%$ of the total cross section. Hence, the initial state is mostly Bose-symmetric. Following the recommendations of \cite{Bernreuther:2013aga}, the value of $\cos\theta_\ell$ is multiplied by the sign of the scattering angle $\vartheta = \hat{\textbf{p}}\cdot\hat{\textbf{p}}_t$ with $\hat{\textbf{p}}_t=\textbf{p}_t/|\textbf{p}_t|$  the top quark direction of flight in the $t\bar{t}$ rest frame and $\hat{\textbf{p}}=(0,0,1)$.}.

Furthermore, we consider the opening angle between the two oppositely charged leptons produced in the decays of the top (anti)quarks which is defined by
\begin{eqnarray}
 \cos\varphi_{\ell_a\ell_b} = \frac{\hat{p}_{\ell^+} \cdot \hat{p}_{\ell^-}}{|\hat{p}_{\ell^+}| |\hat{p}_{\ell^-}|},
\end{eqnarray}
where $\hat{p}_{\ell^+}$($\hat{p}_{\ell^-}$) is the direction of the flight of the charged lepton $\ell^+$($\ell^-$) in the parent top (anti)quark rest frame.

The azimuthal angle $\Delta \phi_{\ell^+ \ell^-}=|\phi_{\ell+} - \phi_{\ell^-}|$ is a clean observable to measure the spin-spin correlations between the top and the antitop quarks. The momenta of the charged leptons are usually measured in the laboratory frame \cite{ATLAS-CONF-2018-027,CMS-PAS-TOP-17-014}. This observable shows a high sensitivity to the degree of  correlations between the top (anti)quarks in $t \bar{t}$ production. However, since we are considering the $t\bar{t} H_1$
production mode, the presence of the Higgs boson may wash out the sensitivity of $\Delta\phi$ to the correlations, though we have found this not to be the case.

In addition to the aforementioned observables, we also study the sensitivity of the following angle  \cite{Boudjema:2015nda}
\begin{eqnarray}
\cos\theta_{\ell H_1} = \frac{(\hat{p}_{\ell^+} \times \hat{p}_{H_1}) \cdot  (\hat{p}_{\ell^-} \times \hat{p}_{H_1})}{|(\hat{p}_{\ell^+} \times \hat{p}_{H_1})|  |(\hat{p}_{\ell^-} \times \hat{p}_{H_1})|},
\label{eq:costhetaHL}
\end{eqnarray}
where $\hat{p}_{\ell^+}$, $\hat{p}_{\ell^-}$ and $\hat{p}_{H_1}$ are the directions of flight of the postively-, negatively charged lepton and of the reconstructed Higgs boson in the laboratory frame. The $\theta_{\ell H_1}$ angle defines the angle spanned by the charged lepton momenta projected
onto the plane perpendicular to the Higgs boson direction of flight. This observable can be redefined to yield better dependence on the $CP$-violating effects in the $t\bar{t} H_1$ coupling. We define the new observable as
\begin{eqnarray}
\cos\tilde{\theta}_{\ell H_1} = \lambda \cos\theta_{\ell H_1},
\end{eqnarray}
with $\lambda = \sign((\vec{p}_{b} - \vec{p}_{\bar{b}})\cdot(\vec{p}_{\ell^-} \times \vec{p}_{\ell^+}))$.

One can obtain the polarization of the (anti)top quark by integrating \autoref{thetadouble} over the angle $\theta_\ell^a$ (or $\theta_\ell^b$):
\begin{equation}
   \frac{1}{\sigma} \frac{\text{d}\sigma}{\text{d}\cos\theta_{\ell^\pm}^a} =
   \frac{1}{2}\bigg(1 + \alpha_{\ell^\pm} P_{t,\bar{t}}^a \cos\theta_{\ell^\pm}^a \bigg),
 \label{thetasingle}
\end{equation}
which applies to all the  spin quantization axes used here.

\begin{figure}[!t]
  \caption{Differential distributions for some selected observables. \emph{Top panels:} we display the differential cross section versus $|\Delta\phi_{\ell^+\ell^-}|$ (\emph{left}) in the laboratory frame, and versus $\cos\theta_\ell^n\cos\theta_\ell^n$ (\emph{right}) with $\cos\theta_\ell$ is defined in the top quark rest frame. \emph{Middle panels:} we show the differential cross sections versus the cosine of the opening angle between the two charged leptons' direction of flight (\emph{left}) and versus $\cos\tilde{\theta}_{\ell H}$ (\emph{right}).  \emph{Bottom panels:} The differential cross section as a function of $\cos\omega_6$ (\emph{left}) and as a function of the anti-symmetric combination of $\cos\theta_\ell^k \cos\theta_\ell^n$ (\emph{right}). Red lines are for the SM,  blue lines are for the signal benchmark point with $\alpha_2=0.27$ while green lines are for the pure pseudoscalar case $\alpha_2=\pi/2$ as a comparison, which is of course excluded. The differential distributions are normalized to their total cross sections.}
\label{fig:observables}
\centering
\includegraphics[width=0.49\linewidth]{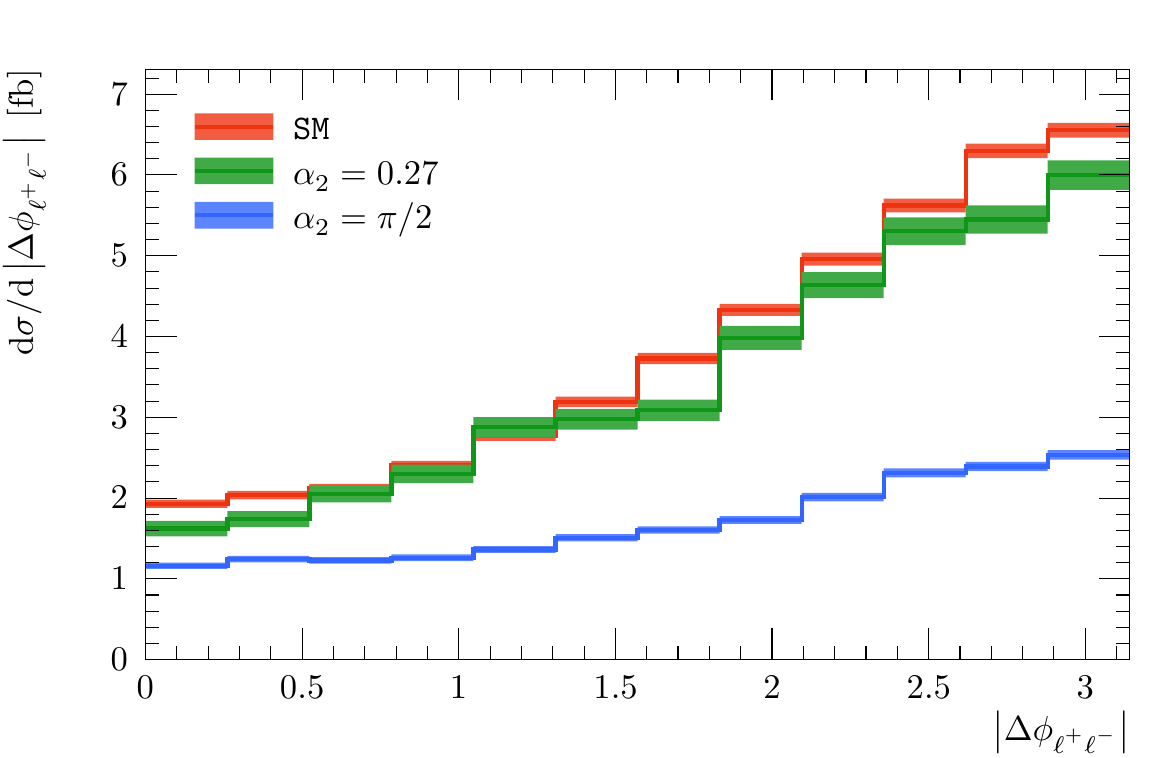}
\hfill
\includegraphics[width=0.49\linewidth]{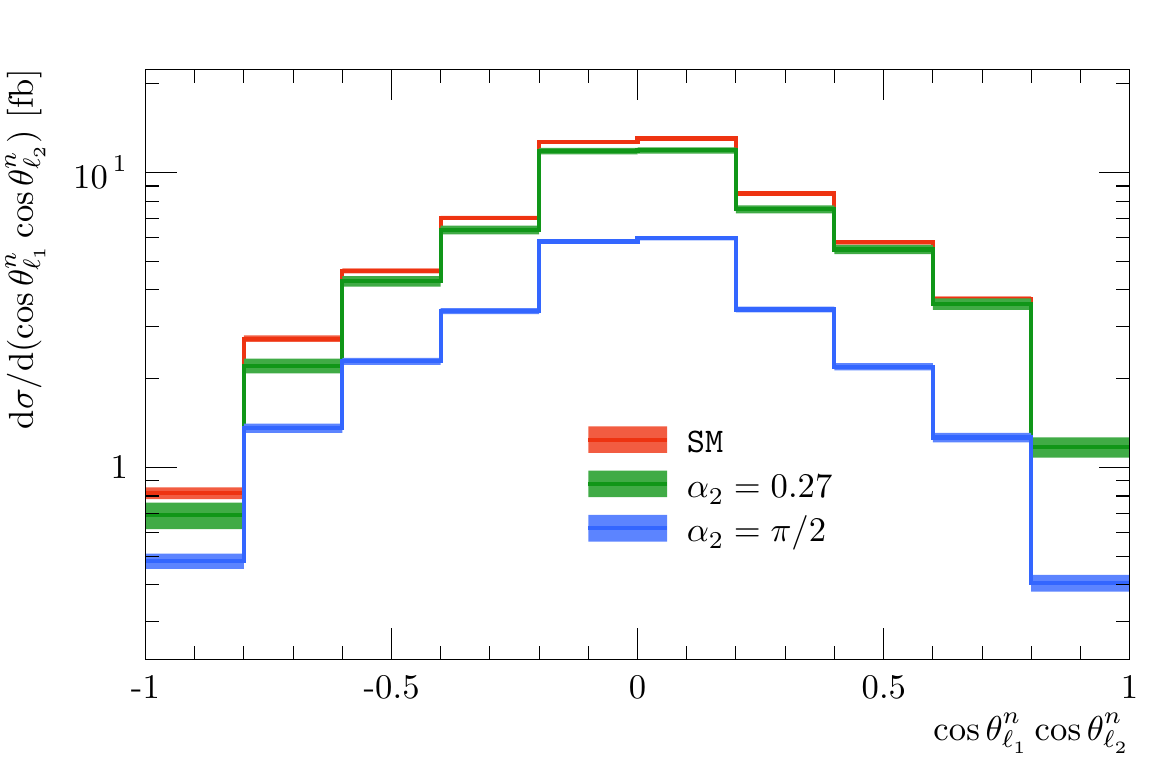}
\vfill
    \includegraphics[width=0.49\linewidth]{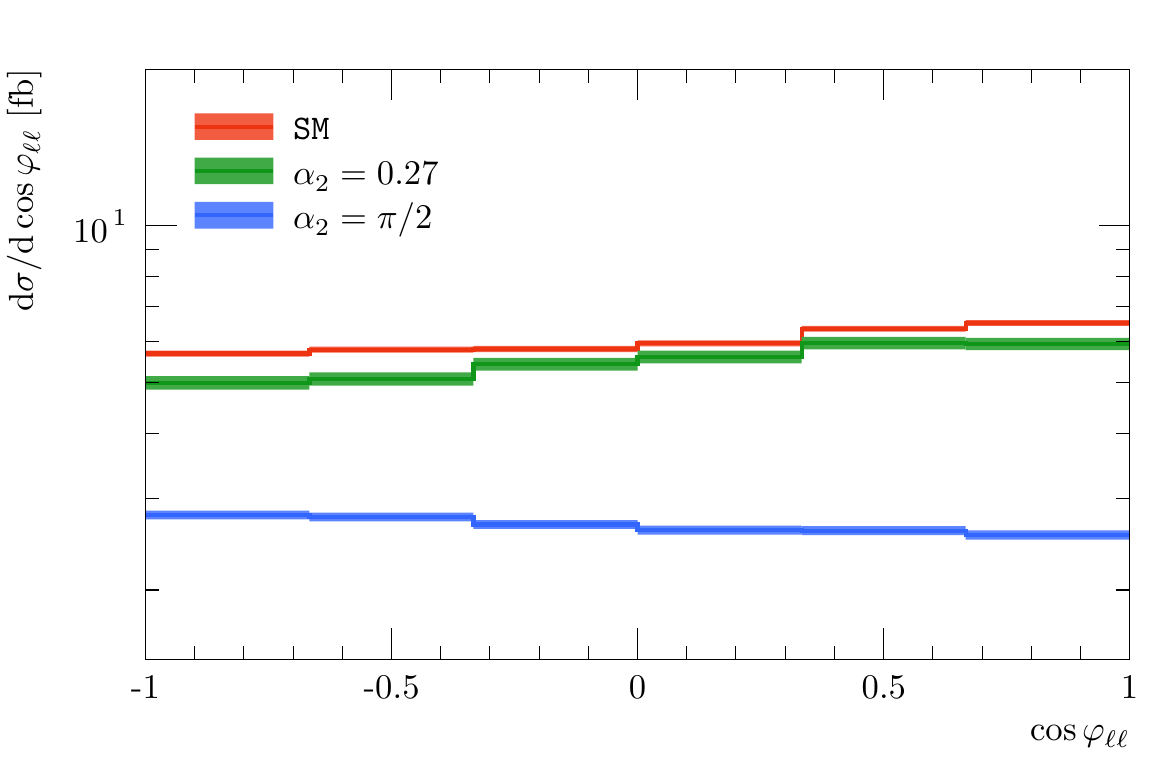}
    \hfill
    \includegraphics[width=0.49\linewidth]{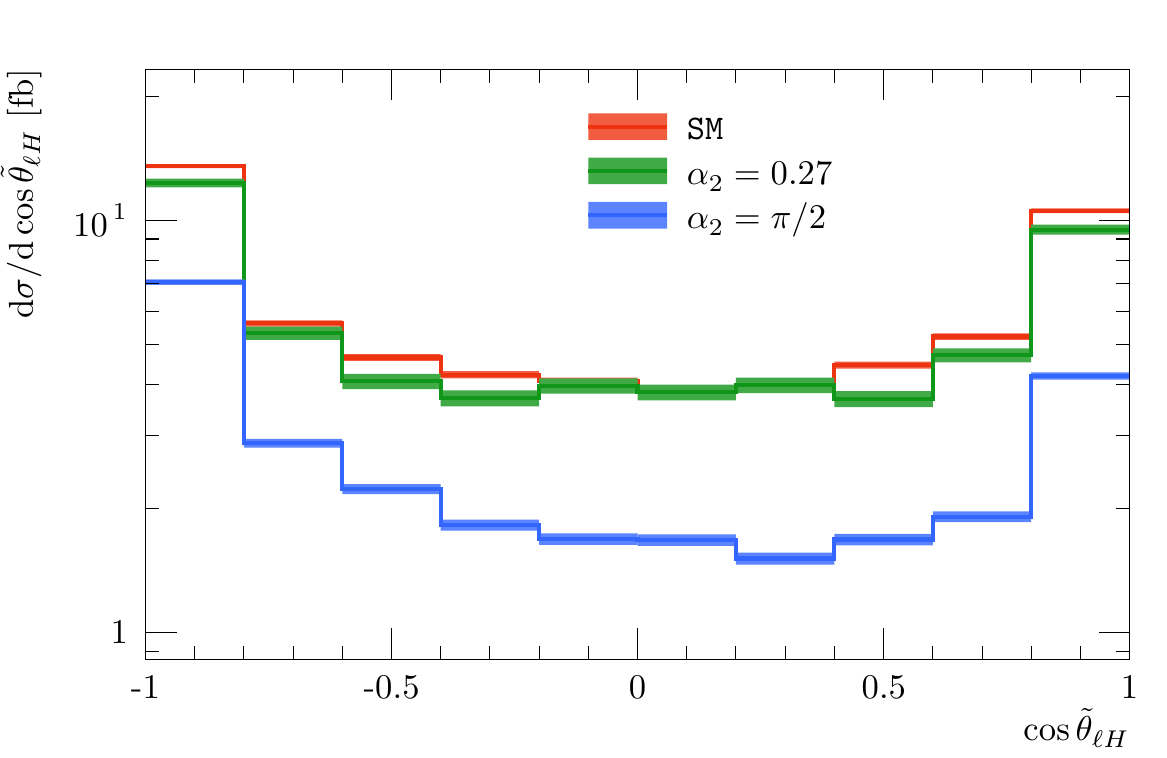}
\vfill
   \includegraphics[width=0.49\linewidth]{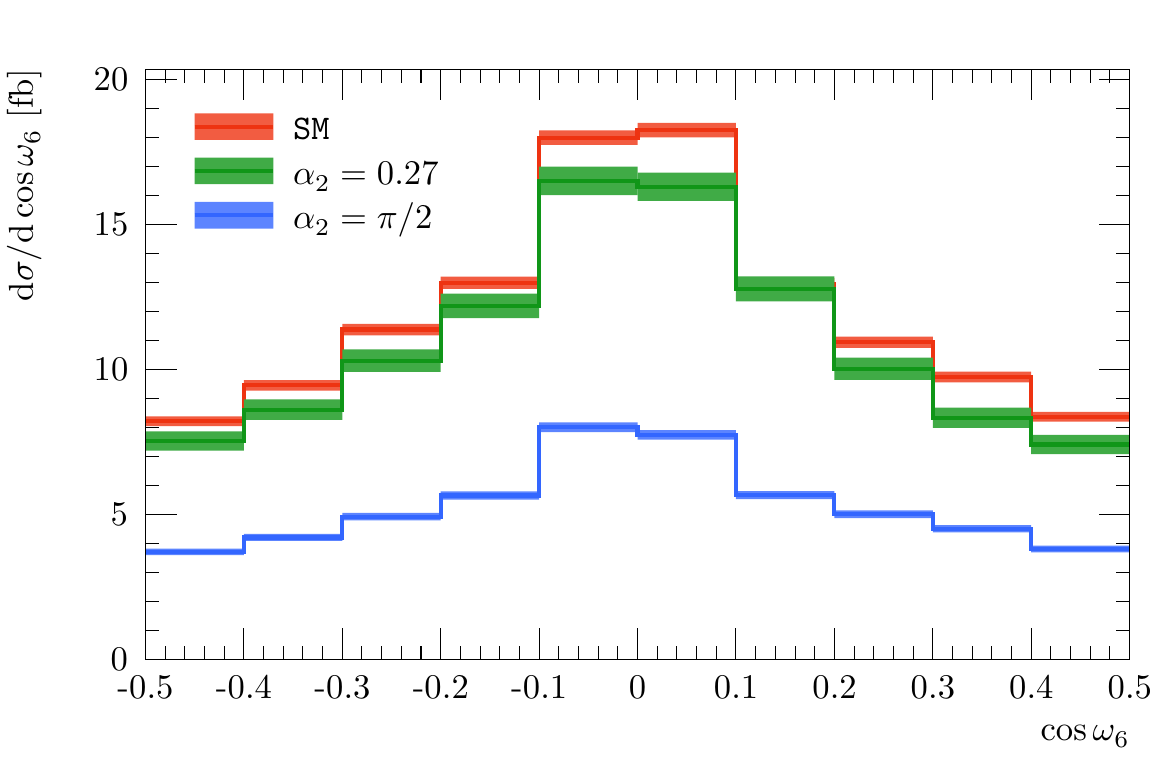}
    \hfill
  \includegraphics[width=0.49\linewidth]{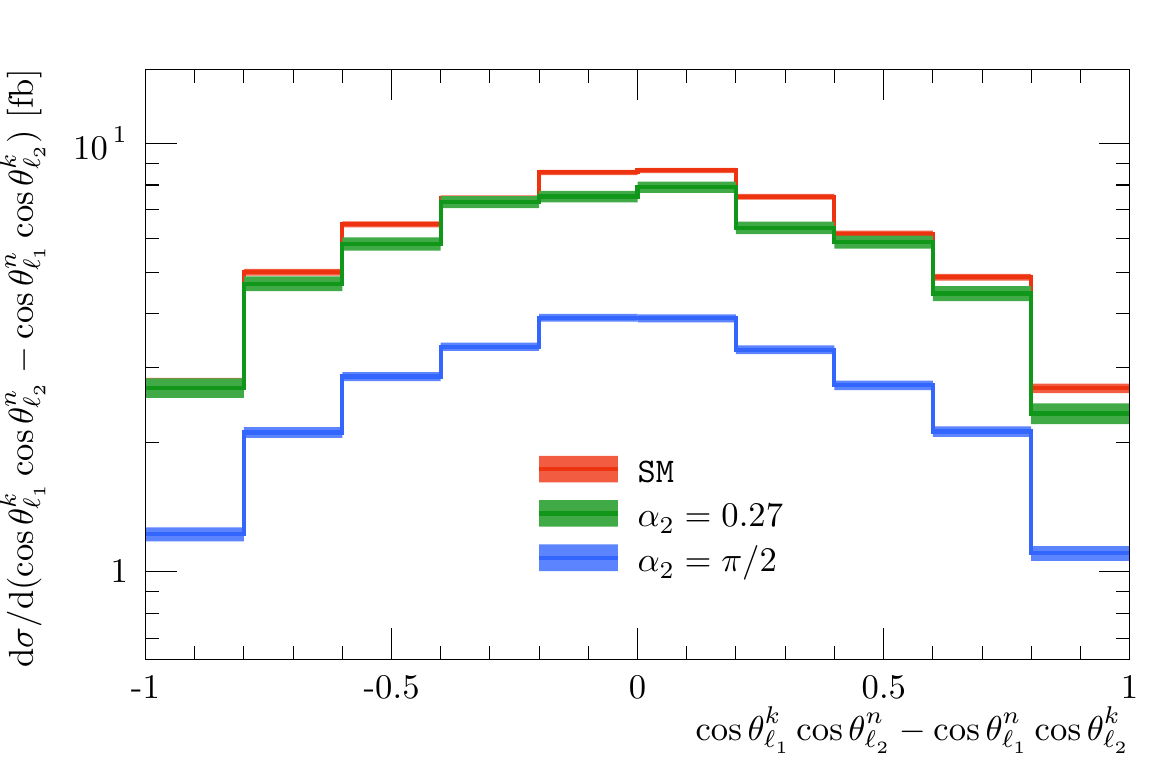}
\end{figure}

It was also found that the energy distributions of the top quark decay products carry some information on the polarization state of the top (anti)quark \cite{Godbole:2011vw, Rindani:2011pk, Prasath:2014mfa, Godbole:2015bda, Jueid:2018wnj, Arhrib:2018bxc, Godbole:2019erb, Arhrib:2019tkr,  Chatterjee:2019brg}. We follow the same definitions used by \cite{Prasath:2014mfa,Jueid:2018wnj} and study the ratios of the different energies. We give the first two observables as follows
\begin{equation}
u = \frac{E_\ell}{E_\ell+E_b}, \qquad
z = \frac{E_b}{E_t},
\end{equation}
where $E_\ell$, $E_b$ and $E_t$ are the energies of the charged lepton,
$b$-jet and top quark in the laboratory frame.
Finally, we consider the energy of the charged leptons in the laboratory frame
\begin{eqnarray}
 x_\ell = \frac{2 E_\ell}{m_t},
\end{eqnarray}
where $m_t = 172.5$ GeV is the  pole mass of the top quark. \\

We complement this analysis by including the sensitivity of some laboratory frame observables introduced in \cite{Faroughy:2019ird}. We found that the observable denoted by $\omega_6$ has higher sensitivity than the others (defined in \cite{Faroughy:2019ird} by $\omega_{b,\ell}^{X,Y}$). We define this angle by
\begin{eqnarray}
\cos\omega_6 = \frac{[(\vec{p}_{\ell^-} \times \vec{p}_{\ell^+})\cdot(\vec{p}_{b} + \vec{p}_{\bar{b}})] [(\vec{p}_{\ell^-} - \vec{p}_{\ell^+}) \cdot (\vec{p}_{b} + \vec{p}_{\bar{b}})] }{|\vec{p}_{\ell^-} \times \vec{p}_{\ell^+}||\vec{p}_{b} + \vec{p}_{\bar{b}}||\vec{p}_{\ell^-} - \vec{p}_{\ell^+}||\vec{p}_{b} + \vec{p}_{\bar{b}}|}
\end{eqnarray}

In \autoref{fig:observables}, we show some observables used in our analysis for the $t\bar{t} H$ in the SM, and in the 2HDM with $\alpha_2=0.27$ (green) and $\alpha_2=\pi/2$. The latter case is shown for comparison only. We can see from \autoref{fig:observables} that the shape of all these observables are slightly changed as we go from the SM to the 2HDM with $\alpha_2 = 0.27$. The only difference between the two cases reside in the total normalization which depends on the cross section.

%%%%%%%%%%%%%%%%%%%%%%%%%%%%%%%%%%%%%%%%%%%%
\subsection{Results}
%%%%%%%%%%%%%%%%%%%%%%%%%%%%%%%%%%%%%%%%%%%%
In this subsection, we show the results of the sensitivity of the observables defined in the previous section. In order to quantify the sensitivity of the various spin observables
to the benchmark points, we compute forward-backward
asymmetries. An asymmetry $\mathcal{A}_\mathcal{O}$ on the observable
$\mathcal{O}$ is defined by
\begin{eqnarray}
\mathcal{A}_\mathcal{O} &=& \frac{N(\mathcal{O} > \mathcal{O}_c ) - N(\mathcal{O} < \mathcal{O}_c)}{N(\mathcal{O} > \mathcal{O}_c) + N(\mathcal{O} < \mathcal{O}_c)} \equiv \frac{N^+ - N^-}{N^+ + N^-},
\end{eqnarray}
where $\mathcal{O}_c$ is a reference point for the observable $\mathcal{O}$ with respect to which the asymmetry is evaluated. For the observable $|\Delta\phi_{\ell^+\ell^-}|$, we choose $\mathcal{O}_c=\pi/2$. While for other angular (energy) observables, we choose $\mathcal{O}_c=0$ ($\mathcal{O}_c=0.5$).

To quantify deviations from the SM expectations, we compute the $\chi^2$ as
\begin{eqnarray}
\chi^2 = \frac{(\mathcal{A}_\mathcal{O} - \mathcal{A}_\mathcal{O}^\textsc{SM})^2}{\sigma_\mathcal{O}^2},
\end{eqnarray}
with $\sigma_\mathcal{O}$  the uncertainty on the measurement of the asymmetry in the SM. We assume that the $N^+$ and $N^-$ are correlated, i.e. measured in the same run of an experiment. In this case, the uncertainty on the asymmetry is given by
\begin{eqnarray}
\sigma_{\mathcal{O}}^2 = \frac{4 N^+ N^-}{N^3},
\end{eqnarray}
where $N = A\times \epsilon \sigma\times \mathcal{L}$. Here, $A\times \epsilon$ is the acceptance times the efficiency of the signal process after full selection, and $\sigma$ is the cross section times the BRs, i.e.,
\begin{equation}
\sigma = \sigma(t\bar{t}H_1) \times \textsc{BR}(H_1\to b\bar{b}) \times \textsc{BR}(t\to b\ell\nu)^2.
\end{equation}
In \autoref{tab:asymmetries}, we show the expected deviations from the SM expectation at $\mathcal{L}=3000~\textrm{fb}^{-1}$.

\begin{table}[!t]
\setlength\tabcolsep{18pt}
\caption{The asymmetries for the SM and 2HDM with $\alpha_2=0.27$. The values of the $\chi^2$ quantifying the deviations from the SM expectations are shown in the fourth column. The $p$-value for different asymmetries, defined in \autoref{eq:pvalue}, are given in the fifth column. The computations are performed for an integrated luminosity of $3000$ fb$^{-1}$. The shorthand notations $c_{\ell^+}^r = \cos\theta_{\ell^+}^k, \cdots$ are used. Details about the calculations are discussed in the text.}
{\begin{tabular}{@{}ccccc@{}}
\hline
Observable &  $\mathcal{A}_\textrm{SM}$ & $\mathcal{A}_{\alpha_2=0.27}$ & $\chi^2$  & $p$-value \\
\hline
\multicolumn{5}{l}{\textbf{Polarisation observables}} \\
$\cos\theta_\ell^k$        &  $4.12 \times 10^{-3}$    &   $5.32 \times 10^{-3}$  &  $6.34 \times 10^{-3}$ & $0.937$\\
$\cos\theta_\ell^n$        &  $4.74 \times 10^{-3}$    &   $4.79 \times 10^{-3}$  &  $8.34\times 10^{-6}$ & $0.997$ \\
$\cos\theta_\ell^r$         &  $-6.54 \times 10^{-4}$   &   $-9.31 \times 10^{-3}$ & $0.33$ & $0.565$ \\
$\cos\varphi_{\ell\ell}$   &  $3.77 \times 10^{-2}$    &   $6.01 \times 10^{-2}$  &  $2.21$ & $0.136$ \\
$u$                                &  $0.232$                       &  $0.237$  & $0.10$ &  $0.751$ \\
$x_\ell$                         &   $-0.832$                     &   $-0.822$ & $1.26$ & $0.259$ \\
$z$                                &  $-0.387$                      &   $-0.401$ & $0.93$ & $0.332$ \\
\texttt{Combined}  & & & & $6.19 \times 10^{-2}$ \\
\hline
\multicolumn{5}{l}{\textbf{Spin-Spin correlations observables}} \\
$\cos\theta_{\ell H}$     &   $-7.84 \times 10^{-2}$ & $-6.68 \times 10^{-2}$ & $0.59$ & $0.43$ \\
$\cos\theta_\ell^k \cos\theta_\ell^k$ &  $6.79 \times 10^{-3}$ & $1.49 \times 10^{-2}$ & $0.28$ & $0.59$ \\
$\cos\theta_\ell^n \cos\theta_\ell^n$ & $7.09 \times 10^{-2}$ & $7.69 \times 10^{-2}$ & $0.15$ & $0.69$ \\
$\cos\theta_\ell^r  \cos\theta_\ell^r$  & $2.98 \times 10^{-2}$ & $3.01 \times 10^{-2}$ & $1.74 \times 10^{-4}$ & $0.98$ \\
$c_\ell^k c_\ell^n - c_\ell^n c_\ell^k$ & $-8.16 \times 10^{-3}$ & $1.22\times 10^{-2}$ & $1.83$ & $0.17$ \\
$c_\ell^k c_\ell^n + c_\ell^n c_\ell^k$ & $-6.28\times 10^{-3}$ & $-1.97 \times 10^{-2}$ & $0.79$ & $0.37$ \\
$c_\ell^k c_\ell^r  - c_\ell^r  c_\ell^k$ &  $3.15\times 10^{-3}$ & $-7.98\times 10^{-3}$ & $0.54$ & $0.46$ \\
$c_\ell^k c_\ell^r  + c_\ell^k c_\ell^k$ & $-3.25\times 10^{-2}$ & $-3.99\times 10^{-2}$ & $0.24$ & $0.62$ \\
$c_\ell^r c_\ell^n  - c_\ell^n c_\ell^r$ & $-8.88 \times 10^{-4}$ & $-1.73\times 10^{-2}$ & $1.18$ & $0.27$ \\
$c_\ell^r c_\ell^n  + c_\ell^n c_\ell^r$ & $-2.53\times 10^{-3}$ & $9.31\times 10^{-3}$ & $0.61$ & $0.43$ \\
$|\Delta\phi_{\ell^+ \ell^-}|$    &  $0.39$ & $0.35$ & $5.81$ & $1.59
\times 10^{-2}$ \\
\texttt{Combined} & & & & $1.21 \times 10^{-2}$ \\
\hline
\multicolumn{5}{l}{\textbf{$CP$-odd laboratory-frame observables}} \\
$\cos\tilde{\theta}_{\ell H}$ & $-1.2\times 10^{-2}$ & $3.99\times 10^{-3}$ & $1.14$  & $0.28$ \\
$\cos\omega_6$ & $-6.11\times 10^{-3}$ & $1.38\times 10^{-2}$ & $1.75$ & $0.18$ \\
\texttt{Combined} & & & & $8.89 \times 10^{-2}$ \\
\hline
\textbf{All the combinations} & & & & $1.87\times 10^{-2}$ \\
\hline
\end{tabular} \label{tab:asymmetries}}
\end{table}

We can see that, for $\alpha_2=0.27$, the $\chi^2$ can be larger than $1$ for seven observables: $\cos\varphi_{\ell\ell}, x_\ell, c_\ell^k c_\ell^n - c_\ell^n c_\ell^k, c_\ell^r c_\ell^n  - c_\ell^n c_\ell^r, |\Delta\phi_{\ell^+ \ell^-}|, \cos\tilde{\theta}_{\ell H}, ~\textrm{and}~\cos\omega_6$.  After combining all the observables in \autoref{tab:asymmetries}, the $\chi^2$ can reach about $19.2$. However, the naive $\chi^2$ combination may become obsolete, or misleading. In order to improve this combination, we compute the $p$-value given by
\begin{eqnarray}
p = \int_{\chi^2_\textrm{min}}^{\infty} f(x, N_{\textrm{DoF}}) \textrm{d}x
\label{eq:pvalue}
\end{eqnarray}
with $f(x, N_{\textrm{DoF}})$ is the $\chi^2$ probability distribution function for $N_\textrm{DoF}$ degrees of freedom\footnote{The number of degrees of freedom ($N_\textrm{DoF})$ is the number of observables used in the fit minus the number of free parameters (i.e., here we have one free parameter, $\alpha_2$).}, and  $\chi^2_{\textrm{min}} \equiv \chi^2_{\alpha_2=0.27}$. The $p$-value gives to what extent the null hypothesis (SM $CP$-conserving case) is excluded. Values of $p$-value smaller than $0.05$ implies that the null hypothesis is excluded. We show the $p$-value as a function of $\alpha_2$ in \autoref{tab:asymmetries}. In \autoref{tab:asymmetries}, the label \texttt{Combined} refers to the combination of different asymmetries for each category removing observables with $\chi^2 < 1$: (i) In the Polarisation observables, \texttt{Combined} refers to the combination of $\cos\varphi_{\ell\ell}$ and $x_\ell$; (ii) while in the Spin-Spin correlation observables, \texttt{Combined} refers to the combination of three observables: $c_\ell^k c_\ell^n - c_\ell^n c_\ell^k$,  $c_\ell^r c_\ell^n  - c_\ell^n c_\ell^r$ and $|\Delta\phi_{\ell^+ \ell^-}|$. As we can see that the important role played by the spin-spin correlations asymmetries for which the $p$-value is about $1.59 \times 10^{-2}$.  The results depends weakly on $\beta$ and $\alpha_1$ in our favored region ($t_{\beta}\sim1$ and $\alpha_1\sim0$), because the observables are sensitive only to the $t\bar{t}H_1$ CP-violating phase $\simeq s_{\alpha_2}/t_{\beta}$ in this region. It is worth to mention that the results can be further improved by using different approaches. On the one hand, the weighted fits, as used in \cite{Faroughy:2019ird}, may improve the results since another important factor which we did not take advantage of is the total cross section for a given value of $\alpha_2$. On the other hand, methods based on Machine Learning may play important role in the determination of the maximum allowed $CP$-violating phase in the $t\bar{t}H_1$ coupling \cite{Ren:2019xhp}.

\section{Conclusions}
\label{sec:dnc}

In this work, we have analyzed  soft CP-violating effects in both EDMs and LHC
phenomenology in  a 2HDM with soft CP-violation. In this scenario, the
mixing angle $\alpha_2$ is the key parameter measuring the size of
CP-violation since the CP-violating phases in $H_1f\bar{f}$ Yukawa
vertices are proportional to $s_{\alpha_2}$.

We have considered all four standard types of Yukawa couplings, named
Type I-IV models, in our analysis. In Type I and IV models, there
is no cancellation mechanism in electron EDM calculations,
leading to a very strict
constraint on the CP-violating phase
$|\arg{c_{t/\tau,1}}|\lesssim8.2\times10^{-4}$, which renders all CP-violating
effects unobservable in further collider studies for these two models.

In Type II and III models, we have discussed two scenarios: (a) $H_{2,3}$
are closed in mass while $\alpha_3$ is away from $0$ or $\pi/2$; and
(b) $H_{2,3}$ have a large mass splitting while $\alpha_3$ must appear
close to $0$ or $\pi/2$. The cancellation behavior in the electron EDM leads
to a larger allowed region for $\alpha_2$ in both scenarios. In such two
models, $t_{\beta}$ is favored to be close to $1$, whose location
depends weakly on the masses of the heavy (pseudo)scalars, with a strong correlation with
$\alpha_1$. The electron EDM alone cannot set constraints on $\alpha_2$
directly. In the Type II model, $|\alpha_2|\lesssim0.09$ is estimated from the
neutron EDM constraint if we consider only the central value
estimation, and this constraint can be as weak as $\lesssim0.15$ if
theoretical uncertainty in neutron EDM estimation is also considered. In the Type III model,
no constraint can be drawn from the neutron EDM and $|\alpha_2|\lesssim0.27$ is
estimated from LHC constraints if $m_2\simeq500~\textrm{GeV}$. Such
results mean that in Type III model, both Scenario (a) and (b), the CP-violation phase
$|\arg(c_{t,1})|\simeq|s_{\alpha_2}/t_{\beta}|$ can reach as large as $\simeq0.28$,
which leads us to consider further phenomenology of the model. This result is
independent on $\alpha_3$, and depends weakly on $\alpha_1$ in its allowed region (close
to zero). Other LHC direct searches do not set further limits for the 2HDM.

Our analysis shows the importance of further neutron EDM measurements
to an accuracy of $\mathcal{O}(10^{-27}~e\cdot\textrm{cm})$. An
$\alpha_2$ of the size $\sim\mathcal{O}(0.1)$ will lead to significant
non-zero results in such experiments. If CP-violation in the Higgs sector
exists, as we have discussed, first evidence of it is expected to appear in the
neutron EDM measurements. Conversely, if there is still a null
result for the neutron EDM, direct constraints on $|\alpha_2|$ can be
pushed to about $4\times10^{-3}$ in the Type II model and
$2\times10^{-2}$ in the Type III model. Such a strict constraint can
exclude the Type II model as an explanation of matter-antimatter asymmetry
in the Universe. Thus, we conclude that, for models
in which a cancellation mechanism can appear in the electron EDM, the neutron EDM
measurements are good supplements to find evidence of
CP-violation or set constraints on the CP-violating angle directly.

We have also performed a phenomenological study of soft CP-violation in the 2HDM  for the case of
$t\bar{t}H_1$ associate production at the LHC with a luminosity of
$3000~\textrm{fb}^{-1}$. With fixed $\beta$ and $\alpha_{1,2}$, its
properties are independent of the mixing angle $\alpha_3$ and the masses
of the heavy (pseudo)scalars $H_{2,3}$ and $H^\pm$. Upon choosing the bencdhmark point $\beta=0.76$,
$\alpha_1=0.02$ and $\alpha_2=0.27$ (corresponding to the case
$m_{2,3}\simeq500~\textrm{GeV}$ with the maximal CP-violation phase $|\arg(c_{t,1})|\simeq0.28$ in Type III model),
we constructed top (anti)quark spin dependent observables  and tested their deviations from the
SM. Amongst these, a single observable, the azimuthal angle between the two leptons from fully leptonic $t\bar t$ decays,
$\Delta\phi_{\ell^+\ell^-}$, is the most sensitive one, with $\chi^2=5.81$. On the other hand, by combining asymmetries constructed out of seven spin-dependent observables, we found that the $p$-value is about $1.87 \times 10^{-2}$  meaning that the null hypothesis (the $CP$-conserving case) can be excluded by the use of these observables and one can probe the maximum allowed $CP$-violating phase in the $t\bar{t}H_1$ coupling obtained for $\alpha_2=0.27$.
Thus the LHC experiments can provide a complementary cross-check of the EDM results.

Finally, we note that we did not perform phenomenological study of the heavy (pseudo)scalars ($H_{2,3}$ or $H^{\pm}$) in this paper. In this case, interference effects with the SM backgrounds may become very important, and thus need a dedicated treatment which we postpone for a forthcoming paper.

\subsection*{Acknowledgements}

We thank Abdesslam Arhrib, Jianqi Chen, Nodoka Yamanaka, Fa Peng Huang,
Qi-Shu Yan, Hao Zhang, and Shou-hua Zhu for helpful discussion.
We also thank Abdesslam Arhrib for collaboration at the beginning of this
project. The work of KC and YNM was supported in part by the MoST of Taiwan under
the grant no. 107-2112-M-007-029-MY3. The
work of AJ was supported by the National Research Foundation of Korea
under grant no. NRF-2019R1A2C1009419. AJ would like to thank the CERN Theory Department and the HECAP Section of the Abdus Salam International Centre for Theoretical Physics for their hospitality where part of this work has been done.  SM is supported in part through
the NExT Institute and the STFC CG ST/L000296/1 award.

\appendix
\section{Yukawa Couplings}
\label{app:Yuk}
Following the parameterization in \autoref{eq:para}, we list the Yukawa couplings in the mass eigenstate basis explicitly \cite{ElKaffas:2006gdt,Osland:2008aw,Arhrib:2010ju} in terms of the mixing angles $\beta,\alpha_{1,2,3}$. By denoting with $c_{f,i}^X$ the Yukawa coupling $c_{f,i}$ in the 2HDM Type $X$ ($X=\textrm{I}-\textrm{IV}$) below, we have the following:
\begin{eqnarray}
\label{eq:cu}
c_{U_i,1}^{\textrm{I}-\textrm{IV}}&=&\frac{c_{\alpha_2}s_{\beta+\alpha_1}}{s_{\beta}}-\textrm{i}\frac{s_{\alpha_2}}{t_{\beta}},\\
c_{U_i,2}^{\textrm{I}-\textrm{IV}}&=&\frac{c_{\beta+\alpha_1}c_{\alpha_3}-s_{\beta+\alpha_1}s_{\alpha_2}s_{\alpha_3}}{s_{\beta}}-\textrm{i}\frac{c_{\alpha_2}s_{\alpha_3}}{t_{\beta}},\\
c_{U_i,3}^{\textrm{I}-\textrm{IV}}&=&-\frac{c_{\beta+\alpha_1}s_{\alpha_3}+s_{\beta+\alpha_1}s_{\alpha_2}c_{\alpha_3}}{s_{\beta}}-\textrm{i}\frac{c_{\alpha_2}c_{\alpha_3}}{t_{\beta}},
\end{eqnarray}
\begin{eqnarray}
c_{D_i,1}^{\textrm{I},\textrm{III}}&=&\frac{c_{\alpha_2}s_{\beta+\alpha_1}}{s_{\beta}}+\textrm{i}\frac{s_{\alpha_2}}{t_{\beta}},\nonumber\\
c_{D_i,1}^{\textrm{II},\textrm{IV}}&=&\frac{c_{\alpha_2}c_{\beta+\alpha_1}}{c_{\beta}}-\textrm{i}s_{\alpha_2}t_{\beta},\\
c_{D_i,2}^{\textrm{I},\textrm{III}}&=&\frac{c_{\beta+\alpha_1}c_{\alpha_3}-s_{\beta+\alpha_1}s_{\alpha_2}s_{\alpha_3}}{s_{\beta}}+\textrm{i}\frac{c_{\alpha_2}s_{\alpha_3}}{t_{\beta}},\nonumber\\ c_{D_i,2}^{\textrm{II},\textrm{IV}}&=&-\frac{s_{\beta+\alpha_1}c_{\alpha_3}+c_{\beta+\alpha_1}s_{\alpha_2}s_{\alpha_3}}{c_{\beta}}-\textrm{i}c_{\alpha_2}s_{\alpha_3}t_{\beta},\\
c_{D_i,3}^{\textrm{I},\textrm{III}}&=&-\frac{c_{\beta+\alpha_1}s_{\alpha_3}+s_{\beta+\alpha_1}s_{\alpha_2}c_{\alpha_3}}{s_{\beta}}+\textrm{i}\frac{c_{\alpha_2}c_{\alpha_3}}{t_{\beta}},\nonumber\\ c_{D_i,3}^{\textrm{II},\textrm{IV}}&=&\frac{s_{\beta+\alpha_1}s_{\alpha_3}-c_{\beta+\alpha_1}s_{\alpha_2}c_{\alpha_3}}{c_{\beta}}-\textrm{i}c_{\alpha_2}c_{\alpha_3}t_{\beta},
\end{eqnarray}
\begin{eqnarray}
c_{\ell_i,1}^{\textrm{I},\textrm{IV}}&=&\frac{c_{\alpha_2}s_{\beta+\alpha_1}}{s_{\beta}}+\textrm{i}\frac{s_{\alpha_2}}{t_{\beta}},\nonumber\\
c_{\ell_i,1}^{\textrm{II},\textrm{III}}&=&\frac{c_{\alpha_2}c_{\beta+\alpha_1}}{c_{\beta}}-\textrm{i}s_{\alpha_2}t_{\beta},\\
c_{\ell_i,2}^{\textrm{I},\textrm{IV}}&=&\frac{c_{\beta+\alpha_1}c_{\alpha_3}-s_{\beta+\alpha_1}s_{\alpha_2}s_{\alpha_3}}{s_{\beta}}+\textrm{i}\frac{c_{\alpha_2}s_{\alpha_3}}{t_{\beta}},\nonumber\\
c_{\ell_i,2}^{\textrm{II},\textrm{III}}&=&-\frac{s_{\beta+\alpha_1}c_{\alpha_3}+c_{\beta+\alpha_1}s_{\alpha_2}s_{\alpha_3}}{c_{\beta}}-\textrm{i}c_{\alpha_2}s_{\alpha_3}t_{\beta},\\
c_{\ell_i,3}^{\textrm{I},\textrm{IV}}&=&-\frac{c_{\beta+\alpha_1}s_{\alpha_3}+s_{\beta+\alpha_1}s_{\alpha_2}c_{\alpha_3}}{s_{\beta}}+\textrm{i}\frac{c_{\alpha_2}c_{\alpha_3}}{t_{\beta}},\nonumber\\
c_{\ell_i,3}^{\textrm{II},\textrm{III}}&=&\frac{s_{\beta+\alpha_1}s_{\alpha_3}-c_{\beta+\alpha_1}s_{\alpha_2}c_{\alpha_3}}{c_{\beta}}-\textrm{i}c_{\alpha_2}c_{\alpha_3}t_{\beta}.
\end{eqnarray}

\section{Scalar Couplings}
\label{app:sca}
The scalar couplings in the potential can be expressed using the physical parameters as \cite{ElKaffas:2006gdt,Osland:2008aw,Arhrib:2010ju}
\begin{eqnarray}
\lambda_1&=&\frac{1}{c^2_{\beta}v^2}\left[c^2_{\beta+\alpha_1}c^2_{\alpha_2}m^2_1
+(c_{\beta+\alpha_1}s_{\alpha_2}s_{\alpha_3}+s_{\beta+\alpha_1}c_{\alpha_3})^2m^2_2\right.\nonumber\\
&&\left.+(c_{\beta+\alpha_1}s_{\alpha_2}c_{\alpha_3}-s_{\beta+\alpha_1}s_{\alpha_3})^2m^2_3-s^2_{\beta}\mu^2\right],\\
\lambda_2&=&\frac{1}{s^2_{\beta}v^2}\left[s^2_{\beta+\alpha_1}c^2_{\alpha_2}m^2_1
+(c_{\beta+\alpha_1}c_{\alpha_3}-s_{\beta+\alpha_1}s_{\alpha_2}s_{\alpha_3})^2m^2_2\right.\nonumber\\
&&\left.+(s_{\beta+\alpha_1}s_{\alpha_2}c_{\alpha_3}+c_{\beta+\alpha_1}s_{\alpha_3})^2m^2_3-c^2_{\beta}\mu^2\right],\\
\lambda_3&=&\frac{1}{s_{2\beta}v^2}\left[s_{2(\beta+\alpha_1)}\left(c^2_{\alpha_2}m^2_1+(s^2_{\alpha_2}s^2_{\alpha_3}-c^2_{\alpha_3})m^2_2
+(s^2_{\alpha_2}c^2_{\alpha_3}-s^2_{\alpha_3})m^2_3\right)\right.\nonumber\\
&&\left.+s_{\alpha_2}s_{2\alpha_3}c_{2(\beta+\alpha_1)}(m^2_3-m^2_2)\right]+\frac{2m^2_{\pm}-\mu^2}{v^2},\\
\lambda_4&=&\frac{1}{v^2}\left(s^2_{\alpha_2}m^2_1+c^2_{\alpha_2}s^2_{\alpha_3}m^2_2+c^2_{\alpha_2}c^2_{\alpha_3}m^2_3+\mu^2-2m^2_{\pm}\right),\\
\lambda_5&=&\frac{1}{v^2}\left(\mu^2-s^2_{\alpha_2}m^2_1-c^2_{\alpha_2}s^2_{\alpha_3}m^2_2-c^2_{\alpha_2}c^2_{\alpha_3}m^2_3\right)\nonumber\\
&&-\frac{\textrm{i}}{s_{2\beta}v^2}
\left[c_{\beta}\left(c_{\beta+\alpha_1}s_{2\alpha_2}m_1^2
-(c_{\beta+\alpha_1}s_{2\alpha_2}s^2_{\alpha_3}+s_{\beta+\alpha_1}c_{\alpha_2}s_{2\alpha_3})m^2_2\right.\right.\nonumber\\
&&\left.+(s_{\beta+\alpha_1}c_{\alpha_2}s_{2\alpha_3}-c_{\beta+\alpha_1}s_{2\alpha_2}c^2_{\alpha_3})m^2_3\right)
+s_{\beta}\left(s_{\beta+\alpha_1}s_{2\alpha_2}m^2_1\right.\nonumber\\
&&\left.\left.+(c_{\beta+\alpha_1}c_{\alpha_2}s_{2\alpha_3}-s_{\beta+\alpha_1}s_{2\alpha_2}s^2_{\alpha_3})m^2_2
-(c_{\beta+\alpha_1}c_{\alpha_2}s_{2\alpha_3}+s_{\beta+\alpha_1}s_{2\alpha_2}c^2_{\alpha_3})m^2_3\right)\right].~~~~
\end{eqnarray}
Consider the-bounded-from-below conditions as \cite{Branco:2011iw}
\begin{equation}
\lambda_1>0,\quad\lambda_2>0,\quad\lambda_3>-\sqrt{\lambda_1\lambda_2},\quad\lambda_3+\lambda_4-\left|\lambda_5\right|>-\sqrt{\lambda_1\lambda_2},
\end{equation}
then
$\mu^2\lesssim(450~\textrm{GeV})^2$ is favored and thus we choose $\mu^2=(450~\textrm{GeV})^2$ in the analysis.

The couplings between neutron and charged scalars $c_{i,\pm}$ are \cite{Arhrib:2010ju}
\begin{eqnarray}
c_{i,\pm}&=&c_{\beta}(s^2_{\beta}(\lambda_1-\lambda_4-\textrm{Re}(\lambda_5))+c^2_{\beta}\lambda_3)R_{i1}\nonumber\\
&&+s_{\beta}(c^2_{\beta}(\lambda_2-\lambda_4-\textrm{Re}(\lambda_5))+s^2_{\beta}\lambda_3)R_{i2}+s_{\beta}c_{\beta}\textrm{Im}(\lambda_5)R_{i3},
\end{eqnarray}
where $R$ is the matrix in \autoref{eq:R}. These couplings are useful in the calculations of fermionic EDMs from the contribution of a charged Higgs boson.

\section{Loop Integrations for EDM}
\label{app:loop}
The loop functions in the calculation of the Barr-Zee diagrams are \cite{Barr:1990vd,Chang:1990sf,Leigh:1990kf,Abe:2013qla,Brod:2013cka,Cheung:2014oaa,Altmannshofer:2015qra}:
\begin{eqnarray}
f(z)&=&\frac{z}{2}\int_0^1dx\frac{1-2x(1-x)}{x(1-x)-z}\ln\left(\frac{x(1-x)}{z}\right),\\
g(z)&=&\frac{z}{2}\int_0^1dx\frac{1}{x(1-x)-z}\ln\left(\frac{x(1-x)}{z}\right),\\
h(z)&=&\frac{z}{2}\int_0^1dx\frac{1}{x(1-x)-z}\left[\frac{z}{x(1-x)-z}\ln\left(\frac{x(1-x)}{z}\right)-1\right],\\
F(x,y)&=&\frac{yf(x)-xf(y)}{y-x};\quad G(x,y)=\frac{yg(x)-xg(y)}{y-x},\\
H^a_i(z)&=&z\int_0^1dx\frac{(1-x)^2(x-4+x(z_{\pm,W}-z_{WH_i}^{-1}))}{x+(1-x)z_{WH_i}-x(1-x)z}\ln\left(\frac{x+(1-x)z_{WH_i}}{x(1-x)z}\right),\\
H^b_i(z)&=&2z\int_0^1dx\frac{x(1-x)^2}{x+(1-x)z_{\pm,i}-x(1-x)z}\ln\left(\frac{x+(1-x)z_{\pm,i}}{x(1-x)z}\right).
\end{eqnarray}
Denoting
\begin{eqnarray}
a_x=x(1-x),\quad b=a_x/z_a,\quad A=x+y/z_a,\quad B=A-a_x,\quad B'=A-a_y,\nonumber\\
C=\frac{A}{B}\ln\frac{A}{a_x}-1,\quad C'=\frac{a_x}{B}\ln\frac{A}{a_x}-1,\quad C''=\frac{a_y}{B'}\ln\frac{A}{a_y}-1\;,
\end{eqnarray}
the loop functions in the non-Barr-Zee type diagrams with a $W$ boson are \cite{Leigh:1990kf}
\begin{eqnarray}
(D_W^a)_i&=&-\frac{1}{2}\mathop{\int}_0^1dx\mathop{\int}_0^{1-x}dy\frac{x}{B}\left[\frac{2C}{B}(3A-2xy)-3+\frac{2xy}{a_x}\right],\\
(D_W^b)_i&=&\mathop{\int}_0^1dx\mathop{\int}_0^{1-x}dyx\left[C'\left(\frac{3A-2xy}{B^2}+\frac{1+\frac{3x}{2a_x}(1-2y+B)}{B}\right)+\frac{3A-2xy}{2a_xB}\right],\\
(D_W^c)_i&=&\mathop{\int}_0^1dx\mathop{\int}_0^{1-x}dy\frac{x^2y}{a_x(1-y-b)}\left[\frac{b}{1-y-b}\ln\frac{1-y}{b}-1\right],\\
(D_W^d)_i&=&-\frac{1}{8}\mathop{\int}_0^1dx\mathop{\int}_0^{1-x}dy\left[\frac{1}{Bz_{WH_i}}\left(1-\frac{2Ca_x}{B}\right)+\frac{x}{B}\left(1-\frac{2CA}{B}\right)\right],\\
(D_W^e)_i&=&\frac{1}{8}\mathop{\int}_0^1dx\mathop{\int}_0^{1-x}dy\frac{x}{a_x}\nonumber\\
&&\times\left[\frac{C'}{B^2}\left(xa_x(2x-1)+Bx(3x-1)-2B^2\right)-2+\frac{x(2x-1)}{2B}\right].
\end{eqnarray}
The loop functions in the non-Barr-Zee type diagrams with a $Z$ boson are instead \cite{Leigh:1990kf}
\begin{eqnarray}
(D_Z^a)_i&=&\mathop{\int}_0^1dx\mathop{\int}_0^{1-x}dy\frac{2x}{a_x}\left[1+C'\left(1+\frac{x(1-x-y)}{2B}\right)\right],\\
(D_Z^b)_i&=&\mathop{\int}_0^1dx\mathop{\int}_0^{1-x}dy\frac{x^2y}{a_x(1-y-b)}\left[\frac{b}{1-y-b}\ln\frac{1-y}{b}-1\right],\\
(D_Z^c)_i&=&\mathop{\int}_0^1dx\mathop{\int}_0^{1-x}dy\frac{1}{a_y}\left[y-x+C''\left(y-x+\frac{y^2(1-x-y)}{B'}\right)\right].
\end{eqnarray}
In the functions $(D_W^p)_i$, we have $z_a\equiv z_{WH_i}$ while, in the functions $(D_Z^p)_i$, we have $z_a\equiv z_{ZH_i}$. Last, the loop function for the Weinberg operator is \cite{Brod:2013cka}
\begin{equation}
W(z)=4z^2\int_0^1dv\int_0^1du\frac{(1-v)(uv)^3}{[zv(1-uv)+(1-u)(1-v)]^2}.
\end{equation}

\section{Loop Integrations for Higgs Production and Decay}
\label{app:loopb}
The loop functions for Higgs production and decay are \cite{Djouadi:2005gi,Djouadi:2005gj}
\begin{eqnarray}
\mathcal{A}_0(x)&=&\frac{x-I(x)}{x^2},\\
\mathcal{A}_1(x)&=&-\frac{x+(x-1)I(x)}{x^2},\\
\mathcal{A}_2(x)&=&\frac{2x^2+3x+3(2x-1)I(x)}{x^2},\\
\mathcal{B}_1(x)&=&-2\frac{I(x)}{x},
\end{eqnarray}
where
\begin{equation}
I(z)=\left\{\begin{array}{cc}\arcsin^2\left(\sqrt{z}\right),&z\leq1,\\
-\frac{1}{4}\left[\ln\left(\frac{1+\sqrt{1-z^{-1}}}{1-\sqrt{1-z^{-1}}}\right)-\textrm{i}\pi\right]^2,&z>1.\end{array}\right.
\end{equation}

\section{Decay of Heavy (Pseudo)scalars}
\label{app:heavy}
For heavy neutral (pseudo)scalars, we  consider the decay channels $H_{2,3}\rightarrow t\bar{t},WW,ZZ$ and $ZH_1$. The partial decay widths are given by
\begin{eqnarray}
\Gamma_{H_i\rightarrow t\bar{t}}&=&\frac{3m_im^2_t}{8\pi v^2}\left[[\textrm{Re}(c_{t,i})]^2\left(1-\frac{4m^2_t}{m^2_i}\right)^{\frac{3}{2}}
+[\textrm{Im}(c_{t,i})]^2\left(1-\frac{4m^2_t}{m^2_i}\right)^{\frac{1}{2}}\right],\\
\Gamma_{H_i\rightarrow WW}&=&\frac{m^3_ic^2_{V,i}}{16\pi v^2}\sqrt{1-\frac{4m^2_W}{m^2_i}}\left(1-\frac{4m^2_W}{m^2_i}+\frac{12m^4_W}{m^4_i}\right),\\
\Gamma_{H_i\rightarrow ZZ}&=&\frac{m^3_ic^2_{V,i}}{32\pi v^2}\sqrt{1-\frac{4m^2_Z}{m^2_i}}\left(1-\frac{4m^2_Z}{m^2_i}+\frac{12m^4_Z}{m^4_i}\right),\\
\Gamma_{H_i\rightarrow ZH_1}&=&\frac{m^3_ic^2_{V,k}}{32\pi v^2}\mathcal{F}_{VS}\left(\frac{m^2_Z}{m^2_i},\frac{m^2_i}{m^2_i}\right).
\end{eqnarray}
Here $k\neq i$ or $1$, and the functions
\begin{equation}
\mathcal{F}_{VS}(x,y)=(1+x^2+y^2-2x-2y-2xy)^{\frac{3}{2}}.
\end{equation}
In Scenario (b), since $H_{2,3}$ have large mass splitting, we should also consider the $H_3\rightarrow ZH_2$ decay. Its partial width is
\begin{equation}
\Gamma_{H_3\rightarrow ZH_2}=\frac{m^3_3c^2_{V,1}}{32\pi v^2}\mathcal{F}_{VS}\left(\frac{m^2_Z}{m^2_3},\frac{m^2_2}{m^2_3}\right).
\end{equation}
Thus numerically the total decay widths $\Gamma_{2,3}$ can reach about $20~\textrm{GeV}$ if $m_{2,3}\simeq500~\textrm{GeV}$,
and they both dominantly decay to $t\bar{t}$. In Scenario (b),
if $m_2=500~\textrm{GeV}$ and $m_3=650~\textrm{GeV}$, $\textrm{Br}_{H_3\rightarrow ZH_2}$ can reach about $10\%$.

The charged Higgs boson $H^+$ decays mainly to $t\bar{b}$ in the small $t_{\beta}$ region. Ignoring the coupling term proportional to $m_b$, we have
\begin{equation}
\Gamma_{H^+\rightarrow t\bar{b}}=\frac{3m_{\pm}}{8\pi v^2}\left(\frac{m_t}{t_{\beta}}\right)^2\left(1-\frac{m^2_t}{m^2_{\pm}}\right)^2.
\end{equation}
Besides this, $H^+$ also have subdominant decay channels, like $W^+H_i$ \cite{Arhrib:2010ju}, yielding
\begin{equation}
\Gamma_{H^+\rightarrow W^+H_i}=\frac{m^3_{\pm}\left(1-c^2_{V,i}\right)}{16\pi v^2}\mathcal{F}_{VS}\left(\frac{m^2_W}{m^2_{\pm}},\frac{m^2_i}{m^2_{\pm}}\right).
\end{equation}
For $\beta=0.76$ and $m_{\pm}=600~\textrm{GeV}$, $\Gamma_{H^+\rightarrow t\bar{b}}=33~\textrm{GeV}$ while  the sum for all three neutral scalars $\sum_i\Gamma_{H^+\rightarrow W^+H_i}\lesssim5~\textrm{GeV}$ for $|\alpha_2|\lesssim0.27$.

\section{Top Quark Reconstruction}
\label{app:toprec}

\begin{figure}[!t]
    \caption{Validation plots for the \textsc{PseudoTop} reconstruction method in $t\bar{t}H_1(\to b\bar{b})$ (green) and the QCD-mediated $t\bar{t}b\bar{b}$ (red). Here, we show the absolute value of the rapidity of the top quark (upper left), the transverse momentum of the $t\bar{t}$ system (upper right), the invariant mass of the $t\bar{t}$ system (middle left), the one of the reconstructed top (anti)quark (middle right), the one of the Higgs boson (lower left) and the one of the $t\bar{t}H_1$ system (lower right).}\label{fig:recons}
    \centering
    \includegraphics[width=0.49\linewidth]{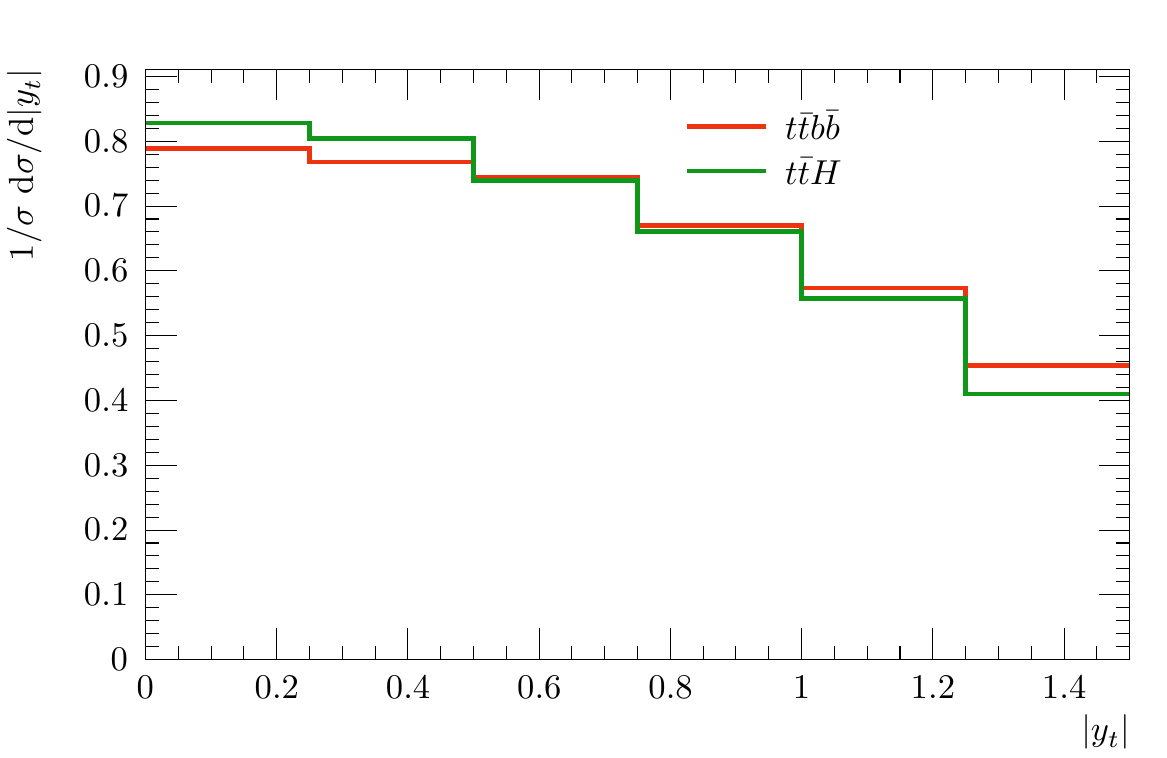}
    \hfill
    \includegraphics[width=0.49\linewidth]{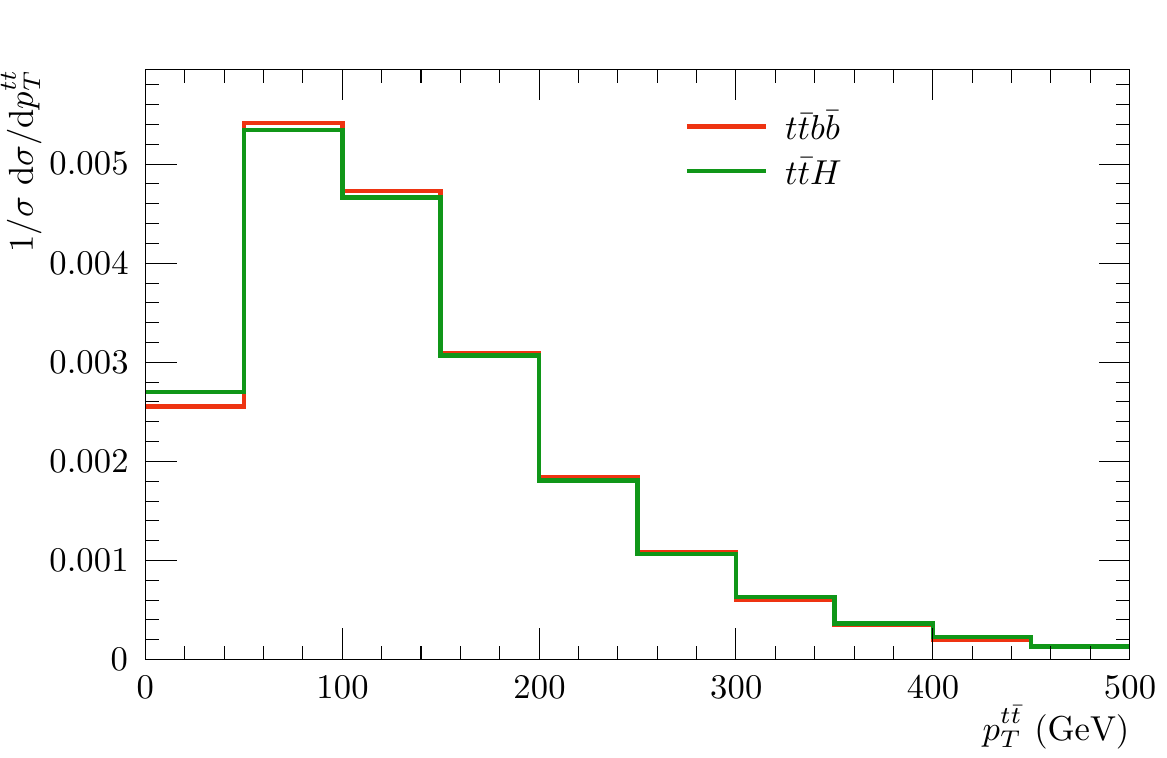}
    \vfill
    \includegraphics[width=0.49\linewidth]{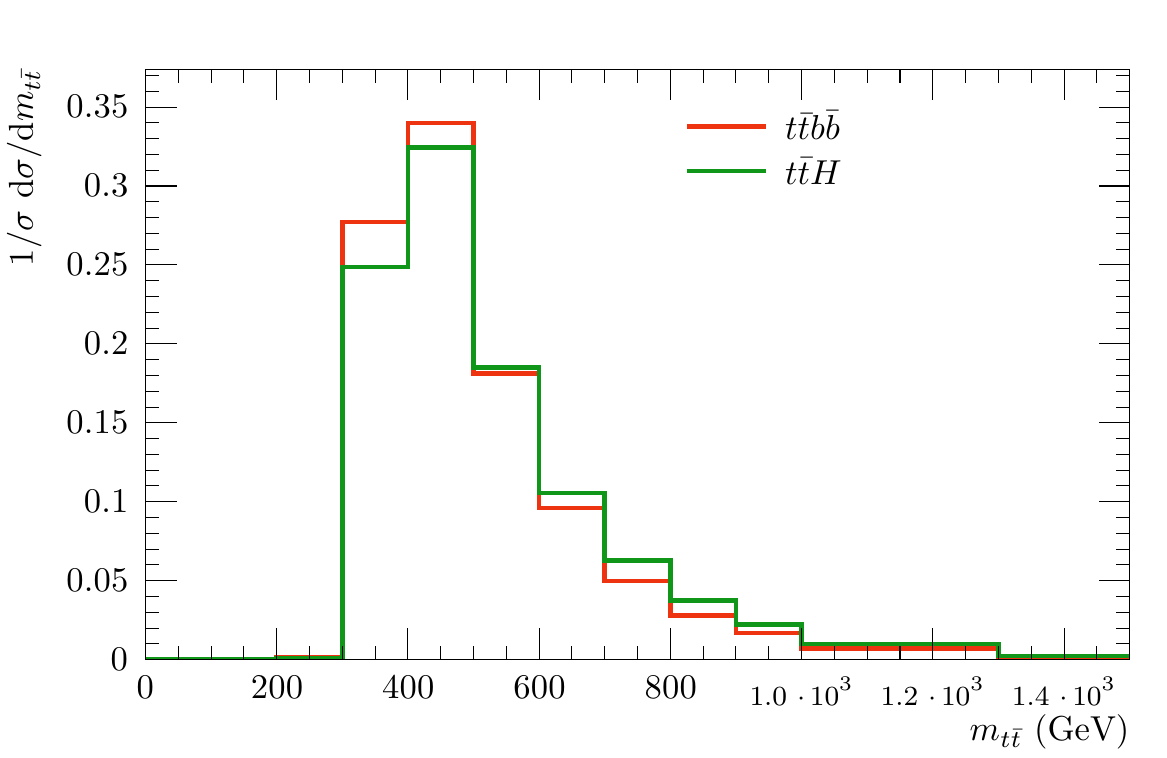}
    \hfill
    \includegraphics[width=0.49\linewidth]{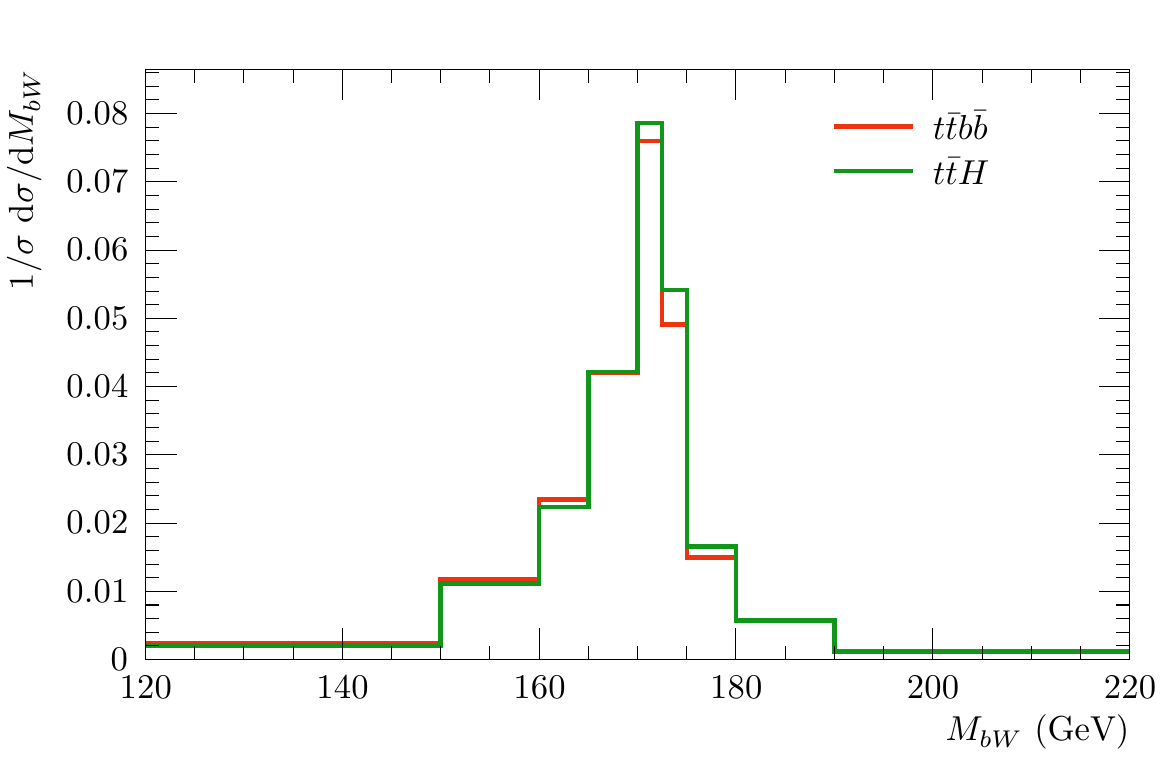}
        \vfill
    \includegraphics[width=0.49\linewidth]{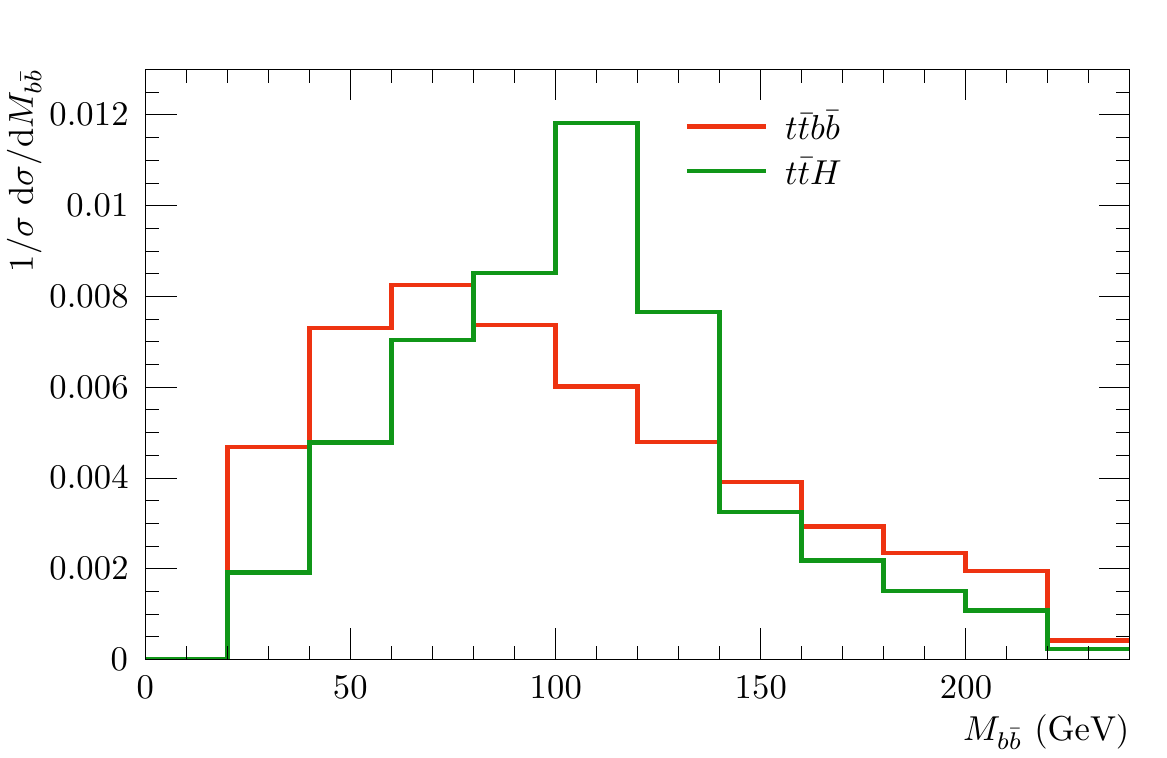}
    \hfill
    \includegraphics[width=0.49\linewidth]{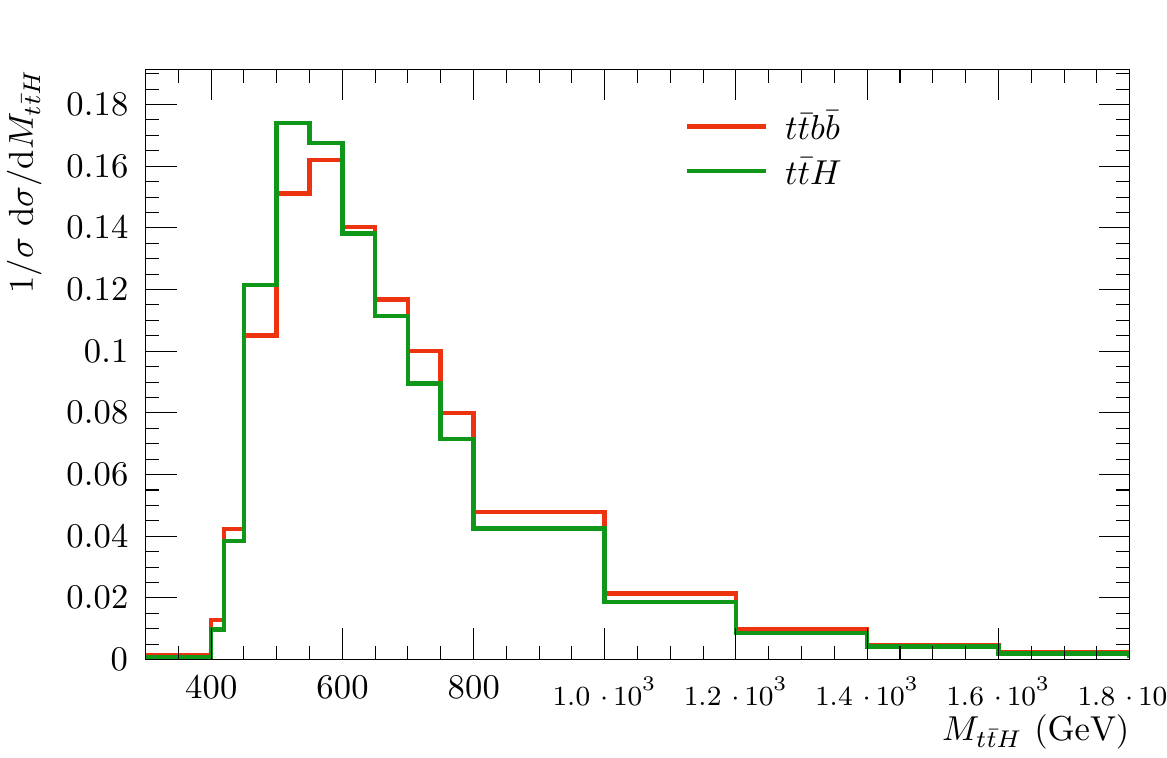}
\end{figure}

For $t\bar{t}$ spin-spin correlation and polarization observables in the top quark rest frame, it is mandatory to fully reconstruct the top (anti)quark four-momentum. In this regard, we employ the \textsc{PseudoTop} definition \cite{Collaboration:2267573} widely used by the ATLAS and the CMS collaborations for, e.g., validation of MC event generators. We slightly modify the \textsc{Rivet} implementation of the CMS measurement of the $t\bar{t}$ differential cross section at $\sqrt{s}=8$ TeV \cite{Khachatryan:2015oqa}. We minimize the following quantity
\begin{equation}
\label{top-reconstruction}
K^2=\left(M_{\Tilde{t}_\ell} - m_t \right)^2 + \left( M_{j_1 j_2} - m_W \right)^2 + \left( M_{\Tilde{t}_h} - m_t \right)^2 + \left(M_{\Tilde{p}_{H_1}} - m_{H_1} \right)^2,
\end{equation}
to select the hadronic, leptonic (anti)top quarks and SM-like Higgs boson decaying into $b\bar{b}$. In \autoref{top-reconstruction}, $m_t, m_W~\mathrm{and}~ m_H$ are the  masses of the top quark,  $W$ boson and the Higgs boson, respectively, while $\Tilde{t}_\ell$($\Tilde{t}_h$) is the momentum of the (anti)top constructed in the leptonic(hadronic) decays of the $W$ boson, with $\Tilde{p}_{H_1}$  the four-momentum of the Higgs boson candidate. In the reconstruction procedure, all  jets and leptons in the event are considered provided they satisfy the selection criteria which was highlighted in \autoref{sec:setup}. Validation plots for the \textsc{PseudoTop} reconstruction method in $t\bar{t}H_1(\to b\bar{b})$ (green) and the QCD-mediated $t\bar{t}b\bar{b}$ (red) are shown in \autoref{fig:recons}.


\bibliographystyle{JHEP}
\bibliography{CPVTHDM}

\end{document}